\documentclass[aps,prb,reprint]{revtex4-2}
\usepackage{amsmath,amssymb,graphicx,numprint,times,adjustbox, multirow, dcolumn,mathtools,xspace}
\usepackage[usenames,dvipsnames,svgnames,table]{xcolor}
\usepackage[colorlinks=true, urlcolor=blue, linkcolor=blue, citecolor=blue, pdftex]{hyperref}
\usepackage{orcidlink}

\npthousandsep{\,}
\npdecimalsign{.}

\newcolumntype{L}{>{$}l<{$}} 
\newcolumntype{u}{D{.}{.}{12}}
\newcolumntype{w}{D{.}{.}{8}}
\newcolumntype{m}{D{.}{.}{7}}
\newcolumntype{q}{D{.}{.}{7}}
\newcolumntype{a}{D{.}{.}{4}}

\def\<{\langle}
\def\>{\rangle}
\newcommand{\varEpsilon}{\mathcal{E}}
\newcommand{\zbar}{\bar{z}}
\newcommand{\chidof}{\ensuremath{\chi^2/\text{d.o.f.}}\xspace}
\newcommand{\Zt}{$\mathbb Z_2$}
\newcommand{\xmin}{\ensuremath{x_{\text{min}}}\xspace}
\newcommand{\xLmax}{\ensuremath{(x/L)_{\text{max}}}\xspace}
\newcommand{\Lmin}{\ensuremath{L_{\text{min}}}\xspace}
\newcommand{\zmin}{\ensuremath{z_{\text{min}}}\xspace}
\newcommand{\zLmax}{\ensuremath{(z/L)_{\text{max}}}\xspace}
\newcommand{\prrlsection}[1]{{\it #1.}}

\newcommand{\switchtoletter}[1]{
\setcounter{equation}{0}
\renewcommand{\theHequation}{#1\arabic{equation}}
\renewcommand{\theequation}{#1\arabic{equation}}

\setcounter{figure}{0}
\renewcommand{\theHfigure}{#1\arabic{figure}}
\renewcommand{\thefigure}{#1\arabic{figure}}

}
\newcounter{appcounter}
\newcommand{\prrlappendix}[1]{
\addtocounter{appcounter}{1}
\switchtoletter{\Alph{appcounter}}
{\it Appendix \Alph{appcounter}: #1.}}

\newcommand{\EPL}{EPL}

\newcommand{\PRL}{Phys.~Rev.~Lett.}

\newcommand{\PRB}{Phys.~Rev. B}

\newcommand{\PRE}{Phys.~Rev. E}
\newcommand{\PRX}{Phys.~Rev. X}

\newcommand{\JPCM}{J.~Phys.: Condens.~Matter}
\newcommand{\JPAOLD}{J.~Phys.~A: Math.~Gen.}
\newcommand{\JPA}{J.~Phys.~A: Math.~Theor.}
\newcommand{\physrep}{Phys.~Rep.}

\newcommand{\ZPB}{Z.~Phys.~B}

\newcommand{\NPB}{Nucl.~Phys.~B}

\newcommand{\RMP}{Rev.~Mod.~Phys}

\begin{document}

\title{Boundary operator product expansion coefficients of the three-dimensional Ising universality class}
\author{\firstname{Dorian} \surname{Przetakiewicz},\orcidlink{0009-0009-4749-0418}}
\email{dorian.przetakiewicz@rwth-aachen.de}
\affiliation{\mbox{Institute for Theoretical Solid State Physics, RWTH Aachen University, Otto-Blumenthal-Str. 26, 52074 Aachen, Germany}}
\author{\firstname{Stefan} \surname{Wessel},\orcidlink{0000-0002-6353-5083}}
\email{wessel@physik.rwth-aachen.de}
\affiliation{\mbox{Institute for Theoretical Solid State Physics, RWTH Aachen University, Otto-Blumenthal-Str. 26, 52074 Aachen, Germany}}
\author{\firstname{Francesco} \surname{Parisen Toldin}\,\orcidlink{0000-0002-1884-9067}}
\email{parisentoldin@physik.rwth-aachen.de}
\affiliation{\mbox{Institute for Theoretical Solid State Physics, RWTH Aachen University, Otto-Blumenthal-Str. 26, 52074 Aachen, Germany}}

\begin{abstract}
Recent advances in conformal field theory and critical phenomena have focused on the characterization of boundary or defects in a conformally invariant system.
In this Letter we study the critical behavior of the three-dimensional Ising universality class in the presence of a surface, realizing the ordinary, the special, and the normal universality classes.
By combining high-precision Monte Carlo simulations of an improved model, where leading scaling corrections are suppressed, with a finite-size scaling analysis informed by conformal field theory,
we determine unbiased, accurate estimates of universal boundary operator product expansion coefficients of experimental relevance.
Furthermore, we improve the value of the scaling dimension of the surface field at the special transition by the estimate $\hat{\Delta}_\sigma = 0.3531(3)$.
\end{abstract}

\maketitle

\prrlsection{Introduction}
The modern theory of critical phenomena is a cornerstone of contemporary physics, bridging key concepts from fundamental physics, statistical mechanics, and condensed matter physics. A central focus is the characterization of universal quantities---properties that assume specific values within a given universality class (UC), regardless of the underlying microscopic details of a system. Well-known examples include critical exponents, amplitude ratios, and coefficients in operator product expansions (OPE). The latter in fact encode the renormalization group (RG) flow near fixed points, and are fundamental ingredients in the characterization of conformal invariance at a critical point \cite{Cardy-book}.
In many realistic scenarios, it is necessary to extend beyond the standard field-theoretical and RG approaches to bulk criticality and consider the effects of boundaries, such as those encountered in surface critical phenomena \cite{Diehl-86}. Recent years have seen significant advances in our understanding of critical
many-body systems with boundaries, such as classical \cite{PT-20,HDL-21,PTM-21,SHL-23,PT-23,PTKM-24,SJ-25} and quantum magnets \cite{ZW-17,DZG-18,WPTW-18,WW-19,WW-20,ZDZG-20,XPXZ-21,DZGZ-21,SLL-22,SL-22,WZG-22,YHSXDZ-22,LSHUO-24}, and gapless quantum systems with topologically protected boundary states \cite{GV-12,BQ-14,BMF-15,CCBCN-15,SPV-17,PSV-18,JXWX-20,TVV-21,Verresen-20,VTJP-21,WP-23,WP-23b,MWX-24}.
While the overall phase diagrams of many such systems have been well characterized through field-theoretical and computational analyses, new and unexpected RG fixed points, such as the extraordinary log surface transition, have recently been predicted \cite{Metlitski-20} and explored both within the RG and numerically \cite{PT-20,HDL-21,PTM-21,PKMGM-21,SLL-22,SHL-23,PTKM-24}.
Simultaneously, conformal field theory (CFT) has provided insights into various universal properties, such as boundary OPE (BOPE) coefficients, for critical three-dimensional systems with surfaces. 
Using advanced approximate techniques, such as the truncated conformal bootstrap (TCB) \cite{Gliozzi-13} and the fuzzy sphere (FZ) construction \cite{AHHHH-23}, some of these OPE coefficients have been estimated recently \cite{GLMR-15,ZZ-24}. 
However,  unbiased estimates for these universal numbers often still remain unknown, even for the fundamental UC of the three-dimensional Ising model, which corresponds to scalar $\phi^4$ theory. In this case, different characteristic surface UCs emerge, depending on the surface enhancement of
interactions.
For surface interactions preserving the \Zt symmetry, the ordinary UC and the extraordinary UC arise for weak and strong surface enhancement, respectively, with the special UC separating them.
A finite surface-ordering field explicitly breaking the \Zt symmetry realizes
the normal UC,
equivalent to the extraordinary UC \cite{Diehl-94}.
The presence of a surface restricts the conformal symmetry to the subgroup of transformations that leave the boundary invariant \cite{Cardy-84}.
Accordingly, one distinguishes between {\it bulk} and {\it surface} operators, the latter admitting a usual OPE.
Close to the surface, bulk operators admit a BOPE \cite{DD-81,*DD-81_erratum,MO-95}, whose form is fixed by CFT.
\begin{figure}[t]
  \centering
  \includegraphics[width=0.9\linewidth,keepaspectratio]{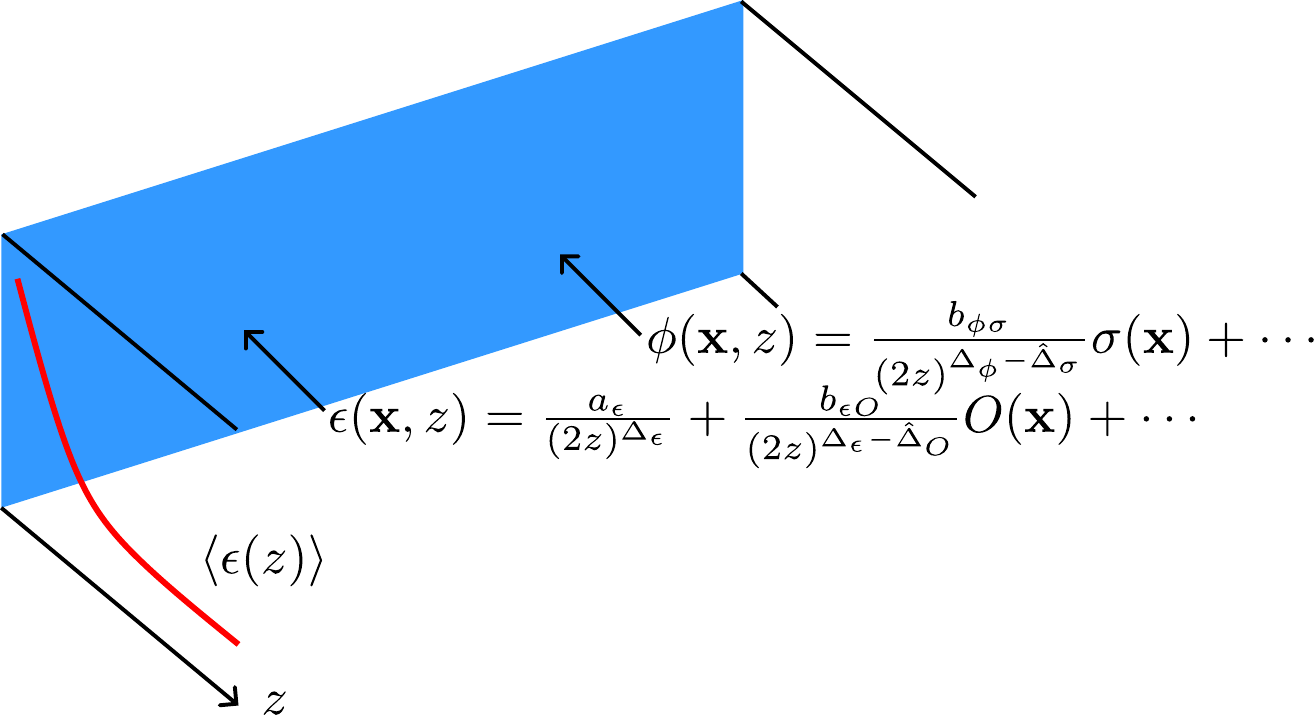}
  \caption{Illustration of the BOPE.}
  \label{fig:bope}
\end{figure}
\renewcommand{\arraystretch}{1.1}
\begin{table*}[t]
\caption{Universal BOPE coefficients.
The quoted uncertanties for TCB\cite{TCB_note} and FZ results are only estimates of their systematic errors.}
    \begin{tabular*}{\textwidth}{@{\extracolsep{\fill}}aaaaaal}
    \hline\hline
    \multicolumn{7}{c}{Ordinary UC} \\
    \multicolumn{1}{c}{$b_{\phi\sigma}$} & \multicolumn{1}{c}{$a_\epsilon$} & \multicolumn{1}{c}{$b_{\epsilon D}$} & \multicolumn{1}{c}{$\beta_{\phi\sigma D}$} & \multicolumn{1}{c}{$\lambda_{\sigma\sigma D}$} & \multicolumn{1}{c}{$C_D$}   & \multicolumn{1}{l}{Method and references}  \\
    \hline
    0.870(3) & -0.753(3) & -0.84(1) & 1.05(3) & 1.44(7) & 0.0101(3) & MC, this work \\
    0.869(7) & -0.750(3) & & & & & TCB \cite{GLMR-15} \\
    0.87(2)  & -0.74(4)  & -0.92(4) & & & 0.0089(2) & FZ \cite{ZZ-24} \\
    \multicolumn{7}{c}{Special UC} \\
    \multicolumn{1}{c}{$b_{\phi\sigma}$} & \multicolumn{1}{c}{$a_\epsilon$} & \multicolumn{1}{c}{$b_{\epsilon \varepsilon}$} & \multicolumn{1}{c}{$\beta_{\phi\sigma\varepsilon}$} & \multicolumn{1}{c}{$\lambda_{\sigma\sigma \varepsilon}$} & \multicolumn{1}{c}{$\lambda_{\varepsilon\varepsilon\varepsilon}$}   & \multicolumn{1}{l}{Method and references}  \\
    \hline
    1.435(3) & 1.160(4) & 3.09(5) & 0.79(1) & 0.86(1) & 1.02(2) & MC, this work \\
    \end{tabular*}
    \begin{tabular*}{\textwidth}{@{\extracolsep{\fill}}aaaaal}
    \multicolumn{6}{c}{Normal UC} \\
    \multicolumn{1}{c}{$a_\phi$} & \multicolumn{1}{c}{$b_{\phi D}$} & \multicolumn{1}{c}{$a_\epsilon$} & \multicolumn{1}{c}{$b_{\epsilon D}$} & \multicolumn{1}{c}{$C_D$}  & \multicolumn{1}{l}{Method and references}  \\
    \hline
    2.6143(5) & 0.242(2)   & 6.679(6) & 1.69(1)  & 0.198(3) & MC, this work \\
    2.60(5)   & 0.244(8)   &          &          & 0.193(5) & MC \cite{PTD-10,PTM-21} \\
    2.599(1)  & 0.25064(6) & 6.607(7) & 1.742(6) & 0.182(1) & TCB \cite{GLMR-15} \\
    2.58(16)  & 0.254(17)  & 6.4(9)   & 1.74(22) & 0.176(2) & FZ \cite{ZZ-24} \\
    \hline\hline
    \end{tabular*}
    \label{tab:results}
\end{table*}
Here, we bridge the gap between such a general structure and the actual universal coefficients of such OPEs for the three-dimensional Ising surface UCs
by combining accurate and unbiased Monte Carlo (MC) simulations of an improved model and a CFT-motivated scaling analysis.
The bulk Ising UC features
two relevant operators: the \Zt-even energy operator $\epsilon$ and the \Zt-odd magnetization operator $\phi$.
Introducing coordinates $({\bf x}, z)$, where $\bf x$ ($z$) is parallel (perpendicular) to the surface, the BOPE for the ordinary and special UC reads \cite{MO-95}
\begin{align}
    \phi({\bf x}, z) &\underset{z\rightarrow 0}{=} \frac{b_{\phi\sigma}}{(2z)^{\Delta_\phi-\hat{\Delta}_\sigma}}\sigma({\bf x}) +\cdots,
    \label{phi_bope}
    \\
    \epsilon({\bf x}, z) &\underset{z\rightarrow 0}{=} \frac{a_\epsilon}{(2z)^{\Delta_\epsilon}}+
    \frac{b_{\epsilon O}}{(2z)^{\Delta_\epsilon-\hat{\Delta}_O}} O({\bf x})+
    \cdots.
    \label{e_bope}
\end{align}
Here, $\sigma$ is the lowest-lying \Zt-odd surface operator, associated with the surface magnetization, with dimension $\hat{\Delta}_\sigma$, while
$O$ is the lowest-lying \Zt-even surface operator with dimension $\hat{\Delta}_O$.
The BOPE is illustrated in Fig.~\ref{fig:bope}.
In addition, by fusing $\sigma({\bf 0})$ with $\phi({\bf 0}, z)$ one obtains \cite{PTM-21}
\begin{equation}
    \sigma({\bf 0}) \phi({\bf 0}, z)\underset{z\rightarrow 0}{=} b_{\phi\sigma}\frac{2^{\hat{\Delta}_\sigma-\Delta_\phi}}{z^{\hat{\Delta}_\sigma+\Delta_\phi}} \left[1 + \beta_{\phi\sigma O} z^{\hat{\Delta}_O} O({\bf 0}) + \cdots\right].
    \label{phi-sigma_bope}
\end{equation}
In Eqs.~(\ref{phi_bope})--(\ref{phi-sigma_bope}) $a_\epsilon$, $b_{\epsilon O}$, $b_{\phi\sigma}$ and $\beta_{\phi\sigma O}$ are universal BOPE coefficients. 
Dots indicate subleading contributions arising from descendants of $\sigma$ and $O$, as well as surface operators of higher dimension.
In the ordinary UC there are no relevant \Zt-even operators, and the leading operator $O$ in Eq.~(\ref{e_bope}) is the displacement operator $D$, with dimension $\hat{\Delta}_D=3$ \cite{BC-87,BGLM-16}.
In the special UC, the leading operator $O$ is instead the relevant surface \Zt-even operator $\varepsilon$, associated with the deviation of the enhancement of surface interactions from its critical value.

In the normal UC there are no relevant operators and the lowest surface operator is the displacement $D$.
The \Zt symmetry is broken on the boundary, so that a classification of surface operators in terms of \Zt symmetry does not apply.
Accordingly, Eq.~(\ref{phi_bope}) is modified as follows:
\begin{equation}
    \phi({\bf x}, z) \underset{z\rightarrow 0}{=} \frac{a_\phi}{(2z)^{\Delta_\phi}} + \frac{b_{\phi D}}{(2z)^{\Delta_\phi-3}} D({\bf x}) +\cdots,
    \label{phi_bope_normal}
\end{equation}
with $a_\phi$ and $b_{\phi D}$ universal BOPE coefficients.
The BOPE of the energy operator holds as in Eq.~(\ref{e_bope}), with $O=D$.

The various BOPE coefficients---the main target of this work---are summarized in Table \ref{tab:results}. They are extracted from the one- and two-point correlations that follow from Eqs.~(\ref{phi_bope})-(\ref{phi_bope_normal}).
In particular, in the ordinary and special UCs, we use the surface-bulk two-point correlator with leading scaling
\begin{equation}
    \< \sigma({\bf 0}) \phi({\bf 0}, z) \> = b_{\phi\sigma}\frac{2^{\hat{\Delta}_\sigma-\Delta_\phi}}{z^{\hat{\Delta}_\sigma+\Delta_\phi}} \left[1 + \beta_{\phi\sigma O} z^{\hat{\Delta}_O} \<O({\bf 0})\> \right],
    \label{surface_bulk_cft}
\end{equation}
while in the normal UC we exploit the one-point function with
\begin{equation}
    \< \phi({\bf x}, z) \> = \frac{a_\phi}{(2z)^{\Delta_\phi}} + \frac{b_{\phi D}}{(2z)^{\Delta_\phi-3}} \< D({\bf x})\>.
    \label{phi_profile_cft}
\end{equation}
Further, for all surface UC considered, the one-point function of the energy operator obeys
\begin{equation}
    \< \epsilon({\bf x}, z) \> = \frac{a_\epsilon}{(2z)^{\Delta_\phi}} + \frac{b_{\epsilon O}}{(2z)^{\Delta_\phi-\hat{\Delta}_O}} \< O({\bf x})\>.
    \label{e_profile_cft}
\end{equation}

\prrlsection{Model}
We simulate the Blume-Capel model \cite{Blume-66,Capel-66} on a three-dimensional lattice. Its reduced Hamiltonian is
\begin{equation}
    {\cal H} = -\beta \sum_{\langle x x'\rangle} S_{x}S_{x'} + \delta\sum_{x}S_{x}^2, \qquad S_{x}=-1,0,1,
    \label{BC}
\end{equation}
such that the Gibbs weight is $\exp(-{\cal H})$.
In Eq.~(\ref{BC}) the first sum extends over nearest-neighbor sites and  the second sum over all lattice sites.
In the limit $\delta\rightarrow -\infty$ the Hamiltonian (\ref{BC}) reduces to the Ising model.
In the $(\beta, \delta)$ plane, the model exhibits a line of continuous phase transitions in the Ising universality class  for $\delta < \delta_{\rm tri}$, terminating at a tricritical point at $\delta_{\rm tri} = 2.006(8)$~\cite{Deserno-97,ZFJ-15}; see also Refs.~~\cite{HB-98,DB-04} for previous determinations of $\delta_{\rm tri}$. For $\delta > \delta_{\rm tri}$ the model undergoes a first-order phase transition.
At $\delta=0.656(20)$~\cite{Hasenbusch-10} the Hamiltonian is ``improved'' \cite{PV-02}, i.e., the leading scaling corrections $\propto L^{-\omega}$, with $\omega=0.832(6)$~\cite{Hasenbusch-10} are suppressed.
As in previous MC studies \cite{Hasenbusch-10c,PTD-10,Hasenbusch-11,PTTD-13,PTTD-14,Hasenbusch-14,PT-13,PTAW-17}, here we fix $\delta=0.655$ and $\beta=\beta_c=\numprint{0.387721735}(25)$~\cite{Hasenbusch-10}, realizing an improved critical model in the Ising UC.

\prrlsection{Results}
In order to determine the normalization of bulk fields, we have first simulated the model on a $L\times L\times L$ lattice, with periodic boundary conditions (BCs).
We sample the local order parameter $S_x$ and the local energy $E_x$, defined as
\begin{equation}
    E_x \equiv S_{x} \sum_{\substack{\text{$x'$ n. n.} \\ \text{of $x$}}} S_{x'} = S_{x} \sum_{n=1}^{3} (S_{x + e_n} + S_{x - e_n}),
    \label{Edef_periodic}
\end{equation}
where $e_n$ is the unit vector in direction $n$, so that the sum extends over all nearest neighbor sites of $x$.
The expectation value of $E_x$ and its finite-size amplitude is obtained by fitting \cite{Young_notes} $E_{\rm bulk} \equiv (1/L^3)\sum_{x}\<E_x\>$ to
\begin{equation}
    E_{\rm bulk} = E_0 + U_E L^{-\Delta_\epsilon},
    \label{E_fit}
\end{equation}
where $\Delta_\epsilon = \numprint{1.412625}(10)$ \cite{KPSDV-16} is the scaling dimension of the relevant even operator, related to the standard exponent $\nu$ by $\Delta_\epsilon = 3-1/\nu$ \cite{Cardy-book}.
See Appendix A for a discussion of the scaling forms used in this work.
Fits of MC data \cite{SM} allow us to infer estimates of the coefficients $E_0$ and $U_E$. In Table \ref{tab:fitresults} we report them, together with other amplitudes discussed below.
We fit the two-point function of $S_x$ and $E_x$ to \cite{PTM-21}
\begin{gather}
    \langle S_{x} S_{0}\rangle = \frac{{\cal N}_S^2}{x^{2\Delta_\phi}} \left[1 + B_{\phi\phi} \left(\frac{x}{L}\right)^{\Delta_\epsilon} + Cx^{-2}\right], \label{2pt_fit_S} \\
    \langle E_{x} E_{0}\rangle_c = \frac{{\cal N}_E^2}{x^{2\Delta_\epsilon}} \left[1{+}B_{\epsilon\epsilon} \left(\frac{x}{L}\right)^{\Delta_\epsilon}{+}B'_{\epsilon\epsilon} \left(\frac{x}{L}\right)^{2\Delta_\epsilon}{+}Cx^{-2}\right], \label{2pt_fit_E}
\end{gather}
where $\Delta_\phi=0.518\,1489(10)$ \cite{KPSDV-16} is the scaling dimension of the relevant odd operator, related to the standard exponent $\eta$ by $\Delta_\phi = (1+\eta)/2$ \cite{Cardy-book} and
the subscript $c$ indicates the connected part of the correlations.
Equations (\ref{2pt_fit_S}) and (\ref{2pt_fit_E}) are valid for $(x/L)\ll 1$ and $x \gtrsim x_0$, with $x_0$ a nonuniversal length, associated with  short-distance behavior.
Incidentally,
from the results in Table \ref{tab:fitresults}
we can compute the universal finite-size amplitude of the energy operator $u_\epsilon = U_E / {\cal N}_E = 3.220(6)$ and the bulk OPE coefficients $\lambda_{\phi\phi\epsilon} = B_{\phi\phi} / u_\epsilon=1.052(3)$, $\lambda_{\epsilon\epsilon\epsilon} = B_{\epsilon\epsilon}/u_\epsilon = 1.55(2)$, in line with MC $\lambda_{\phi\phi\epsilon} = 1.051(1)$, $\lambda_{\epsilon\epsilon\epsilon} = 1.533(5)$ \cite{Hasenbusch-18}, and conformal bootstrap $\lambda_{\phi\phi\epsilon} = 1.051\,8537(41)$, $\lambda_{\epsilon\epsilon\epsilon} = \numprint{1.532435}(19)$ \cite{KPSDV-16} estimates.

\begin{table}[t]
  \centering
  \caption{Estimates of the scaling amplitudes as obtained from fits of MC data \cite{SM}. For every set of amplitudes we indicate the corresponding UC and the scaling form.}
    \begin{ruledtabular}
    \begin{tabular}{lLLr}
    UC & \multicolumn{2}{c}{Fit results} & Equation \\
    \hline
    \multirow{4}{*}{Bulk}         & E_0 = \numprint{1.204234}(8)           & U_E = 3.496(6)                 & (\ref{E_fit}) \\
                                  & {\cal N}_S^2 = \numprint{0.18566}(7)   & B_{\phi\phi} = 3.386(7)        & (\ref{2pt_fit_S}) \\
                                  & {\cal N}_E^2 = 1.1790(8)      & B_{\epsilon\epsilon} = 5.00(8) & \multirow{2}{*}{(\ref{2pt_fit_E})} \\
                                  & B'_{\epsilon\epsilon} = -9(1)  &                              & \\
    \hline
    \multirow{5}{*}{Ordinary}     & {\cal N}_\sigma^2 = 0.383(1)  & B_{\sigma\sigma} = 4.3(2)      & (\ref{2pt_fit_Ssurf}) \\
                                  & A_\epsilon = -0.545(2)        & B_\epsilon=-26.8(8)            & \multirow{2}{*}{(\ref{fss_1ptE_open})} \\
                                  & z_0=1.01(3)                   &                                & \\
                                  & M_{\phi\sigma} = 0.392(1)     & B_{\phi\sigma}=3.15(9)         & \multirow{2}{*}{(\ref{fss_2pt_open})} \\
                                  & z_0=0.99(2)                   &                                & \\
    \hline
    \multirow{7}{*}{Special}      & {\cal N}_\sigma^2 = 0.3544(6) & B_{\sigma\sigma} = 1.49(1)     & (\ref{2pt_fit_Ssurf}) \\
                                  & A_\epsilon = 0.840(3)         & B_\epsilon=5.92(7)             & \multirow{2}{*}{(\ref{fss_1ptE_open})} \\
                                  & z_0=0.53(2)                   &                                & \\
                                  & M_{\phi\sigma} = 0.3282(6)     & B_{\phi\sigma}=1.38(1)         & \multirow{2}{*}{(\ref{fss_2pt_open})} \\
                                  & z_0=0.50(2)                   &                                & \\
                                  & \varEpsilon_0 = \numprint{1.32291}(1)  & U_{\varEpsilon} =2.12(2)       & (\ref{Esurf_fit}) \\
                                  & {\cal N}_\varepsilon^2 = 1.48(2) & B_{\varepsilon\varepsilon} = 1.77(3) & (\ref{2pt_fit_Esurf}) \\
    \hline
    \multirow{4}{*}{Normal $(+,o)$} & A_\epsilon = 4.835(4)       & B_\epsilon =-2.68(8)           & \multirow{2}{*}{(\ref{fss_1ptE_open})} \\
                                  & z_0 = 1.43(1)                 &                                & \\
                                  & A_\phi = \numprint{1.12644}(6) & B_\phi = -1.000(15)            & \multirow{2}{*}{(\ref{fss_1ptS_normal})} \\
                                  & z_0 = 1.437(3)                &                                & \\
    \end{tabular}
    \end{ruledtabular}
    \label{tab:fitresults}
\end{table}
To realize the various boundary UCs we generalize the Hamiltonian (\ref{BC}), imposing open BCs on one direction, and periodic BC in the remaining two, thus realizing a $L\times L\times L$ lattice with two surfaces ${\cal S}_\downarrow$ and ${\cal S}_\uparrow$, each of size $L^2$.
The reduced Hamiltonian thus reads
\begin{equation}
\begin{split}
    {\cal H} = &-\beta \sideset{}{’}\sum_{\langle x\ x'\rangle} S_{x}S_{x'} + \delta\sum_{x}S_{x}^2\\
    &-\beta_s \sum_{\mathclap{\langle x\ x'\rangle \in {\cal S}_\downarrow \cup {\cal S}_\uparrow}} S_{x}S_{x'}
    +h_\downarrow \sum_{x\in {\cal S}_\downarrow}S_x+h_\uparrow \sum_{x\in {\cal S}_\uparrow}S_x,
    \label{BCopen}
\end{split}
\end{equation}
where the first sum extends over the nearest-neighbor sites for which at least one site belongs to the bulk, the second over all sites, the third sum over all nearest-neighbor pairs on the boundaries, and the fourth and fifth sums each extend over a single surface.
In Eq.~(\ref{BCopen}) we have imposed the same coupling $\beta_s$ on both surfaces, while we allow for different boundary fields $h_\downarrow$ and $h_\uparrow$ on the two surfaces.
The various surface UC studied here are realized by setting the bulk couplings $\beta$, $\delta$ to the improved critical point $\delta=0.655$, $\beta=\beta_c=\numprint{0.387721735}(25)$~\cite{Hasenbusch-10}, and then tuning the boundary parameters $\beta_s$, $h_\downarrow$, $h_\uparrow$.
Due to the reduced translation symmetry, we adopt a slightly different definition for the local energy observable $E_{{\bf x}, z}$, summing over nearest-neighbor sites along the directions parallel to the surfaces only:
\begin{equation}
    E_{{\bf x}, z} \equiv S_{{\bf x}, z} \sum_{n=1}^{2} (S_{{\bf x} + e_n, z} + S_{{\bf x} - e_n, z}).
    \label{Edef_open}
\end{equation}

We first consider the ordinary UC, which is realized by setting $\beta_s=\beta$ and $h_\downarrow = h_\uparrow = 0$.
To extract the normalization of the boundary field, we fit the surface two-point function to
\begin{equation}
   \langle S_{{\bf x}, 0} S_{{\bf 0}, 0}\rangle = \frac{{\cal N}_\sigma^2}{x^{2\hat{\Delta}_\sigma}} \left[1 + B_{\sigma\sigma} \left(\frac{x}{L}\right)^{\hat{\Delta}} + C x^{-2}\right],
   \label{2pt_fit_Ssurf}
\end{equation}
where $\hat{\Delta}_\sigma = 2 - 0.7249(6)$ \cite{Hasenbusch-11} is the scaling dimension of the lowest odd surface operator $\sigma$ and
$\hat{\Delta}$ is the dimension of the leading even surface operator appearing in the BOPE $\sigma \times \sigma$, which is the displacement $D$, with 
$\hat{\Delta}_D=3$ \cite{BC-87,BGLM-16}.
We fit the energy profile to
\begin{equation}
    \langle E_{{\bf x}, z} \rangle = \frac{2}{3}E_0
    + \frac{{A_\epsilon}}{(2\zbar)^{\Delta_\epsilon}}\left[1+B_\epsilon\left(\frac{\zbar}{L}\right)^{\hat{\Delta}} + \frac{C}{\zbar^{2}}\right],
  \label{fss_1ptE_open}
\end{equation}
where $\zbar \equiv z+z_0$, with $z_0$ a model-dependent correction-to-scaling amplitude \cite{PTM-21},
$\hat{\Delta}=3$ and we fix $E_0$ to the estimate given in Table \ref{tab:fitresults}; the factor $2/3$ in front of $E_0$ accounts for the different definitions used in the bulk [Eq.~(\ref{Edef_periodic})] and with open BCs [Eq.~(\ref{Edef_open})].
The BOPE coefficient $b_{\phi\sigma}$ is extracted from the surface-bulk two-point function, which we fit to
\begin{equation}
  \langle S_{{\bf x}, 0} S_{{\bf x}, z}\rangle
  = \frac{M_{\phi\sigma}}{\zbar^{\hat{\Delta}_\sigma+\Delta_\phi}}\left[1 + B_{\phi\sigma}\left(\frac{\zbar}{L}\right)^{\hat{\Delta}} + \frac{C}{\zbar^{2}}\right],
  \label{fss_2pt_open}
\end{equation}
where $\zbar \equiv z+z_0$, $\hat{\Delta}_\sigma = 2 - 0.7249(6)$ \cite{Hasenbusch-11} is the dimension of the lowest-lying odd surface operator, and as in Eq.~(\ref{fss_1ptE_open}) $\hat{\Delta}=3$.

The special UC is realized in the model (\ref{BCopen}) by setting $\beta_s= \numprint{0.54914}(2)$ \cite{Hasenbusch-11b}.
The analysis of the MC data is analogous to the ordinary UC, with the important difference in the presence
of a {\it relevant} surface \Zt-even operator $\varepsilon$,
so that in Eqs.~(\ref{2pt_fit_Ssurf})--(\ref{fss_2pt_open})
we have
$\hat{\Delta}=\hat{\Delta}_{\varepsilon}=2-0.718(2)$ \cite{Hasenbusch-11b}.
In this work we have determined a new estimate of the scaling dimension of the lowest odd operator \cite{SM}
\begin{equation}
    \hat{\Delta}_\sigma = 0.3531(3),
    \label{delta_sigma_special}
\end{equation}
improving an available estimate $\hat{\Delta}_\sigma = 0.3535(6)$ \cite{Hasenbusch-11b}.
In fitting the surface-surface and surface-bulk correlations, we use in Eqs. ~(\ref{2pt_fit_Ssurf}) and (\ref{fss_2pt_open}) the value given in Eq.~(\ref{delta_sigma_special}).
For the special UC, we also study the scaling behavior of the surface energy.
We fit $E_{\text{surf}} \equiv (1/L^2)\sum_{\bf x} \< E_{{\bf x}, 0}\>$ to
\begin{equation}
    E_{\rm surf} = \varEpsilon_0 + U_{\varEpsilon} L^{-\hat{\Delta}_{\varepsilon}}.
    \label{Esurf_fit}
\end{equation}
Further, to extract the normalization of the surface energy we fit its two-point function to
\begin{equation}
   \langle E_{{\bf x}, 0} E_{{\bf 0}, 0}\rangle = \frac{{\cal N}_\varepsilon^2}{x^{2\hat{\Delta}_\epsilon}} \left[1 + B_{\varepsilon\varepsilon} \left(\frac{x}{L}\right)^{\hat{\Delta}_\varepsilon} + C x^{-2}\right].
   \label{2pt_fit_Esurf}
\end{equation}
To realize the normal UC, we set $\beta_s=\beta$ and $h_\downarrow=\infty$ in Eq.~(\ref{BCopen}), thereby polarizing a surface with fixed spins $S=1$.
For the opposite surface, we have considered two cases: a surface with $h_\uparrow=0$, realizing the ordinary UC, and a surface identical to the lower one, with $h_\uparrow=\infty$.
We refer to these BCs as $(+,o)$, and $(+,+)$, respectively.
The determination of the BOPE coefficients in the $(+,+)$ BCs delivers results somewhat less precise, though in full agreement with those obtained with $(+,o)$ BCs. Here, we discuss the analysis of the $(+,o)$ case \cite{SM}.
In the normal UC there are no relevant surface operators, and the lowest-lying operator entering in Eq.~(\ref{fss_1ptE_open}) is the displacement $D$, with $\hat{\Delta}=3$.
Furthermore, 
due to the broken \Zt symmetry, a  nonvanishing order-parameter profile emerges in this case, which we fit to
\begin{equation}
  \langle S_{{\bf x}, z} \rangle
     = \frac{{A_\phi}}{(2\zbar)^{\Delta_\phi}}\left[1{+}B_\phi\left(\frac{\zbar}{L}\right)^{3} + \frac{C}{\zbar^{2}}\right],\quad \zbar \equiv z+z_0.
    \label{fss_1ptS_normal}
\end{equation}
\prrlsection{Discussion}
As we discuss in Appendix A, using the fitted amplitudes of Table \ref{tab:fitresults} one can determine the  BOPE coefficients introduced above. In particular, $b_{\phi\sigma} = M_{\phi\sigma} 2^{\Delta_\phi-\hat{\Delta}_\sigma} / ({\cal N}_S{\cal N}_\sigma)$ [Eq.~(\ref{fss_1ptE_open})], $a_\epsilon = (3/2)A_\epsilon / {\cal N}_E$ [Eq.~(\ref{fss_2pt_open})], and, for the normal UC, $a_\phi = A_\phi / {\cal N}_\phi$ [Eq.~(\ref{fss_1ptS_normal})].
In the ordinary and normal UC, the finite-size corrections are proportional to the amplitude $u_D$ of the displacement operator $D$.
This enters also in the boundary expansion of the energy-momentum tensor $T_{zz}  \underset{z\rightarrow 0}{=}-\sqrt{C_D} D$, where $C_D$ is the universal norm coefficient \cite{Cardy-90}.
On a finite geometry, $\< T_{zz} \> = \Theta / L^3$, where $\Theta$ is the universal amplitude of the critical Casimir force \cite{FG-78,Krech-94,Krech-99,BDT-00,Gambassi-09,GD-11,MD-18,DD-23}.
The universal norm $C_D$ can be related to the finite-size correction of the one-point function of the energy and, for the normal UC, of the order parameter \cite{Cardy-90,PTM-21,normalization}
\begin{align}
    & C_D = -\frac{2^d \Delta_\epsilon\Theta}{S_d B_\epsilon} = -\frac{2\Delta_\epsilon\Theta}{\pi B_\epsilon},
    \label{CD_thetaE}\\
    & C_D = -\frac{2^d \Delta_\phi\Theta}{S_d B_\phi} = -\frac{2\Delta_\phi\Theta}{\pi B_\phi},
    \label{CD_thetaS}
\end{align}
where $S_d = 2\pi^{d/2}/\Gamma(d/2)$ and $d=3$.
Employing the value of $\Theta$ reported in Appendix B, one can determine $C_D$, as well as the amplitude $u_D = \Theta / \sqrt{C_D}$; for the ordinary UC we obtain $u_D = 3.00(5)$.
In the case of the normal UC, Eqs.~(\ref{CD_thetaE}) and (\ref{CD_thetaS}) provide two different estimators for $C_D$.
In Table \ref{tab:results} we quote $C_D$ as obtained from Eq.~(\ref{CD_thetaS}), which is more precise and in full agreement with that obtained from Eq.~(\ref{CD_thetaE}) $C_D=0.202(6)$. The finite-size amplitude $u_D$ is found to be $u_D=1.34(2)$ [using Eq.~(\ref{CD_thetaE})] and $u_D=1.35(1)$ [using Eq.~(\ref{CD_thetaS})].
The BOPE coefficients $b_{\epsilon D}$ and $b_{\phi D}$ are extracted from $a_\epsilon$ and $a_\phi$ with the identities \cite{Cardy-90}
\begin{align}
    &\frac{b_{\epsilon D}}{\Delta_\epsilon a_\epsilon} = \frac{1}{S_d \sqrt{C_D}},
    \label{wardE}\\
    &\frac{b_{\phi D}}{\Delta_\phi a_\phi} = \frac{1}{S_d \sqrt{C_D}}.
    \label{wardS}
\end{align}
Again, in the case of the normal UC, employing on the right-hand side of Eqs.~(\ref{wardE}) and (\ref{wardS}) $C_D$ as extracted from the energy operator [Eq.~(\ref{CD_thetaE})] gives results in full agreement with, though slightly less precise than, those quoted in Table \ref{tab:results}.
Other BOPE coefficients can be extracted as $\beta_{\phi\sigma D} = B_{\phi\sigma} / u_D$ and $\lambda_{\sigma\sigma D} = B_{\sigma\sigma} / u_D$.
In the special UC, finite-size corrections are unrelated to the displacement operator and are due to the lowest-lying \Zt-even operator $\varepsilon$.
Its finite-size amplitude $u_\varepsilon$ can be conveniently obtained from results in Table \ref{tab:fitresults} as $u_\varepsilon = U_\varEpsilon / {\cal N}_\varepsilon = 1.74(2)$.
Then, BOPE coefficients are obtained as $b_{\epsilon\varepsilon} = 2^{-\hat{\Delta}_O}B_\epsilon a_\epsilon / u_\varepsilon$, $\beta_{\phi\sigma\varepsilon} = B_{\phi\sigma} / u_\varepsilon$ and $\lambda_{\sigma\sigma\varepsilon} = B_{\sigma\sigma} / u_\varepsilon$.
The finite-size scaling (FSS) analysis of the surface energy correlations [Eq.~(\ref{2pt_fit_Esurf})] allows us to extract a further OPE coefficient $\lambda_{\varepsilon\varepsilon\varepsilon} = B_{\varepsilon\varepsilon} / u_\varepsilon$. The remaining BOPE coefficients $a_\epsilon$ and $b_{\phi\sigma}$ are obtained as for the ordinary UC. 
In Table \ref{tab:results} we summarize our results and compare them with available results in the literature.
For several quantities we are not aware of previous calculations.
Whenever a
comparison with previous works is available, we generally observe a reasonable agreement, with a small but significant deviation especially in the estimates of $C_D$.
In the case of the normal UC, our results significantly improve the precision of the universal coefficients obtained in Ref.~\cite{PTM-21} by reanalyzing some MC results of Ref.~\cite{PTD-10}.
We stress that, different from the TCB and FZ techniques, our method, based on a combination of MC simulations and a FSS analysis, is numerically exact.
As outlined above, and in the Supplemental Material \cite{SM}, for some BOPE independent estimates deliver perfectly consistent values, thus underscoring their robustness and the reliability of the quoted uncertainties: We expect our results to provide a benchmark for future studies.
Future improvements might be obtained by searching for an improved model where also the leading boundary irrelevant operator, the displacement $D$, is suppressed.
Such a realization, if at all possible, requires however a fine-tuning of surface interactions.
Some of the coefficients of Table \ref{tab:results} are experimentally accessible, in particular those of the normal UC, obtained in critical adsorption experiments \cite{FD-95,Law-01,FYP-05,BGL-15,normalization_exp}.
We are not aware of experimental determinations for other UCs, though in principle coefficients could be accessed by spatially-resolved probes, e.g., x-ray scattering under grazing incidence \cite{DH-95}.
The technique used here can be clearly employed for other boundary UCs.
In particular, it would be interesting to extend the study to the $O(N)$ model, $N>1$.
While a surface ordinary UC is always present, the special transition exists only for $N \le N_c$ \cite{Metlitski-20}, with $N_c \approx 5$ \cite{PKMGM-21}.
For the $O(N)$ UC, precise estimates of the critical exponents at the ordinary UC are available \cite{PT-23} and the exponents at the special transition have been numerically determined in Refs.~\cite{DBN-05,PT-20,Deng-06}, whereas
a numerically reliable determination of BOPE coefficients is presently not available.

\prrlsection{Acknowledgments}
  We are grateful to Marco Meineri for useful discussions and comments on the manuscript.
  We thank Max Metlitski for insightful comments on the manuscript.
  F.P.T. thanks Martin Hasenbusch for useful communications.
  F.P.T. is funded by the Deutsche Forschungsgemeinschaft (DFG, German Research Foundation), Project No. 414456783.
  We gratefully acknowledge the Gauss Centre for Supercomputing e.V. \footnote{\href{www.gauss-centre.eu}{www.gauss-centre.eu}} for funding this project by providing computing time through the John von Neumann Institute for Computing (NIC) on the GCS Supercomputer JUWELS at Jülich Supercomputing Centre (JSC).
  We gratefully acknowledge computing time granted by the IT Center of RWTH Aachen University and used for the calculations of critical Casimir amplitude.

\prrlsection{Data availability}
The data that support the findings of this article are openly available \cite{PWPT-25_dataset}.
  
\prrlappendix{Scaling forms}
A crucial ingredient for a numerically reliable determination of universal quantities is the formulation of precise scaling ansatzes, where one identifies
the leading corrections to the scaling behavior of critical observables.
Here, this is accomplished by a scaling analysis motivated by the CFT describing the UCs studied here, and its connection to the simulated lattice model.
Most of the scaling forms used in this work in order to fit the MC data are generalizations of those derived in the Supplemental Material of Ref.~\cite{PTM-21}.
Here, we briefly recall the essential steps in this derivation.
Lattice observables are generally expanded in terms of CFT fields belonging to the same symmetry sector.
In particular, for bulk observables,
\begin{equation}
  \begin{split}
    S_{x} &= {\cal N}_S \left(1 + c_S \vec{\nabla}^2 \right) \phi,\\
    S^2_{x} - \langle S^2_{x} \rangle &= {\cal N}_{S^2} \left(1 + c_{S^2} \vec{\nabla}^2 \right) \epsilon,\\
    E_{x}- \langle E_{x} \rangle &= {\cal N}_E \left(1 + c_E \vec{\nabla}^2 \right) \epsilon,
  \end{split}
\label{cft_exp_bulk}
\end{equation}
where the constants ${\cal N}_S$, ${\cal N}_{S^2}$, ${\cal N}_E$, $c_S$, $c_{S^2}$ and $c_E$ are model-dependent.
With this expansion, Eq.~(\ref{E_fit}) follows immediately by taking the expectation value on a finite volume. Moreover, the finite-size amplitude $U_E$ is given by $U_E={\cal N}_E u_\epsilon$,  
where $u_\epsilon$ is the universal finite-size amplitude of the energy operator,
\begin{equation}
    \<\epsilon\> = \frac{u_\epsilon}{L^{\Delta_\epsilon}}\, .
\end{equation}
We notice that, using Eq.~(\ref{cft_exp_bulk}), one can in principle expect an additional correction $\propto L^{-2-\Delta_\epsilon}$ to the right-hand side of Eq.~(\ref{E_fit}): Our fit analysis shows that such a term is negligible within the precision of MC data.
Given that also $S^2$ couples to the energy operator, a FSS form for $S^2_{\rm bulk} \equiv (1/V) \sum_x S^2_s$ analogous to Eq.~(\ref{E_fit}) holds
\begin{equation}
    S^2_{\rm bulk} = S^2_0 + U_{S^2} L^{-\Delta_\epsilon},
    \label{S2_fit}
\end{equation}
where $U_{S^2} = {\cal N}_{S^2} u_\epsilon$.

Using the bulk OPEs  $\phi \times \phi = \lambda_{\phi\phi\epsilon} \epsilon$, and $\epsilon \times \epsilon = \lambda_{\epsilon\epsilon\epsilon} \epsilon$,
one obtains Eqs.~(\ref{2pt_fit_S}) and (\ref{2pt_fit_E}) \cite{PTM-21}, as well as an analogous formula for the correlations of $S^2$:
\begin{equation}
  \langle S^2_{x} S^2_{0}\rangle_c = \frac{{\cal N}_{S^2}^2}{x^{2\Delta_\epsilon}} \left[1 + B_{\epsilon\epsilon} \left(\frac{x}{L}\right)^{\Delta_\epsilon} + C x^{-2}\right].
  \label{2pt_fit_S2}
\end{equation}
We notice that the subleading correction $\propto (x/L)^{2\Delta_\epsilon}$ in Eq.~(\ref{2pt_fit_E}) is due to the fact that we consider the connected part of the correlations: Such a term is absent in Eq.~(\ref{2pt_fit_S}).
In principle, a subleading term  $\propto (x/L)^{2\Delta_\epsilon}$ is also present in Eq.~(\ref{2pt_fit_S2}): Due to a reduced precision of the $S^2$ correlations, our analysis shows that such a correction is not relevant \cite{SM}.
In Eqs.~(\ref{2pt_fit_S}), (\ref{2pt_fit_E}) and (\ref{2pt_fit_S2}), the finite-size coefficients are related to the OPE constants by
\begin{equation}
    \begin{split}
        B_{\phi\phi} &= \lambda_{\phi\phi\epsilon}u_\epsilon,\\
        B_{\epsilon\epsilon} &= \lambda_{\epsilon\epsilon\epsilon} u_\epsilon.
    \end{split}
    \label{B_from_OPE}
\end{equation}
Moreover, the subleading amplitude $B'_{\epsilon\epsilon}$ is related to the finite-size amplitude of the energy by $B'_{\epsilon\epsilon} = - u_\epsilon^2$, which holds within the available numerical precision.
Combining the expansion (\ref{cft_exp_bulk}) with the BOPE (\ref{phi_bope_normal}) we derive
in Ref.~\cite{PTM-21} Eq.~(\ref{fss_1ptS_normal}) and show that $A_\phi = {\cal N}_S a_\phi$.
The scaling form of the energy profile (\ref{fss_1ptE_open}) follows analogously, and we have $A_\epsilon = (2/3){\cal N}_E a_\epsilon$; here, as in  Eq.~(\ref{fss_1ptE_open}), the factor $2/3$ is due to the fact that in the definition of energy profile [Eq.~(\ref{Edef_open})] we sum over two lattice directions, as opposed to the definition for the bulk observable [Eq.~(\ref{Edef_periodic})], where the sum is over all three directions.
Further, the finite-size correction is related to BOPEs as
$B_{\epsilon} = 2^{\hat{\Delta}_O}b_{\epsilon O} u_O/ a_\epsilon$, with
$u_O$ the universal finite-size amplitude of the leading boundary operator $O$ in Eq.~(\ref{e_bope}):
\begin{equation}
    \<O\> = \frac{u_O}{L^{\hat{\Delta}_O}}.
    \label{fss_1pt_surf}
\end{equation}
The profile of the $S^2$ observable, being coupled to the energy operator [Eq.~(\ref{cft_exp_bulk})] satisfies a scaling form identical to 
Eq.~(\ref{fss_1ptE_open})
\begin{multline}
    \langle S^2_{{\bf x}, z} \rangle = S^2_0\\
    + \frac{{A_{S^2}}}{(2(z+ z_0))^{\Delta_\epsilon}}\left[1+B_\epsilon\left(\frac{z+z_0}{L}\right)^{\hat{\Delta}} + C (z+z_0)^{-2}\right],
    \label{fss_1ptS2_open}
\end{multline}
with $S^2_0$ extracted by Eq.~(\ref{S2_fit}).
Analogous to the above discussion, we have $A_{S^2} = {\cal N}_{S^2} a_\epsilon$.
We notice that $B_\epsilon$ in Eqs.~(\ref{fss_1ptE_open}) and (\ref{fss_1ptS2_open}) are the same universal amplitude.
At the special transition, the observable $S^2$ on the surface couples to the \Zt-even operator $\varepsilon$; hence its surface average $S^2_{\text{surf}} \equiv (1/L^2)\sum_{\vec{x}} \<S^2_{\vec{x},0}\>$
satisfies a finite-size behavior analogous to Eq.~(\ref{Esurf_fit})
\begin{equation}
S^2_{\text{surf}} = \Sigma_0 + U_{\sigma^2} L^{-\hat{\Delta}_{\varepsilon}}.
\label{S2surf_fit}
\end{equation}

Correlations involving surface quantities follow from a similar expansion of the lattice observables in boundary CFT fields:
\begin{equation}
  \begin{split}
    S_{{\bf x}, 0} &= {\cal N}_\sigma \left(1 + c_\sigma \vec{\nabla}^2 \right) \sigma,\\
    S^2_{{\bf x}, 0} - \langle S^2_{{\bf x}, 0} \rangle &= {\cal N}_{\sigma^2} \left(1 + c_{\sigma^2} \vec{\nabla}^2 \right) O,\\
    E_{{\bf x}, 0}- \langle E_{{\bf x}, 0} \rangle &= {\cal N}_O \left(1 + c_O \vec{\nabla}^2 \right) O,
  \end{split}
\label{cft_exp_surf}
\end{equation}
where $O$ is the lowest-lying \Zt-even surface operator, $O=D$ the displacement operator in the ordinary and in the normal UC, and $O=\varepsilon$ the relevant boundary \Zt-even operator in the special UC.
From this expansion and the OPE of $\sigma\times\sigma$, it follows that $B_{\sigma\sigma}$ in Eq.~(\ref{2pt_fit_Ssurf}) is related to the OPE coefficient $\lambda_{\sigma\sigma O}$ by
\begin{equation}
    B_{\sigma\sigma} = \lambda_{\sigma\sigma O} u_O.
\end{equation}
Further, using Eq.~(\ref{cft_exp_surf}) one obtains Eq.~(\ref{2pt_fit_Esurf}) and an analogous scaling form for the two-point function of $S^2$ on the surface at the special UC
\begin{equation}
   \langle S^2_{{\bf x}, 0} S^2_{{\bf 0}, 0}\rangle = \frac{{\cal N}_{\sigma^2}^2}{x^{2\hat{\Delta}_\epsilon}} \left[1 + B_{\varepsilon\varepsilon} \left(\frac{x}{L}\right)^{\hat{\Delta}_\varepsilon} + C x^{-2}\right].
   \label{2pt_fit_S2surf}
\end{equation}
We notice that the $B_{\varepsilon\varepsilon}$ in Eqs.~(\ref{2pt_fit_Esurf}) and (\ref{2pt_fit_S2surf}) are the same amplitude, related to the OPE coefficient $\lambda_{\varepsilon\varepsilon\varepsilon}$ by
\begin{equation}
    B_{\varepsilon\varepsilon} = \lambda_{\varepsilon\varepsilon\varepsilon} u_\varepsilon,
\end{equation}
where $u_\varepsilon$ is the universal finite-size amplitude of $\varepsilon$ [see Eq.~(\ref{fss_1pt_surf})].
Finally, the scaling behavior of the surface-bulk correlation (\ref{fss_2pt_open}) follows from Eqs.~(\ref{cft_exp_bulk}) and (\ref{surface_bulk_cft}), with $M_{\phi\sigma} = b_{\phi\sigma} 2^{\hat{\Delta}_\sigma-\Delta_\phi}{\cal N}_S{\cal N}_\sigma $ and $B_{\phi\sigma} = \beta_{\phi\sigma O} u_O$ \cite{PTM-21}.

\prrlappendix{Critical Casimir amplitude}
Given a three-dimensional system $L_\parallel\times L_\parallel\times L$ with lateral size $L_\parallel$, the critical Casimir force $F_C$ is defined by
\begin{equation}
    F_C \equiv -\frac{\partial \left[L\left({\cal F}^{\text{(s)}}-f_{\text{bulk}}^{\text{(s)}}\right)\right]}{\partial L}\Bigg|_{L_\parallel,T},
\end{equation}
where ${\cal F}^{\text{(s)}}$ is the singular part of the free energy per volume $L_\parallel^2 L$ and in units of $k_B T$, and $f_{\text{bulk}}^{\text{(s)}}$ its thermodynamic limit $f_{\text{bulk}}^{\text{(s)}} = \lim_{L_\parallel,L\rightarrow \infty} {\cal F}^{\text{(s)}}$.
At a critical point $T=T_c$ and for homogeneous BCs, or in the absence of a relevant length,
\begin{equation}
    F_C = \frac{\Theta}{L^3}.
\end{equation}
At the critical point, ${\cal F}^{\text{(s)}}(T=T_c) = A/L^3$.
The critical Casimir amplitude $\Theta$ is related to the finite-size amplitude $A$ of ${\cal F}^{\text{(s)}}$ by $\Theta=2A$.

To compute the critical Casimir amplitude we employ the so-called coupling-parameter approach \cite{VGMD-07}.
In essence, this consists in a combination of MC sampling and a numerical integration, yielding the free energy difference between systems with thicknesses $L$ and $L-1$ and identical lateral sizes.
We refer to Refs.~\cite{PTD-10,PTTD-13} for a discussion on the implementation.
For the numerical integration we use the Gauss-Kronrod method with 41/20 nodes, allowing for an accurate convergence of the integral; by comparing the resulting quadrature with 41 and 20 nodes we checked that the systematic error of the quadrature is negligible with respect to the statistical error bar.
The free energy difference $I(L)$ at criticality is then fitted to
\begin{equation}
    I(L) = F_{\rm bulk}(\beta_c) + \frac{\Theta}{(L-1/2)^3} + \frac{C}{(L-1/2)^4},
    \label{fit_casimir}
\end{equation}
where $F_{\rm bulk}(\beta_c)$ is the bulk free energy density at criticality, $\Theta$ is the desired Casimir amplitude, and the last term is due to $1/L$ scaling corrections.
We note that for the BCs considered here, previous MC studies have considered the limit of infinite lateral size $L_\parallel$, whereas in this work we have chosen a geometry with equal sizes on all directions.
Indeed, like any other FSS quantity, the Casimir amplitude depends on the aspect ratio $\rho=L/L_\parallel$ of the system.
The dependence on $\rho$ for three-dimensional models in the Ising universality class and periodic BCs has been studied in Ref.~\cite{Dohm-09} with field-theoretical methods and in Ref.~\cite{HGS-11} with MC simulations.
Furthermore, the aspect-ratio dependence has been investigated for nonhomogenous BCs, in the presence of a chemical step and in the limit $\rho\rightarrow 0$ in Refs.~\cite{PTD-10,PTTD-14}.
In this situation, the chemical step forms a line defect and  gives rise to a linear $\rho$ dependence of the critical Casimir force on $\rho$, for $\rho\rightarrow 0$, allowing to extract its contribution to the Casimir force.
In two dimensions, the aspect-ratio dependence of the Casimir force in Ising models with various BCs has been investigated in Refs.~\cite{HH-17,Hucht-17,HH-19,COBH-24}.

\begin{table}
    \caption{Fits of the critical Casimir amplitude [Eq.~(\ref{fit_casimir})] as a function of the minimum lattice size \Lmin taken into account, for three BCs considered here; $\text{d.o.f.}$ denotes the number of degrees of freedom of a fit.}
    \begin{ruledtabular}
    \begin{tabular}{lumma}
    \multicolumn{5}{c}{BCs: (o,o)}\\
    \multicolumn{1}{l}{\Lmin} & \multicolumn{1}{c}{$F_{\rm bulk}(\beta_c)$} & \multicolumn{1}{c}{$\Theta$} & \multicolumn{1}{c}{$C$} & \multicolumn{1}{c}{\chidof} \\
    \hline
     8 &  0.075736842(44) &  0.3005(14) &  -0.184(11) & 0.8 \\
    12 &  0.075736839(51) &  0.3009(37) &  -0.189(44) & 1.1 \\
    16 &  0.075736895(57) &  0.2907(66) &  -0.01(11)  & 0.1 \\
    \hline \\[0.1em]
    \multicolumn{5}{c}{BCs: (+,o)}\\
    \multicolumn{1}{l}{\Lmin} & \multicolumn{1}{c}{$F_{\rm bulk}(\beta_c)$} & \multicolumn{1}{c}{$\Theta$} & \multicolumn{1}{c}{$C$} & \multicolumn{1}{c}{\chidof} \\
    \hline
     8 &  0.075737047(23) &  0.5818(11) &  -1.1665(89) & 20.9 \\
    12 &  0.075736947(25) &  0.5959(19) &  -1.382(26)  & 1.9 \\
    16 &  0.075736923(29) &  0.6013(34) &  -1.496(64)  & 1.0 \\
    \hline \\[0.1em]
    \multicolumn{5}{c}{BCs: (+,+)}\\
    \multicolumn{1}{l}{\Lmin} & \multicolumn{1}{c}{$F_{\rm bulk}(\beta_c)$} & \multicolumn{1}{c}{$\Theta$} & \multicolumn{1}{c}{$C$} & \multicolumn{1}{c}{\chidof} \\
    \hline
     8 &  0.075736644(41) &  -0.7794(16) &  1.561(13) & 2.7 \\
    12 &  0.075736715(47) &  -0.7896(37) &  1.695(45) & 0.5 \\
    16 &  0.075736740(54) &  -0.7945(65) &  1.78(11)  & 0.3 \\
    \end{tabular}    
    \end{ruledtabular}
    \label{tab:casimir}
\end{table}
For three of the BCs studied here, we have computed $I(L)$ for lattice sizes $L=8,12,16,24,32,48$, and $64$.
Corresponding fits to Eq.~(\ref{fit_casimir}) are reported in Table \ref{tab:casimir}.
Inspecting the fit results we infer the following estimates of the critical Casimir amplitude $\Theta$:
\begin{align}
    \Theta &= 0.301(4) \qquad &\text{$(o,o)$ BCs},\\
    \Theta &= 0.601(4) &\text{$(+,o)$ BCs},\\
    \Theta &= -0.790(4) &\text{$(+,+)$ BCs}.
\end{align}
As noted above, all these amplitudes pertain to a model with aspect ratio $\rho=1$.
Interestingly, for $(o,o)$ BCs we obtain a positive, i.e., repulsive critical Casimir amplitude.
This is in stark contrast with slab limit $\rho=0$, where the amplitude is negative $\Theta=-0.030(5)$ \cite{PTTD-13}.
For $(+,o)$ BCs $\Theta$ is significantly larger than for $\rho=0$, where $\Theta=0.492(5)$ \cite{PTTD-13}.
In the case of $(+,+)$ BCs we find instead that the critical Casimir amplitude $\Theta$ is rather close to the $\rho=0$ limit, where $\Theta = -0.820(15)$ \cite{Hasenbusch-10c}.
The data in Table \ref{tab:casimir} allow us to estimate the critical bulk free energy density.
Considering the spread in the fitted values for different BCs, we conservatively estimate $F_{\rm bulk}(\beta_c) = 0.075\,7368(1)$, improving the estimate $F_{\rm bulk}(\beta_c) = 0.075\,7368(4)$ of Ref.~\cite{Hasenbusch-10c}.

\bibliography{francesco,sm}

\begin{thebibliography}{97}
\bibitem [{nor()}]{normalization}%
  \BibitemOpen
  \href@noop {} {}\bibinfo {note} {We adopt the normalization of the
  energy-momentum tensor used in Ref.~\cite{PTM-21}, which differs from that of
  Ref.~\cite{Cardy-90}. In our convention, $\<T_{zz}\>$ equals the critical
  Casimir force $\Theta/L^3$ (see also Appendix B). Compared to
  Ref.~\cite{Cardy-90}, our definition of the energy-momentum tensor
  incorporates a factor $-S_d^{-1}$, and the universal coefficient $c$ of
  Eq.~(1) in Ref.~\cite{Cardy-90} is related to the universal norm $C_D$ by
  $C_D = 4(d-1)/(S_d^2 d) c$.}\BibitemShut {Stop}%
\end{thebibliography}

\clearpage


\onecolumngrid
  \parbox[c][3em][t]{\textwidth}{\centering \large\bf Supplemental Material}
\smallskip
\twocolumngrid

\switchtoletter{S}

\section{Additional results}
\begin{table}[b]
  \centering
  \caption{Estimates of amplitudes involving the $S^2$ observable, as well as those obtained with $(+,+)$ BCs.}
    \begin{ruledtabular}
    \begin{tabular}{lLLr}
    UC & \multicolumn{2}{c}{Fit results} & Eq. \\
    \hline
    \multirow{2}{*}{bulk}         & S^2_0 = 0.6070252(7)          & U_{S^2} = 0.2573(5)           & (\ref{S2_fit}) \\
                                  & {\cal N}_{S^2}^2 = 0.00640(2) & B_{\epsilon\epsilon} = 4.9(1) & (\ref{2pt_fit_S2}) \\
    \hline
    \multirow{2}{*}{ordinary}     & A_{S^2} = -0.0606(5)          & B_\epsilon = -26.6(8)         & \multirow{2}{*}{(\ref{fss_1ptS2_open})} \\
                                  & z_0=1.04(5) & & \\
    \hline
    \multirow{4}{*}{special}      & A_{S^2} = 0.0927(4)           & B_\epsilon = 5.93(8)          & \multirow{2}{*}{(\ref{fss_1ptS2_open})} \\
                                  & z_0 =0.51(2) & & \\
                                  & \Sigma_0 = 0.669447(1)        & U_{\sigma^2} =  0.231(2)      & (\ref{S2surf_fit}) \\
                                  & {\cal N}_{\sigma^2}^2 = 0.0176(2) & B_{\varepsilon\varepsilon} = 1.76(3)  &  (\ref{2pt_fit_S2surf}) \\
    \hline
    normal $(+,o)$                & A_{S^2} = 0.5339(5)           & B_\epsilon = -2.70(8)         & (\ref{fss_1ptS2_open}) \\
    \hline
    \multirow{6}{*}{normal $(+,+)$} & A_\epsilon =4.838(6)          & B_\epsilon =3.6(1)            & \multirow{2}{*}{(\ref{fss_1ptE_open})} \\
                                    & z_0 =1.44(2) & & \\
                                    & A_{S^2} = 0.5342(7)           & B_\epsilon = 3.6(1)           & \multirow{2}{*}{(\ref{fss_1ptS2_open})} \\
                                    & z_0 =1.44(2) & & \\
                                    & A_\phi = 1.1268(2)            & B_\phi = 1.32(3)              & \multirow{2}{*}{(\ref{fss_1ptS_normal})} \\
                                    & z_0 = 1.455(5) & & \\
    \end{tabular}
    \end{ruledtabular}
    \label{tab:additional_fitresults}
\end{table}
As we discuss in Appendix A, both lattice observables $E_x$, defined in Eq.~(\ref{Edef_periodic}) and $S^2$ couple to the energy operator $\epsilon$, hence the analysis of the OPE coefficients involving  $\epsilon$ can be repeated using $S^2$.
In Table \ref{tab:additional_fitresults} we report estimates of the amplitudes extracted from fits of MC data involving the $S^2$ observable, for the bulk and boundary UCs discussed in this work.
We observe that universal finite-size amplitudes $B_{\epsilon\epsilon}$, $B_\epsilon$, $B_{\varepsilon\varepsilon}$ are in nice agreement with the estimates of Table \ref{tab:fitresults}, thus underscoring the reliability of the results.
Furthermore, for all surface UC the fitted value of $z_0$ matches that of Table \ref{tab:fitresults}: as expected, $z_0$ is a model-dependent amplitude, which, for a given model, is independent of the observable \cite{PTM-21}.

Using the amplitudes related to $S^2$ we obtain alternative estimates of some of the universal coefficients discussed in the text.
For the bulk UC, we compute $u_\epsilon=3.216(8)$ and $\lambda_{\epsilon\epsilon\epsilon}=1.52(3)$.
For the ordinary UC we obtain $a_\epsilon = -0.757(6)$, $C_D=0.0102(3)$, $u_D = 2.98(5)$, $b_{\epsilon D} = -0.84(2)$.
For the special UC we extract $a_\epsilon = 1.159(5)$, $u_\varepsilon = 1.74(2)$, $b_{\epsilon\varepsilon} = 3.09(5)$, $\lambda_{\varepsilon\varepsilon\varepsilon} = 1.01(2)$.
For the normal UC, as realized with $(+,o)$ BCs, we obtain $a_\epsilon=6.67(1)$, $u_D=1.34(2)$, $C_D=0.200(6)$; employing the latter in Eqs.~(\ref{wardE})-(\ref{wardS}), we obtain $b_{\epsilon D}=1.68(3)$ and $b_{\phi D}=0.241(4)$.
All these results are in full agreement with those in Table \ref{tab:results}, obtained using the energy observable.

Moreover, as mentioned in the main text, we have also realized the normal UC by setting an infinite field on both surface $h_\uparrow=h_\downarrow=\infty$, i.e., with $(+,+)$ BCs.
Together with the critical Casimir amplitude $\Theta_{++}$ determined in Appendix B, this set of simulations provides an independent set of estimates of the various universal BOPE coefficients.
Following the same procedure,
in Table \ref{tab:additional_fitresults} we report the fitted amplitudes.
We estimate $a_\phi=2.6151(7)$, $a_\epsilon=6.683(9)$ (using the energy observable), $a_\epsilon=6.68(1)$ (using the $S^2$ observable).
The universal norm $C_D$ and the finite-size amplitude $u_D$ are found $C_D=0.197(6)$, $u_D=1.78(3)$ [using the energy observable in Eq.~(\ref{wardE})],
$C_D=0.197(5)$, $u_D=1.78(2)$ [using Eq.~(\ref{wardS})],
$C_D=0.197(6)$, $u_D=1.78(3)$ [using the $S^2$ observable in Eq.~(\ref{wardE})].
Using any of these values for $C_D$, we obtain $b_{\epsilon D}=1.69(2)$ and $b_{\phi D}=0.243(3)$.
All these estimates are perfectly consistent with each other, and in full agreement with the estimates of Table \ref{tab:results} obtained with the $(+,o)$ BCs, thus underscoring the reliability of our results.

\section{Critical exponent at the special transition}
\begin{table}
\caption{Fits of surface two-point function at half-lattice distance in the special UC to Eq.~(\ref{fss_exp_special}), as a function of the minimum lattice size \Lmin taken into account.
In the upper part of the table we fit the data ignoring the scaling-correction term $cL^{-1}$.
}
\begin{ruledtabular}
\begin{tabular}{cclww}
\multicolumn{1}{c}{$L_{min}$} & \multicolumn{1}{c}{$A$} & \multicolumn{1}{c}{$\hat{\Delta}_\sigma$} & \multicolumn{1}{c}{$c$} & \multicolumn{1}{c}{\chidof} \\
\hline
32 & 0.99848(84) & 0.353877(91) &  & 1.87 \\ 
48 & 0.9965(12) & 0.35369(12) &  & 1.05 \\ 
64 & 0.9951(15) & 0.35355(15) &  & 0.66 \\
96 & 0.9935(20) & 0.35341(19) & & 0.39 \\
\hline
32 & 0.9896(31) & 0.35310(28) & 0.135(46) & 0.49 \\
48 & 0.9884(49) & 0.35300(42) &  0.161(95) & 0.59 \\
64 & 0.9905(69) & 0.35318(58) & 0.11(16) & 0.72 \\
\end{tabular}
\end{ruledtabular}
\label{tab:special_sigma-sigma_l2}
\end{table}
Using the MC data we improve the critical exponent of the surface magnetization at the special transition.
By a standard FSS technique,
we fit the two-point function of the surface at midsize distance $L/2$ to
\begin{equation}
    \langle S_{L/2, 0, 0} S_{{\bf 0}, 0}\rangle = AL^{-2\hat{\Delta}_\sigma}\left(1 + cL^{-1}\right),
    \label{fss_exp_special}
\end{equation}
leaving $A$, $\hat{\Delta}_\sigma$ and $c$ as free parameters.
Fit results are reported in Table \ref{tab:special_sigma-sigma_l2}.
In Eq.~(\ref{fss_exp_special}) we have included the expected leading scaling correction $\propto L^{-1}$ arising from the irrelevant displacement operator.
To monitor the relevance of the corrections we have repeated the fits setting $c=0$.
From the results of Table ~\ref{tab:special_sigma-sigma_l2} we infer the estimate reported in Eq.~(\ref{delta_sigma_special}),
in agreement but more precise than a previous estimate $\hat{\Delta}_\sigma = 0.3535(6)$ \cite{Hasenbusch-11b}.

\section{Tables of fits}
In this section we provide the tables of fits used to extract the various amplitudes.
As discussed in Ref.~\cite{PTM-21}, to correctly take into account the statistical correlation in the MC data, we employ the Jackknife method \cite{Young_notes} in the fits.
We generally vary the critical exponents and other parameters employed in the fits by the quoted error bars, and sum the resulting variation in the fitted parameters to the statistical error of the fit.
Under such a variation, the minimum $\chidof$ of the fits may changes significantly. In this case, we quote an uncertainty to the minimum \chidof, reflecting this variation.

In Tables \ref{tab:bulk_energy} and \ref{tab:bulk_S2} we report fit results of bulk observables $E_{\text{bulk}}$ and $S^2_{\text{bulk}}$ to Eq.~(\ref{E_fit}) and Eq.~(\ref{S2_fit}), respectively.
In Tables \ref{tab:bulk_S-S}, \ref{tab:bulk_energy-energy}, \ref{tab:bulk_S2-S2} we report the fits of the two-point functions of bulk observables.
In Table \ref{tab:bulk_energy-energy} we have fitted the energy correlations to
Eq.~(\ref{2pt_fit_E}) setting $B'_{\epsilon\epsilon} = 0$, i.e., considering the leading finite-size correction only.
We have observed a somewhat large residual $\chidof$ in the fits, and a small drift in the fitted values of $B_{\epsilon\epsilon}$.
 Including the next-to-leading correction $\propto (x/L)^{2\Delta_\epsilon}$ significantly improve the quality of the fits, leading to a good $\chidof$ for $\xLmax = 1/8$.
Nevertheless, the additional coefficient $B'_{\epsilon\epsilon}$ is rather unstable with respect to the maximum value of $(x/L)$ used in the fits: we have therefore quoted in Table \ref{tab:fitresults} a rather conservative estimate for it.
For the fitted amplitude $B_{\epsilon\epsilon}$ we quote an uncertainty compatible with the fits obtained for $\xmin=6$ and $\xmin=8$.
Due to a reduced precision, fits of the $S^2$ correlations do not necessitate the additional correction term $\propto (x/L)^{2\Delta_\epsilon}$.

In Tables \ref{tab:ordinary_sigma-sigma}-\ref{tab:ordinary_sigma-phi} we report fits pertaining to the ordinary UC.
In the fits of the energy and $S^2$ profile to Eq.~(\ref{fss_1ptE_open}) and Eq.~(\ref{fss_1ptS2_open}) (Tables \ref{tab:ordinary_E} and \ref{tab:ordinary_S2}) we observed that the fitted coefficient $C$ vanishes within error bars.
Therefore we have repeated the fits setting $C=0$ (Tables \ref{tab:ordinary_E_noC} and \ref{tab:ordinary_S2_noC}).

In Tables \ref{tab:special_sigma-sigma}-\ref{tab:special_S2surf-S2surf} we report fits pertaining to the special UC.
In the fits of the surface two-point function of the order parameter to Eq.~(\ref{2pt_fit_Ssurf}) (Table \ref{tab:special_sigma-sigma}), of the energy profile to Eq.~(\ref{fss_1ptE_open}) (Table \ref{tab:special_E}), and of the $S^2$ profile to Eq.~(\ref{fss_1ptS2_open}) (Table \ref{tab:special_S2}) we observed that the fitted coefficient $C$ vanishes within error bars.
As  for the ordinary UC, we report in Tables \ref{tab:special_sigma-sigma_noC}, \ref{tab:special_E_noC}, \ref{tab:special_S2_noC} fits setting $C=0$.

In Tables \ref{tab:normal_po_E}-\ref{tab:normal_pp_S_noC} we report fits pertaining to the normal UC.
In the case of $(+,+)$, fits of the order parameter profile to Eq.~(\ref{fss_1ptS_normal}) reported in Table \ref{tab:normal_pp_S} give a rather small coefficient $C$, marginally compatible with 0.
As in previous cases, we report in Table \ref{tab:normal_pp_S_noC}  fits setting $C=0$.


\begin{table}
\caption{Fit of $E_{\text{bulk}}$ to Eq.~(\ref{E_fit}), as a function of the minimum lattice size \Lmin taken into account.
The quoted error bars are the sum of the statistical uncertainty obtained from the fit procedure and the variation of the fitted parameters on varying the exponent $\Delta_{\epsilon} = 1.412625(10)$ \cite{KPSDV-16} within one error bar.}
\begin{ruledtabular}
\begin{tabular}{lwwa}
\multicolumn{1}{c}{\Lmin} & \multicolumn{1}{c}{$E_{0}$} & \multicolumn{1}{c}{$U_E$} & \multicolumn{1}{c}{\chidof} \\
\hline
32 & 1.2042336(80) & 3.4956(58) & 0.5 \\ 
48 & 1.2042334(88) & 3.4958(76) & 0.6 \\ 
64 & 1.2042348(98) & 3.4938(98) & 0.8 \\ 
96 & 1.2042404(120) & 3.4844(152) & 0.8 \\ 
128 & 1.2042278(174) & 3.51(30) & 0.7 \\
\end{tabular}
\end{ruledtabular}
\label{tab:bulk_energy}
\end{table}
\begin{table}
\caption{Same as Table \ref{tab:bulk_energy} for the fits of $S^2_{\text{bulk}}$ to Eq.~(\ref{S2_fit}).}
\begin{ruledtabular}
\begin{tabular}{lwwa}
\multicolumn{1}{c}{\Lmin} & \multicolumn{1}{c}{$S^2_{0}$} & \multicolumn{1}{c}{$U_{S^2}$} & \multicolumn{1}{c}{\chidof} \\
\hline
32 & 0.60702515(63) & 0.25726(46) & 0.5 \\ 
48 & 0.60702515(69) & 0.25726(60) & 0.7 \\ 
64 & 0.60702540(76) & 0.25689(76) & 0.7 \\ 
96 & 0.60702583(94) & 0.2562(12) & 0.7 \\ 
128 & 0.6070248(14) & 0.2583(23) & 0.4 \\ 
\end{tabular}
\end{ruledtabular}
\label{tab:bulk_S2}
\end{table}

\clearpage

\begin{table*}
\caption{
Fits of the two-point function of bulk observable $S$ to Eq.~(\ref{2pt_fit_S}), as a function of the minimum distance \xmin, the maximum value \xLmax, and the minimum lattice size \Lmin considered in the fits.
Variation of the critical exponent $\Delta_\phi=\numprint{0.5181489}(10)$ \cite{KPSDV-16} within its uncertainty gives a rather small contribution to the uncertainty.}
\begin{ruledtabular}
\begin{tabular}{cclwmmq}
\xmin & \xLmax & \Lmin & \multicolumn{1}{c}{$\mathcal{N}_S^2$} & \multicolumn{1}{c}{$B_{\phi\phi}$} & \multicolumn{1}{c}{$C$} & \multicolumn{1}{c}{\chidof} \\
\hline\multirow{10}{*}{4} & \multirow{5}{*}{$\frac{1}{4}$} & 32 & 0.185359(33) & 3.4220(32) & 0.3382(29) & 5.64 \\ 
& \multirow{5}{*}{} & 48 & 0.185337(32) & 3.4261(38) & 0.3405(28) & 5.16 \\ 
& \multirow{5}{*}{} & 64 & 0.185320(29) & 3.4301(42) & 0.3426(25) & 4.67 \\ 
& \multirow{5}{*}{} & 96 & 0.185304(24) & 3.4353(53) & 0.3448(21) & 3.97 \\ 
& \multirow{5}{*}{} & 128 & 0.185286(19) & 3.4482(76) & 0.3475(16) & 1.99 \\ 
\cline{2-7}
 & \multirow{5}{*}{$\frac{1}{8}$} & 32 & 0.185431(36) & 3.3976(47) & 0.3347(29) & 7.18 \\ 
& \multirow{5}{*}{} & 48 & 0.185421(37) & 3.4000(53) & 0.3356(30) & 7.15 \\ 
& \multirow{5}{*}{} & 64 & 0.185399(36) & 3.4063(57) & 0.3374(29) & 7.02 \\ 
& \multirow{5}{*}{} & 96 & 0.185360(34) & 3.4187(70) & 0.3407(28) & 6.6 \\ 
& \multirow{5}{*}{} & 128 & 0.185282(28) & 3.4508(92) & 0.3477(23) & 3.77 \\ 
\hline
\hline
\multirow{9}{*}{6} & \multirow{5}{*}{$\frac{1}{4}$} & 32 & 0.185548(61) & 3.4147(39) & 0.278(12) & 2.19 \\ 
& \multirow{5}{*}{} & 48 & 0.185542(61) & 3.4152(44) & 0.279(12) & 2.16 \\ 
& \multirow{5}{*}{} & 64 & 0.185528(57) & 3.4170(47) & 0.283(11) & 2.06 \\ 
& \multirow{5}{*}{} & 96 & 0.185508(51) & 3.4198(56) & 0.289(10) & 1.83 \\ 
& \multirow{5}{*}{} & 128 & 0.185452(40) & 3.4329(76) & 0.3047(80) & 0.67 \\ 
\cline{2-7}
& \multirow{4}{*}{$\frac{1}{8}$} & 48 & 0.185656(64) & 3.3864(67) & 0.269(11) & 0.84 \\ 
& \multirow{4}{*}{} & 64 & 0.185656(65) & 3.3864(72) & 0.269(11) & 0.86 \\ 
& \multirow{4}{*}{} & 96 & 0.185653(68) & 3.3871(92) & 0.269(12) & 0.9 \\ 
& \multirow{4}{*}{} & 128 & 0.185560(67) & 3.409(12) & 0.287(12) & 0.3 \\ 
\hline
\hline
\multirow{8}{*}{8} & \multirow{5}{*}{$\frac{1}{4}$} & 32 & 0.185568(92) & 3.4171(50) & 0.250(31) & 1.72 \\ 
& \multirow{5}{*}{} & 48 & 0.185577(93) & 3.4161(54) & 0.246(32) & 1.71 \\ 
& \multirow{5}{*}{} & 64 & 0.185573(91) & 3.4163(57) & 0.249(32) & 1.69 \\ 
& \multirow{5}{*}{} & 96 & 0.185554(86) & 3.4179(67) & 0.259(30) & 1.58 \\ 
& \multirow{5}{*}{} & 128 & 0.185442(72) & 3.4340(84) & 0.309(26) & 0.64 \\ 
\cline{2-7}
& \multirow{3}{*}{$\frac{1}{8}$} & 64 & 0.185716(96) & 3.3852(92) & 0.231(28) & 0.46 \\ 
& \multirow{3}{*}{} & 96 & 0.18572(10) & 3.384(11) & 0.228(31) & 0.46 \\ 
& \multirow{3}{*}{} & 128 & 0.18562(12) & 3.403(17) & 0.262(37) & 0.18 \\ 
\end{tabular}
\end{ruledtabular}
\label{tab:bulk_S-S}
\end{table*}

\begin{table*}
\caption{
Fits of the two-point function of bulk observable $E$ to Eq.~(\ref{2pt_fit_E}), fixing $B'_{\epsilon\epsilon}=0$, as a function of the minimum distance \xmin, the maximum value \xLmax, and the minimum lattice size \Lmin considered in the fits.
Variation of the critical exponent $\Delta_\epsilon=\numprint{1.412625}(10)$ \cite{KPSDV-16} within its uncertainty gives a rather small contribution to the uncertainty.}
\begin{ruledtabular}
\begin{tabular}{cclwmmq}
\xmin & \xLmax & \Lmin & \multicolumn{1}{c}{$\mathcal{N}_{E}^2$} & \multicolumn{1}{c}{$B_{\epsilon\epsilon}$} & \multicolumn{1}{c}{$C$} & \multicolumn{1}{c}{\chidof} \\
\hline\multirow{14}{*}{4} & \multirow{5}{*}{$\frac{1}{4}$} & 32 & 1.17834(37) & 4.6515(84) & 1.3622(43) & 24.78 \\ 
& \multirow{5}{*}{} & 48 & 1.17830(38) & 4.663(10) & 1.3602(45) & 21.78 \\ 
& \multirow{5}{*}{} & 64 & 1.17822(39) & 4.683(12) & 1.3567(46) & 17.34 \\ 
& \multirow{5}{*}{} & 96 & 1.17821(42) & 4.707(18) & 1.3497(49) & 12.09 \\ 
& \multirow{5}{*}{} & 128 & 1.17745(45) & 4.789(28) & 1.3498(51) & 7.74 \\ 
\cline{2-7}
 & \multirow{5}{*}{$\frac{1}{8}$} & 32 & 1.17598(38) & 4.8248(94) & 1.3792(43) & 20.79 \\ 
& \multirow{5}{*}{} & 48 & 1.17611(39) & 4.818(11) & 1.3776(46) & 20.96 \\ 
& \multirow{5}{*}{} & 64 & 1.17609(41) & 4.829(13) & 1.3758(47) & 17.95 \\ 
& \multirow{5}{*}{} & 96 & 1.17618(44) & 4.844(19) & 1.3705(52) & 13.73 \\ 
& \multirow{5}{*}{} & 128 & 1.17555(48) & 4.928(30) & 1.3710(55) & 9.11 \\ 
\cline{2-7}
& \multirow{4}{*}{$\frac{1}{12}$} &   48    &  1.17460(39)   &  4.963(13)   &  1.3865(44)   &  16.6   \\
& \multirow{4}{*}{} & 64    &  1.17476(41)   &  4.953(15)   &  1.3845(46)   &  16.8   \\
& \multirow{4}{*}{} & 96    &  1.17495(47)   &  4.949(21)   &  1.3809(53)   &  14.9   \\
& \multirow{4}{*}{} & 128    &  1.17391(54)   &  5.064(34)   &  1.3878(61)   &  8.8   \\
\hline
\hline
\multirow{11}{*}{6} & \multirow{5}{*}{$\frac{1}{4}$} & 32 & 1.18596(65) & 4.490(10) & 1.147(17) & 7.12 \\ 
& \multirow{5}{*}{} & 48 & 1.18566(67) & 4.501(12) & 1.154(17) & 7.09 \\ 
& \multirow{5}{*}{} & 64 & 1.18532(71) & 4.521(15) & 1.156(18) & 6.33 \\ 
& \multirow{5}{*}{} & 96 & 1.18521(78) & 4.542(21) & 1.145(19) & 4.84 \\ 
& \multirow{5}{*}{} & 128 & 1.18496(87) & 4.581(33) & 1.125(21) & 3.0 \\ 
\cline{2-7}
& \multirow{4}{*}{$\frac{1}{8}$} & 48 & 1.18318(70) & 4.649(14) & 1.176(17) & 4.36 \\ 
& \multirow{4}{*}{} & 64 & 1.18291(74) & 4.659(17) & 1.183(18) & 4.32 \\ 
& \multirow{4}{*}{} & 96 & 1.18273(82) & 4.671(23) & 1.184(20) & 3.97 \\ 
& \multirow{4}{*}{} & 128 & 1.18229(96) & 4.715(37) & 1.181(23) & 2.63 \\ 
\cline{2-7}
& \multirow{2}{*}{$\frac{1}{12}$} &  96    &  1.18140(82)   &  4.778(25)   &  1.192(19)   &  1.83 \\
& \multirow{2}{*}{} &  128    &  1.1802(11)   &  4.848(45)   &  1.215(25)   &  0.84   \\
\hline
\hline
\multirow{10}{*}{8} & \multirow{5}{*}{$\frac{1}{4}$} & 32 & 1.18969(99) & 4.410(13) & 0.995(45) & 3.38 \\ 
& \multirow{5}{*}{} & 48 & 1.1898(10) & 4.408(14) & 0.991(46) & 3.39 \\ 
& \multirow{5}{*}{} & 64 & 1.1891(11) & 4.428(18) & 1.017(50) & 3.27 \\ 
& \multirow{5}{*}{} & 96 & 1.1887(12) & 4.449(25) & 1.010(54) & 2.8 \\ 
& \multirow{5}{*}{} & 128 & 1.1887(14) & 4.480(40) & 0.959(61) & 2.0 \\ 
\cline{2-7}
& \multirow{3}{*}{$\frac{1}{8}$} & 64 & 1.1863(11) & 4.571(21) & 1.044(50) & 2.23 \\ 
& \multirow{3}{*}{} & 96 & 1.1860(13) & 4.580(27) & 1.057(54) & 2.24 \\ 
& \multirow{3}{*}{} & 128 & 1.1851(16) & 4.622(47) & 1.071(66) & 1.76 \\ 
\cline{2-7}
& \multirow{2}{*}{$\frac{1}{12}$} &   96    &  1.1840(12)   &  4.715(33)   &  1.060(51)   &  0.54 \\
& \multirow{2}{*}{} &  128    &  1.1827(17)   &  4.766(57)   &  1.113(67)   &  0.32 \\
\end{tabular}
\end{ruledtabular}
\label{tab:bulk_energy-energy}
\end{table*}

\begin{table*}
\caption{
  Fits of the two-point function of bulk observable $E$ to Eq.~(\ref{2pt_fit_E}), including the correction term $B'_{\epsilon\epsilon}(x/L)^{2\Delta_\epsilon}$, as a function of the minimum distance \xmin, the maximum value \xLmax, and the minimum lattice size \Lmin considered in the fits.
Variation of the critical exponent $\Delta_\epsilon=\numprint{1.412625}(10)$ \cite{KPSDV-16} within its uncertainty gives a rather small contribution to the uncertainty.}
\begin{ruledtabular}
\begin{tabular}{cclwmmmq}
\xmin & \xLmax & \Lmin & \multicolumn{1}{c}{$\mathcal{N}_{E}^2$} & \multicolumn{1}{c}{$B_{\epsilon\epsilon}$} & \multicolumn{1}{c}{$B'_{\epsilon\epsilon}$} & \multicolumn{1}{c}{$C$} & \multicolumn{1}{c}{\chidof} \\
\hline
\multirow{14}{*}{4} & \multirow{5}{*}{$\frac{1}{4}$} &   32    &  1.17426(40)   &  5.015(14)   &  -4.65(14)   &  1.3933(45)   &   8.36   \\
& \multirow{5}{*}{} &   48    &  1.17441(42)   &  5.029(16)   &  -5.25(17)   &  1.3888(47)   &   6.77   \\
& \multirow{5}{*}{} &   64    &  1.17450(43)   &  5.037(18)   &  -5.47(22)   &  1.3857(49)   &   5.75   \\
& \multirow{5}{*}{} &   96    &  1.17466(48)   &  5.045(25)   &  -5.65(32)   &  1.3808(54)   &   4.85   \\
& \multirow{5}{*}{} &  128    &  1.17369(56)   &  5.164(40)   &  -6.84(55)   &  1.3866(62)   &   3.54   \\
\cline{2-8}
& \multirow{5}{*}{$\frac{1}{8}$} &   32    &  1.17275(42)   &  5.191(22)   &  -8.16(38)   &  1.4030(46)   &  11.35   \\
& \multirow{5}{*}{} &   48    &  1.17266(42)   &  5.256(21)   &  -10.56(41)   &  1.3991(47)   &   7.94   \\
& \multirow{5}{*}{} &   64    &  1.17304(44)   &  5.240(23)   &  -10.69(49)   &  1.3937(48)   &   7.35   \\
& \multirow{5}{*}{} &   96    &  1.17338(51)   &  5.219(32)   &  -10.41(67)   &  1.3887(55)   &   6.94   \\
& \multirow{5}{*}{} &  128    &  1.17230(66)   &  5.341(56)   &  -11.8(1.3)   &  1.3982(68)   &   5.15   \\
\cline{2-8}
& \multirow{4}{*}{$\frac{1}{12}$} &  48    &  1.17252(47)   &  5.277(36)   &  -11.1(1.0)   &  1.3998(48)   &  13.35   \\
& \multirow{4}{*}{} &   64    &  1.17242(47)   &  5.347(35)   &  -14.8(1.0)   &  1.3971(49)   &  11.77   \\
& \multirow{4}{*}{} &   96    &  1.17301(53)   &  5.293(38)   &  -13.6(1.1)   &  1.3902(55)   &  11.26   \\
& \multirow{4}{*}{} &  128    &  1.17257(71)   &  5.285(67)   &  -9.0(2.0)   &  1.3969(71)   &   8.14   \\
\hline
\hline
\multirow{11}{*}{6} & \multirow{5}{*}{$\frac{1}{4}$} &   32    &  1.18150(74)   &  4.773(21)   &  -2.94(18)   &  1.209(18)   &   2.63   \\
& \multirow{5}{*}{} &   48    &  1.18124(76)   &  4.815(22)   &  -3.67(22)   &  1.205(18)   &   1.91   \\
& \multirow{5}{*}{} &   64    &  1.18103(79)   &  4.836(25)   &  -4.01(26)   &  1.205(19)   &   1.72   \\
& \multirow{5}{*}{} &   96    &  1.18110(87)   &  4.846(31)   &  -4.26(35)   &  1.198(20)   &   1.58   \\
& \multirow{5}{*}{} &  128    &  1.1804(11)   &  4.911(53)   &  -4.92(62)   &  1.204(25)   &   1.24   \\
\cline{2-8}
& \multirow{4}{*}{$\frac{1}{8}$} &   48    &  1.17907(82)   &  5.011(39)   &  -7.14(65)   &  1.226(18)   &   1.17   \\
& \multirow{4}{*}{} &   64    &  1.17899(82)   &  5.040(40)   &  -8.03(75)   &  1.222(19)   &   0.95   \\
& \multirow{4}{*}{} &   96    &  1.17913(89)   &  5.064(44)   &  -9.11(88)   &  1.211(20)   &   0.74   \\
& \multirow{4}{*}{} &  128    &  1.1782(13)   &  5.124(77)   &  -9.9(1.6)   &  1.229(26)   &   0.68   \\
\cline{2-8}
& \multirow{2}{*}{$\frac{1}{12}$} &  96    &  1.17867(99)   &  5.128(74)   &  -11.2(2.1)   &  1.213(20)   &   0.67   \\
& \multirow{2}{*}{} &  128    &  1.1784(13)   &  5.09(10)   &  -8.4(2.9)   &  1.228(26)   &   0.43   \\
\hline
\hline
\multirow{10}{*}{8} & \multirow{5}{*}{$\frac{1}{4}$} &   32    &  1.1848(12)   &  4.651(29)   &  -2.17(22)   &  1.112(49)   &   1.85   \\
& \multirow{5}{*}{} &   48    &  1.1844(12)   &  4.695(31)   &  -2.77(26)   &  1.098(49)   &   1.38   \\
& \multirow{5}{*}{} &   64    &  1.1841(12)   &  4.723(32)   &  -3.14(31)   &  1.102(51)   &   1.29   \\
& \multirow{5}{*}{} &   96    &  1.1840(14)   &  4.738(39)   &  -3.40(40)   &  1.095(56)   &   1.26   \\
& \multirow{5}{*}{} &  128    &  1.1831(18)   &  4.798(70)   &  -3.98(72)   &  1.104(70)   &   1.10   \\
\cline{2-8}
& \multirow{3}{*}{$\frac{1}{8}$} &   64    &  1.1805(14)   &  4.990(68)   &  -7.5(1.1)   &  1.154(53)   &   0.79   \\
& \multirow{3}{*}{} &   96    &  1.1803(14)   &  5.048(71)   &  -9.0(1.4)   &  1.134(55)   &   0.59   \\
& \multirow{3}{*}{} &  128    &  1.1798(21)   &  5.08(11)   &  -9.5(2.0)   &  1.152(71)   &   0.64   \\
\cline{2-8}
& \multirow{2}{*}{$\frac{1}{12}$} &  96    &  1.1804(24)   &  5.05(19)   &  -9.0(4.6)   &  1.119(62)   &   0.27   \\
& \multirow{2}{*}{} &  128    &  1.1806(23)   &  4.99(18)   &  -6.5(4.9)   &  1.135(69)   &   0.21   \\
\end{tabular}
\end{ruledtabular}
\label{tab:bulk_energy-energy_Bp}
\end{table*}

\begin{table*}
\caption{Same as Table \ref{tab:bulk_energy-energy} for the fits of bulk $S^2$ correlations to Eq.~(\ref{2pt_fit_S2}).}
\begin{ruledtabular}
\begin{tabular}{cclwmmq}
\xmin & \xLmax & \Lmin & \multicolumn{1}{c}{$\mathcal{N}_{S^2}^2$} & \multicolumn{1}{c}{$B_{\epsilon\epsilon}$} & \multicolumn{1}{c}{$C$} & \multicolumn{1}{c}{\chidof} \\
\hline\multirow{14}{*}{4} & \multirow{5}{*}{$\frac{1}{4}$} & 32 & 0.0063737(35) & 4.757(18) & 1.3321(88) & 2.38 \\ 
& \multirow{5}{*}{} & 48 & 0.0063715(36) & 4.791(21) & 1.3351(89) & 2.18 \\ 
& \multirow{5}{*}{} & 64 & 0.0063695(38) & 4.835(26) & 1.3362(94) & 1.98 \\ 
& \multirow{5}{*}{} & 96 & 0.0063697(41) & 4.863(35) & 1.3309(99) & 1.59 \\ 
& \multirow{5}{*}{} & 128 & 0.0063688(51) & 4.912(62) & 1.328(12) & 1.4 \\ 
\cline{2-7}
 & \multirow{5}{*}{$\frac{1}{8}$} & 32 & 0.0063676(35) & 4.866(22) & 1.3383(89) & 2.1 \\ 
& \multirow{5}{*}{} & 48 & 0.0063667(36) & 4.876(25) & 1.3401(90) & 2.11 \\ 
& \multirow{5}{*}{} & 64 & 0.0063643(38) & 4.918(29) & 1.3432(94) & 1.76 \\ 
& \multirow{5}{*}{} & 96 & 0.0063648(43) & 4.940(39) & 1.339(10) & 1.33 \\ 
& \multirow{5}{*}{} & 128 & 0.0063627(53) & 5.011(66) & 1.338(12) & 1.12 \\ 
\cline{2-7}
& \multirow{4}{*}{$\frac{1}{12}$} &   48    &  0.0063628(36)   &  4.975(27)   &  1.3417(90)   &  1.33 \\
& \multirow{4}{*}{} &   64    &  0.0063609(39)   &  4.999(32)   &  1.3454(95)   &  1.29 \\
& \multirow{4}{*}{} &   96    &  0.0063625(44)   &  4.988(43)   &  1.341(10)   &  1.12 \\
& \multirow{4}{*}{} &  128    &  0.0063581(57)   &  5.096(76)   &  1.346(12)   &  0.84 \\
\hline
\hline
\multirow{12}{*}{6} & \multirow{5}{*}{$\frac{1}{4}$} & 32 & 0.0064211(90) & 4.514(30) & 1.132(48) & 1.34 \\ 
& \multirow{5}{*}{} & 48 & 0.0064177(91) & 4.537(34) & 1.146(49) & 1.33 \\ 
& \multirow{5}{*}{} & 64 & 0.0064130(98) & 4.573(43) & 1.165(51) & 1.36 \\ 
& \multirow{5}{*}{} & 96 & 0.006407(11) & 4.629(56) & 1.184(54) & 1.29 \\ 
& \multirow{5}{*}{} & 128 & 0.006404(12) & 4.653(85) & 1.188(60) & 1.24 \\ 
\cline{2-7}
& \multirow{4}{*}{$\frac{1}{8}$} & 48 & 0.0064097(96) & 4.641(45) & 1.156(50) & 1.08 \\ 
& \multirow{4}{*}{} & 64 & 0.006404(10) & 4.683(49) & 1.181(51) & 1.02 \\ 
& \multirow{4}{*}{} & 96 & 0.006396(11) & 4.743(62) & 1.212(55) & 0.99 \\ 
& \multirow{4}{*}{} & 128 & 0.006389(13) & 4.803(91) & 1.240(61) & 0.94 \\
\cline{2-7}
& \multirow{3}{*}{$\frac{1}{12}$} &   96    &  0.006394(11)   &  4.803(70)   &  1.205(56)   &  0.67 \\
& \multirow{3}{*}{} &  128    &  0.006380(13)   &  4.93(11)   &  1.263(61)   &  0.63 \\
& \multirow{3}{*}{} &  192    &  0.006357(22)   &  5.18(21)   &  1.358(92)   &  0.57 \\
\hline
\hline
\multirow{11}{*}{8} & \multirow{5}{*}{$\frac{1}{4}$} & 32 & 0.006427(18) & 4.477(49) & 1.11(19) & 1.31 \\ 
& \multirow{5}{*}{} & 48 & 0.006427(18) & 4.479(51) & 1.12(19) & 1.32 \\ 
& \multirow{5}{*}{} & 64 & 0.006423(20) & 4.494(65) & 1.14(20) & 1.36 \\ 
& \multirow{5}{*}{} & 96 & 0.006420(21) & 4.535(83) & 1.14(20) & 1.29 \\ 
& \multirow{5}{*}{} & 128 & 0.006436(24) & 4.48(12) & 0.93(22) & 1.25 \\ 
\cline{2-7}
& \multirow{3}{*}{$\frac{1}{8}$} & 64 & 0.006402(22) & 4.704(93) & 1.18(20) & 1.06 \\ 
& \multirow{3}{*}{} & 96 & 0.006400(23) & 4.72(10) & 1.19(21) & 1.08 \\ 
& \multirow{3}{*}{} & 128 & 0.006405(27) & 4.71(14) & 1.12(23) & 1.01 \\ 
\cline{2-7}
& \multirow{3}{*}{$\frac{1}{12}$} &  96    &  0.006395(24)   &  4.88(13)   &  1.09(21)   &  0.70 \\
& \multirow{3}{*}{} &  128    &  0.006383(29)   &  4.96(19)   &  1.18(24)   &  0.72 \\
& \multirow{3}{*}{} &  192    &  0.006316(42)   &  5.48(32)   &  1.63(31)   &  0.56 \\
\hline
\hline
\multirow{6}{*}{10}& \multirow{4}{*}{$\frac{1}{4}$} & 48 & 0.006455(29) & 4.361(63) & 0.87(48) & 1.26 \\ 
& \multirow{4}{*}{} & 64 & 0.006463(30) & 4.330(73) & 0.78(48) & 1.27 \\ 
& \multirow{4}{*}{} & 96 & 0.006462(36) & 4.36(11) & 0.72(55) & 1.23 \\ 
& \multirow{4}{*}{} & 128 & 0.006476(48) & 4.34(17) & 0.42(68) & 1.27 \\ 
\cline{2-7}
& \multirow{2}{*}{$\frac{1}{8}$} & 96 & 0.006447(36) & 4.55(12) & 0.57(56) & 1.05 \\ 
& \multirow{2}{*}{} & 128 & 0.006431(51) & 4.65(20) & 0.70(70) & 1.08 \\ 
\end{tabular}
\end{ruledtabular}
\label{tab:bulk_S2-S2}
\end{table*}


\begin{table*}
\caption{
Fits of the two-point function of surface order parameter S at the ordinary UC to Eq.~(\ref{2pt_fit_Ssurf}), as a function of the minimum distance \xmin, the maximum value \xLmax, and the minimum lattice size \Lmin considered in the fits.
The uncertainty due to the variation of the boundary scaling dimension $\hat{\Delta}_\sigma = 2 - 0.7249(6)$ \cite{Hasenbusch-11} within one quoted error bar largely dominates over the statistical error bars of the fits. 
As a reference, a fit employing only the central value of $\hat{\Delta}_\sigma$ for $\xmin = 4$, $\xLmax=1/4$, $\Lmin=48$ results in ${\cal N}_\sigma^2=0.385279(25)$.
}
\begin{ruledtabular}
\begin{tabular}{cclwmmq}
\xmin & \xLmax & \Lmin & \multicolumn{1}{c}{${\cal N}_\sigma^2$} & \multicolumn{1}{c}{$B_{\sigma\sigma}$} & \multicolumn{1}{c}{$C$} & \multicolumn{1}{c}{\chidof} \\
\hline\multirow{10}{*}{4} & \multirow{5}{*}{$\frac{1}{4}$} & 32 & 0.3853(11) & 4.397(85) & -0.611(17) & 44.0(6.8) \\ 
& \multirow{5}{*}{} & 48 & 0.3853(11) & 4.15(13) & -0.611(17) & 44.0(6.7) \\ 
& \multirow{5}{*}{} & 64 & 0.3853(11) & 3.75(21) & -0.612(17) & 44.4(6.7) \\ 
& \multirow{5}{*}{} & 96 & 0.3853(11) & 3.14(38) & -0.612(17) & 45.1(6.7) \\ 
& \multirow{5}{*}{} & 128 & 0.3853(11) & 2.57(57) & -0.612(17) & 47.3(7.0) \\ 
\cline{2-7}
 & \multirow{5}{*}{$\frac{1}{8}$} & 32 & 0.3853(11) & 3.83(16) & -0.612(17) & 92.0(14.0) \\ 
& \multirow{5}{*}{} & 48 & 0.3853(11) & 3.67(21) & -0.612(17) & 93.0(14.0) \\ 
& \multirow{5}{*}{} & 64 & 0.3853(11) & 2.81(35) & -0.614(17) & 93.0(14.0) \\ 
& \multirow{5}{*}{} & 96 & 0.3854(11) & 0.33(86) & -0.616(17) & 91.0(14.0) \\ 
& \multirow{5}{*}{} & 128 & 0.3854(11) & -2.3(1.5) & -0.617(17) & 93.0(13.0) \\ 
\hline
\hline
\multirow{9}{*}{6} & \multirow{5}{*}{$\frac{1}{4}$} & 32 & 0.3833(13) & 4.723(62) & -0.404(44) & 1.53(29) \\ 
& \multirow{5}{*}{} & 48 & 0.3834(13) & 4.661(89) & -0.405(43) & 1.45(27) \\ 
& \multirow{5}{*}{} & 64 & 0.3834(13) & 4.54(15) & -0.406(43) & 1.36(25) \\ 
& \multirow{5}{*}{} & 96 & 0.3834(13) & 4.37(28) & -0.407(43) & 1.26(23) \\ 
& \multirow{5}{*}{} & 128 & 0.3834(13) & 4.38(42) & -0.406(43) & 1.34(25) \\ 
\cline{2-7}
& \multirow{4}{*}{$\frac{1}{8}$} & 48 & 0.3834(13) & 4.34(19) & -0.408(43) & 1.19(43) \\ 
& \multirow{4}{*}{} & 64 & 0.3834(13) & 4.31(24) & -0.408(43) & 1.20(43) \\ 
& \multirow{4}{*}{} & 96 & 0.3834(13) & 4.23(56) & -0.408(43) & 1.20(42) \\ 
& \multirow{4}{*}{} & 128 & 0.3834(13) & 4.0(1.0) & -0.409(42) & 1.22(41) \\ 
\hline
\hline
\multirow{8}{*}{8} & \multirow{5}{*}{$\frac{1}{4}$} & 32 & 0.3831(15) & 4.778(63) & -0.348(87) & 1.216(96) \\ 
& \multirow{5}{*}{} & 48 & 0.3831(15) & 4.722(78) & -0.352(87) & 1.172(91) \\ 
& \multirow{5}{*}{} & 64 & 0.3831(15) & 4.61(13) & -0.356(87) & 1.106(77) \\ 
& \multirow{5}{*}{} & 96 & 0.3832(15) & 4.45(23) & -0.359(87) & 1.015(61) \\ 
& \multirow{5}{*}{} & 128 & 0.3831(15) & 4.50(35) & -0.357(86) & 1.076(70) \\ 
\cline{2-7}
& \multirow{3}{*}{$\frac{1}{8}$} & 64 & 0.3832(15) & 4.37(30) & -0.364(86) & 0.657(68) \\ 
& \multirow{3}{*}{} & 96 & 0.3832(15) & 4.32(49) & -0.365(87) & 0.661(67) \\ 
& \multirow{3}{*}{} & 128 & 0.3832(15) & 4.26(86) & -0.367(87) & 0.692(67) \\ 
\end{tabular}
\end{ruledtabular}
\label{tab:ordinary_sigma-sigma}
\end{table*}

\begin{table*}
\caption{Fits of the energy profile at the ordinary UC to Eq.~(\ref{fss_1ptE_open}), as a function of the minimum distance \zmin from the surface, the maximum ratio \zLmax, and the minimum lattice size \Lmin considered in the fits.
We employ the fitted value of $E_0$ reported in Table \ref{tab:fitresults}.
Varying $E_0$ within one error bar quoted in Table \ref{tab:fitresults} results in a significant additional uncertainty in the results.
As a reference, a fit employing only the central value of $E_0$ for $\zmin = 4$, $\zLmax=1/8$, $\Lmin=48$ results in $A_\epsilon = -0.54614(80)$.
}
\begin{ruledtabular}
\begin{tabular}{cclwmmqm}
\zmin & \zLmax & \Lmin & \multicolumn{1}{c}{$A_\epsilon$} & \multicolumn{1}{c}{$z_0$} & \multicolumn{1}{c}{$B_{\epsilon}$} & \multicolumn{1}{c}{$C$} & \multicolumn{1}{c}{\chidof} \\
\hline\multirow{12}{*}{4} & \multirow{6}{*}{$\frac{1}{4}$} & 32 & -0.5446(33) & 1.015(54) & -22.61(45) & 0.20(25) & 7.44(40) \\ 
& \multirow{6}{*}{} & 48 & -0.5442(33) & 1.008(55) & -22.57(47) & 0.18(25) & 6.11(46) \\ 
& \multirow{6}{*}{} & 64 & -0.5438(32) & 1.000(53) & -22.53(52) & 0.13(25) & 5.31(53) \\ 
& \multirow{6}{*}{} & 96 & -0.5434(31) & 0.990(50) & -22.42(63) & 0.08(23) & 4.85(54) \\ 
& \multirow{6}{*}{} & 128 & -0.5429(29) & 0.978(47) & -22.12(78) & 0.02(21) & 4.36(42) \\ 
& \multirow{6}{*}{} & 192 & -0.5425(27) & 0.966(42) & -21.6(1.0) & -0.04(18) & 3.27(11) \\ 
\cline{2-8}
 & \multirow{6}{*}{$\frac{1}{8}$} & 32 & -0.5461(28) & 1.031(44) & -26.80(69) & 0.25(20) & 1.25(27) \\ 
& \multirow{6}{*}{} & 48 & -0.5461(28) & 1.030(44) & -26.79(78) & 0.25(20) & 1.26(27) \\ 
& \multirow{6}{*}{} & 64 & -0.5462(28) & 1.031(44) & -26.82(88) & 0.25(20) & 1.28(28) \\ 
& \multirow{6}{*}{} & 96 & -0.5468(29) & 1.042(46) & -27.7(1.1) & 0.31(21) & 1.06(32) \\ 
& \multirow{6}{*}{} & 128 & -0.5471(29) & 1.048(46) & -28.2(1.4) & 0.33(21) & 1.09(37) \\ 
& \multirow{6}{*}{} & 192 & -0.5465(27) & 1.036(43) & -27.1(2.2) & 0.28(19) & 0.93(32) \\ 
\hline
\hline
\multirow{11}{*}{6} & \multirow{6}{*}{$\frac{1}{4}$} & 32 & -0.5447(46) & 1.03(10) & -22.32(55) & 0.41(69) & 6.29(31) \\ 
& \multirow{6}{*}{} & 48 & -0.5453(48) & 1.05(11) & -22.42(55) & 0.53(74) & 5.80(30) \\ 
& \multirow{6}{*}{} & 64 & -0.5452(49) & 1.05(11) & -22.45(56) & 0.54(77) & 5.21(35) \\ 
& \multirow{6}{*}{} & 96 & -0.5447(49) & 1.03(11) & -22.42(62) & 0.43(76) & 4.85(41) \\ 
& \multirow{6}{*}{} & 128 & -0.5433(47) & 0.99(11) & -22.13(74) & 0.13(67) & 4.41(39) \\ 
& \multirow{6}{*}{} & 192 & -0.5413(42) & 0.929(89) & -21.54(96) & -0.31(53) & 3.28(18) \\ 
\cline{2-8}
& \multirow{5}{*}{$\frac{1}{8}$} & 48 & -0.5473(41) & 1.069(87) & -26.60(86) & 0.55(55) & 1.17(14) \\ 
& \multirow{5}{*}{} & 64 & -0.5474(41) & 1.071(87) & -26.73(93) & 0.56(56) & 1.16(14) \\ 
& \multirow{5}{*}{} & 96 & -0.5485(43) & 1.096(92) & -27.8(1.1) & 0.71(60) & 0.82(14) \\ 
& \multirow{5}{*}{} & 128 & -0.5493(45) & 1.114(98) & -28.5(1.3) & 0.83(64) & 0.77(16) \\ 
& \multirow{5}{*}{} & 192 & -0.5486(46) & 1.10(10) & -27.9(2.0) & 0.71(65) & 0.71(17) \\ 
\hline
\hline
\multirow{10}{*}{8} & \multirow{6}{*}{$\frac{1}{4}$} & 32 & -0.5446(57) & 1.05(17) & -22.10(66) & 0.7(1.4) & 5.85(28) \\ 
& \multirow{6}{*}{} & 48 & -0.5458(60) & 1.09(18) & -22.20(66) & 1.1(1.5) & 5.44(22) \\ 
& \multirow{6}{*}{} & 64 & -0.5466(64) & 1.12(19) & -22.29(65) & 1.3(1.7) & 5.12(21) \\ 
& \multirow{6}{*}{} & 96 & -0.5466(68) & 1.12(20) & -22.34(66) & 1.3(1.8) & 4.80(27) \\ 
& \multirow{6}{*}{} & 128 & -0.5444(66) & 1.04(19) & -22.12(73) & 0.6(1.6) & 4.45(32) \\ 
& \multirow{6}{*}{} & 192 & -0.5401(58) & 0.88(16) & -21.48(91) & -0.8(1.2) & 3.32(25) \\ 
\cline{2-8}
& \multirow{4}{*}{$\frac{1}{8}$} & 64 & -0.5485(54) & 1.11(14) & -26.9(1.1) & 0.9(1.2) & 1.147(97) \\ 
& \multirow{4}{*}{} & 96 & -0.5494(55) & 1.13(15) & -27.7(1.2) & 1.1(1.2) & 0.814(92) \\ 
& \multirow{4}{*}{} & 128 & -0.5504(59) & 1.16(16) & -28.4(1.3) & 1.3(1.3) & 0.722(90) \\ 
& \multirow{4}{*}{} & 192 & -0.5499(64) & 1.14(17) & -28.1(1.9) & 1.2(1.4) & 0.68(11) \\ 
\end{tabular}
\end{ruledtabular}
\label{tab:ordinary_E}
\end{table*}

\begin{table*}
\caption{Same as Table \ref{tab:ordinary_E}, fixing $C=0$.
A variation of $E_0$ within the uncertanty quoted in Table \ref{tab:fitresults} gives a rather significant contribution to the final uncertainty of the fitted parameters.
As a reference, a fit employing only the central value of $E_0$ for $\zmin = 4$, $\zLmax=1/8$, $\Lmin=48$ results in $A_\epsilon = -0.54396(32)$.
}
\begin{ruledtabular}
\begin{tabular}{cclwmqm}
\zmin & \zLmax & \Lmin & \multicolumn{1}{c}{$A_\epsilon$} & \multicolumn{1}{c}{$z_0$} & \multicolumn{1}{c}{$B_{\epsilon}$} & \multicolumn{1}{c}{\chidof} \\
\hline\multirow{12}{*}{4} & \multirow{6}{*}{$\frac{1}{4}$} & 32 & -0.5429(14) & 0.9766(92) & -22.67(40) & 7.84(82) \\ 
& \multirow{6}{*}{} & 48 & -0.5428(13) & 0.9750(88) & -22.57(47) & 6.43(80) \\ 
& \multirow{6}{*}{} & 64 & -0.5427(13) & 0.9741(85) & -22.49(56) & 5.54(76) \\ 
& \multirow{6}{*}{} & 96 & -0.5426(12) & 0.9735(80) & -22.36(72) & 4.98(67) \\ 
& \multirow{6}{*}{} & 128 & -0.5427(12) & 0.9735(78) & -22.10(90) & 4.44(45) \\ 
& \multirow{6}{*}{} & 192 & -0.5429(12) & 0.9746(78) & -21.7(1.1) & 3.27(12) \\ 
\cline{2-7}
 & \multirow{6}{*}{$\frac{1}{8}$} & 32 & -0.5440(12) & 0.9831(79) & -26.81(68) & 2.01(87) \\ 
& \multirow{6}{*}{} & 48 & -0.5440(12) & 0.9828(78) & -26.68(84) & 2.00(86) \\ 
& \multirow{6}{*}{} & 64 & -0.5439(12) & 0.9823(76) & -26.5(1.1) & 2.01(85) \\ 
& \multirow{6}{*}{} & 96 & -0.5440(11) & 0.9827(72) & -26.8(1.5) & 1.98(95) \\ 
& \multirow{6}{*}{} & 128 & -0.5439(11) & 0.9826(68) & -26.6(1.9) & 2.08(98) \\ 
& \multirow{6}{*}{} & 192 & -0.5438(10) & 0.9814(64) & -25.1(2.7) & 1.57(72) \\ 
\hline
\hline
\multirow{11}{*}{6} & \multirow{6}{*}{$\frac{1}{4}$} & 32 & -0.5431(22) & 0.980(20) & -22.44(40) & 6.56(60) \\ 
& \multirow{6}{*}{} & 48 & -0.5432(21) & 0.981(20) & -22.52(45) & 6.15(66) \\ 
& \multirow{6}{*}{} & 64 & -0.5430(21) & 0.979(19) & -22.49(52) & 5.54(69) \\ 
& \multirow{6}{*}{} & 96 & -0.5428(20) & 0.976(18) & -22.38(66) & 5.07(65) \\ 
& \multirow{6}{*}{} & 128 & -0.5427(19) & 0.974(17) & -22.10(83) & 4.50(45) \\ 
& \multirow{6}{*}{} & 192 & -0.5428(18) & 0.973(16) & -21.7(1.1) & 3.34(11) \\ 
\cline{2-7}
& \multirow{5}{*}{$\frac{1}{8}$} & 48 & -0.5449(19) & 0.995(17) & -26.67(82) & 1.60(52) \\ 
& \multirow{5}{*}{} & 64 & -0.5449(19) & 0.995(17) & -26.73(93) & 1.61(53) \\ 
& \multirow{5}{*}{} & 96 & -0.5452(19) & 0.999(16) & -27.5(1.2) & 1.45(62) \\ 
& \multirow{5}{*}{} & 128 & -0.5453(18) & 0.999(16) & -27.6(1.6) & 1.53(67) \\ 
& \multirow{5}{*}{} & 192 & -0.5449(17) & 0.995(14) & -26.3(2.4) & 1.20(52) \\ 
\hline
\hline
\multirow{10}{*}{8} & \multirow{6}{*}{$\frac{1}{4}$} & 32 & -0.5430(30) & 0.981(36) & -22.25(43) & 6.06(51) \\ 
& \multirow{6}{*}{} & 48 & -0.5434(30) & 0.987(36) & -22.40(45) & 5.75(55) \\ 
& \multirow{6}{*}{} & 64 & -0.5434(29) & 0.987(35) & -22.47(50) & 5.51(60) \\ 
& \multirow{6}{*}{} & 96 & -0.5432(28) & 0.982(33) & -22.39(62) & 5.13(61) \\ 
& \multirow{6}{*}{} & 128 & -0.5427(27) & 0.975(31) & -22.10(78) & 4.58(46) \\ 
& \multirow{6}{*}{} & 192 & -0.5424(25) & 0.967(29) & -21.6(1.0) & 3.43(13) \\ 
\cline{2-7}
& \multirow{4}{*}{$\frac{1}{8}$} & 64 & -0.5459(26) & 1.013(29) & -27.03(98) & 1.41(34) \\ 
& \multirow{4}{*}{} & 96 & -0.5464(26) & 1.019(30) & -27.8(1.1) & 1.13(37) \\ 
& \multirow{4}{*}{} & 128 & -0.5467(26) & 1.023(30) & -28.3(1.4) & 1.14(42) \\ 
& \multirow{4}{*}{} & 192 & -0.5462(25) & 1.015(28) & -27.3(2.1) & 0.97(36) \\ 
\end{tabular}
\end{ruledtabular}
\label{tab:ordinary_E_noC}
\end{table*}

\begin{table*}
\caption{
Fits of the $S^2$ profile at the ordinary UC to Eq.~(\ref{fss_1ptS2_open}), as a function of the minimum distance \zmin from the surface, the maximum ratio \zLmax, and the minimum lattice size \Lmin considered in the fits.
We employ the fitted value of $S^2_0$ reported in Table \ref{tab:additional_fitresults}.
Varying $S^2_0$ within one error bar quoted in Table \ref{tab:additional_fitresults} results in a significant additional uncertainty in the results.
As a reference, a fit employing only the central value of $S^2_0$ for $\zmin = 4$, $\zLmax=1/4$, $\Lmin=32$ results in $A_{S^2} = -0.060223(80)$.
}
\begin{ruledtabular}
\begin{tabular}{cclwmmqm}
\zmin & \zLmax & \Lmin & \multicolumn{1}{c}{$A_{S^2}$} & \multicolumn{1}{c}{$z_0$} & \multicolumn{1}{c}{$B_{\epsilon}$} & \multicolumn{1}{c}{$C$} & \multicolumn{1}{c}{\chidof} \\
\hline\multirow{12}{*}{4} & \multirow{6}{*}{$\frac{1}{4}$} & 32 & -0.06022(42) & 1.027(65) & -22.43(51) & 0.45(31) & 7.79(62) \\ 
& \multirow{6}{*}{} & 48 & -0.06019(43) & 1.022(65) & -22.41(53) & 0.42(32) & 6.50(70) \\ 
& \multirow{6}{*}{} & 64 & -0.06015(42) & 1.014(63) & -22.38(59) & 0.38(30) & 5.71(78) \\ 
& \multirow{6}{*}{} & 96 & -0.06010(40) & 1.004(59) & -22.26(71) & 0.33(28) & 5.25(80) \\ 
& \multirow{6}{*}{} & 128 & -0.06004(37) & 0.990(55) & -21.93(88) & 0.26(25) & 4.71(64) \\ 
& \multirow{6}{*}{} & 192 & -0.05998(35) & 0.976(49) & -21.4(1.2) & 0.18(22) & 3.46(22) \\ 
\cline{2-8}
 & \multirow{6}{*}{$\frac{1}{8}$} & 32 & -0.06038(35) & 1.041(51) & -26.51(75) & 0.48(24) & 1.39(38) \\ 
& \multirow{6}{*}{} & 48 & -0.06039(36) & 1.042(51) & -26.56(84) & 0.49(24) & 1.40(39) \\ 
& \multirow{6}{*}{} & 64 & -0.06039(36) & 1.042(52) & -26.59(95) & 0.49(24) & 1.42(40) \\ 
& \multirow{6}{*}{} & 96 & -0.06047(37) & 1.054(53) & -27.5(1.2) & 0.55(25) & 1.20(45) \\ 
& \multirow{6}{*}{} & 128 & -0.06050(37) & 1.060(54) & -28.0(1.5) & 0.58(26) & 1.23(51) \\ 
& \multirow{6}{*}{} & 192 & -0.06042(34) & 1.046(49) & -26.7(2.3) & 0.51(23) & 1.05(45) \\ 
\hline
\hline
\multirow{11}{*}{6} & \multirow{6}{*}{$\frac{1}{4}$} & 32 & -0.06026(59) & 1.06(12) & -22.13(63) & 0.73(84) & 6.64(46) \\ 
& \multirow{6}{*}{} & 48 & -0.06035(62) & 1.08(13) & -22.24(63) & 0.88(90) & 6.10(44) \\ 
& \multirow{6}{*}{} & 64 & -0.06035(63) & 1.08(13) & -22.29(64) & 0.90(93) & 5.51(50) \\ 
& \multirow{6}{*}{} & 96 & -0.06030(63) & 1.06(13) & -22.27(70) & 0.80(92) & 5.17(59) \\ 
& \multirow{6}{*}{} & 128 & -0.06013(60) & 1.02(12) & -21.96(82) & 0.47(81) & 4.72(56) \\ 
& \multirow{6}{*}{} & 192 & -0.05989(53) & 0.95(10) & -21.3(1.1) & -0.03(63) & 3.46(30) \\ 
\cline{2-8}
& \multirow{5}{*}{$\frac{1}{8}$} & 48 & -0.06054(51) & 1.09(10) & -26.35(94) & 0.85(65) & 1.24(20) \\ 
& \multirow{5}{*}{} & 64 & -0.06056(52) & 1.09(10) & -26.5(1.0) & 0.87(66) & 1.23(20) \\ 
& \multirow{5}{*}{} & 96 & -0.06069(54) & 1.12(11) & -27.6(1.2) & 1.03(70) & 0.86(20) \\ 
& \multirow{5}{*}{} & 128 & -0.06077(57) & 1.13(11) & -28.3(1.4) & 1.16(75) & 0.81(22) \\ 
& \multirow{5}{*}{} & 192 & -0.06069(58) & 1.12(11) & -27.7(2.1) & 1.02(76) & 0.75(23) \\ 
\hline
\hline
\multirow{10}{*}{8} & \multirow{6}{*}{$\frac{1}{4}$} & 32 & -0.06028(73) & 1.09(19) & -21.91(75) & 1.2(1.7) & 6.20(41) \\ 
& \multirow{6}{*}{} & 48 & -0.06043(77) & 1.13(20) & -22.01(75) & 1.6(1.8) & 5.72(32) \\ 
& \multirow{6}{*}{} & 64 & -0.06053(82) & 1.16(22) & -22.10(75) & 1.9(2.0) & 5.36(31) \\ 
& \multirow{6}{*}{} & 96 & -0.06054(87) & 1.17(24) & -22.16(75) & 1.9(2.2) & 5.04(38) \\ 
& \multirow{6}{*}{} & 128 & -0.06030(85) & 1.08(23) & -21.95(82) & 1.2(2.0) & 4.71(45) \\ 
& \multirow{6}{*}{} & 192 & -0.05980(74) & 0.91(18) & -21.3(1.0) & -0.4(1.4) & 3.51(37) \\ 
\cline{2-8}
& \multirow{4}{*}{$\frac{1}{8}$} & 64 & -0.06071(67) & 1.14(16) & -26.7(1.1) & 1.4(1.4) & 1.16(13) \\ 
& \multirow{4}{*}{} & 96 & -0.06081(69) & 1.16(17) & -27.5(1.3) & 1.5(1.4) & 0.82(12) \\ 
& \multirow{4}{*}{} & 128 & -0.06092(73) & 1.19(18) & -28.2(1.4) & 1.7(1.5) & 0.73(12) \\ 
& \multirow{4}{*}{} & 192 & -0.06086(79) & 1.17(19) & -27.9(2.0) & 1.6(1.6) & 0.69(15) \\ 
\end{tabular}
\end{ruledtabular}
\label{tab:ordinary_S2}
\end{table*}

\begin{table*}
\caption{Same as Table \ref{tab:ordinary_S2}, fixing $C=0$.
A variation of $S^2_0$ within the uncertanty quoted in Table \ref{tab:additional_fitresults} gives a rather significant contribution to the final uncertainty of the fitted parameters.
As a reference, a fit employing only the central value of $S^2_0$ for $\zmin = 4$, $\zLmax=1/4$, $\Lmin=96$ results in $A_\epsilon = -0.059791(33)$.
}
\begin{ruledtabular}
\begin{tabular}{cclwmqm}
\zmin & \zLmax & \Lmin & \multicolumn{1}{c}{$A_{S^2}$} & \multicolumn{1}{c}{$z_0$} & \multicolumn{1}{c}{$B_{\epsilon}$} & \multicolumn{1}{c}{\chidof} \\
\hline\multirow{12}{*}{4} & \multirow{6}{*}{$\frac{1}{4}$} & 32 & -0.05983(17) & 0.946(10) & -22.55(45) & 9.2(1.7) \\ 
& \multirow{6}{*}{} & 48 & -0.05981(17) & 0.9438(99) & -22.41(54) & 7.7(1.7) \\ 
& \multirow{6}{*}{} & 64 & -0.05980(16) & 0.9427(95) & -22.28(64) & 6.7(1.6) \\ 
& \multirow{6}{*}{} & 96 & -0.05979(16) & 0.9418(89) & -22.05(83) & 6.0(1.4) \\ 
& \multirow{6}{*}{} & 128 & -0.05979(15) & 0.9417(87) & -21.7(1.0) & 5.2(1.1) \\ 
& \multirow{6}{*}{} & 192 & -0.05982(15) & 0.9429(87) & -21.2(1.3) & 3.74(51) \\ 
\cline{2-7}
 & \multirow{6}{*}{$\frac{1}{8}$} & 32 & -0.05993(15) & 0.9509(88) & -26.54(75) & 3.9(1.8) \\ 
& \multirow{6}{*}{} & 48 & -0.05993(15) & 0.9505(87) & -26.35(92) & 3.9(1.8) \\ 
& \multirow{6}{*}{} & 64 & -0.05991(15) & 0.9497(84) & -25.9(1.2) & 3.9(1.8) \\ 
& \multirow{6}{*}{} & 96 & -0.05991(14) & 0.9494(80) & -25.8(1.6) & 4.0(1.8) \\ 
& \multirow{6}{*}{} & 128 & -0.05990(13) & 0.9488(75) & -25.2(2.1) & 4.1(1.8) \\ 
& \multirow{6}{*}{} & 192 & -0.05988(13) & 0.9473(70) & -23.1(3.0) & 3.1(1.4) \\ 
\hline
\hline
\multirow{11}{*}{6} & \multirow{6}{*}{$\frac{1}{4}$} & 32 & -0.05995(27) & 0.962(23) & -22.34(45) & 7.3(1.1) \\ 
& \multirow{6}{*}{} & 48 & -0.05996(27) & 0.962(22) & -22.41(51) & 6.9(1.2) \\ 
& \multirow{6}{*}{} & 64 & -0.05994(26) & 0.960(21) & -22.36(59) & 6.3(1.2) \\ 
& \multirow{6}{*}{} & 96 & -0.05992(24) & 0.957(20) & -22.19(75) & 5.7(1.1) \\ 
& \multirow{6}{*}{} & 128 & -0.05990(24) & 0.954(19) & -21.85(94) & 4.99(85) \\ 
& \multirow{6}{*}{} & 192 & -0.05991(23) & 0.953(18) & -21.3(1.2) & 3.55(29) \\ 
\cline{2-7}
& \multirow{5}{*}{$\frac{1}{8}$} & 48 & -0.06013(23) & 0.975(19) & -26.48(89) & 2.22(93) \\ 
& \multirow{5}{*}{} & 64 & -0.06013(23) & 0.975(19) & -26.5(1.0) & 2.23(94) \\ 
& \multirow{5}{*}{} & 96 & -0.06017(23) & 0.978(18) & -27.2(1.3) & 2.1(1.0) \\ 
& \multirow{5}{*}{} & 128 & -0.06016(22) & 0.978(17) & -27.1(1.8) & 2.3(1.1) \\ 
& \multirow{5}{*}{} & 192 & -0.06011(20) & 0.973(16) & -25.3(2.6) & 1.72(85) \\ 
\hline
\hline
\multirow{10}{*}{8} & \multirow{6}{*}{$\frac{1}{4}$} & 32 & -0.06000(37) & 0.972(40) & -22.14(49) & 6.63(86) \\ 
& \multirow{6}{*}{} & 48 & -0.06004(37) & 0.977(40) & -22.29(51) & 6.31(91) \\ 
& \multirow{6}{*}{} & 64 & -0.06005(37) & 0.978(39) & -22.35(57) & 6.08(98) \\ 
& \multirow{6}{*}{} & 96 & -0.06001(35) & 0.972(37) & -22.24(70) & 5.67(98) \\ 
& \multirow{6}{*}{} & 128 & -0.05996(33) & 0.963(35) & -21.90(88) & 5.01(77) \\ 
& \multirow{6}{*}{} & 192 & -0.05991(31) & 0.954(33) & -21.3(1.2) & 3.61(29) \\ 
\cline{2-7}
& \multirow{4}{*}{$\frac{1}{8}$} & 64 & -0.06030(32) & 1.000(32) & -26.9(1.1) & 1.68(57) \\ 
& \multirow{4}{*}{} & 96 & -0.06035(32) & 1.006(33) & -27.6(1.2) & 1.42(61) \\ 
& \multirow{4}{*}{} & 128 & -0.06038(32) & 1.009(33) & -28.0(1.5) & 1.46(67) \\ 
& \multirow{4}{*}{} & 192 & -0.06031(30) & 1.000(31) & -26.7(2.3) & 1.22(58) \\ 
\end{tabular}
\end{ruledtabular}
\label{tab:ordinary_S2_noC}
\end{table*}

\begin{table*}
\caption{
Fits of the surface-bulk two-point function of the observable $S$
at the ordinary UC to Eq.~(\ref{fss_2pt_open}) with $\hat{\Delta}=3$ as a function of the minimum distance \zmin from the surface, the maximum ratio \zLmax, and the minimum lattice size \Lmin considered in the fits.
The uncertainty due to the variation of the boundary scaling dimension $\hat{\Delta}_\sigma = 2 - 0.7249(6)$ \cite{Hasenbusch-11} within one quoted error bar largely dominates over the statistical error bars of the fits. 
As a reference, a fit employing only the central value of $\hat{\Delta}_\sigma$ for $\zmin = 4$, $\zLmax=1/4$, $\Lmin=128$ results in $M_{\phi\sigma} = 0.39501(10)$.}
\begin{ruledtabular}
\begin{tabular}{cclwmmqm}
\zmin & \zLmax & \Lmin & \multicolumn{1}{c}{$M_{\phi\sigma}$} & \multicolumn{1}{c}{$z_0$} & \multicolumn{1}{c}{$B_{\phi\sigma}$} & \multicolumn{1}{c}{$C$} & \multicolumn{1}{c}{\chidof} \\
\hline\multirow{14}{*}{4} & \multirow{5}{*}{$\frac{1}{4}$} & 32 & 0.3947(11) & 1.094(11) &  2.679(17) & 3.022(78) & 30.32(86) \\ 
& \multirow{5}{*}{} & 48 & 0.3947(11) & 1.094(11) & 2.690(20) & 3.017(79) & 30.55(88) \\ 
& \multirow{5}{*}{} & 64 & 0.3948(11) & 1.096(11) & 2.627(31) & 3.037(78) & 30.56(85) \\ 
& \multirow{5}{*}{} & 96 & 0.3949(11) & 1.099(11) & 2.548(39) & 3.058(77) & 30.80(81) \\ 
& \multirow{5}{*}{} & 128 & 0.3950(11) & 1.102(11) & 2.402(57) & 3.086(77) & 30.45(71) \\ 
\cline{2-8}
 & \multirow{5}{*}{$\frac{1}{8}$} & 32 & 0.3950(11) & 1.101(11) &  2.843(46) & 3.086(75) & 56.6(1.1) \\ 
& \multirow{5}{*}{} & 48 & 0.3950(11) & 1.102(11) & 2.806(57) & 3.088(75) & 57.0(1.1) \\ 
& \multirow{5}{*}{} & 64 & 0.3950(11) & 1.101(11) & 2.822(77) & 3.085(75) & 58.2(1.1) \\ 
& \multirow{5}{*}{} & 96 & 0.3950(11) & 1.101(11) & 2.80(11) & 3.082(75) & 60.1(1.2) \\ 
& \multirow{5}{*}{} & 128 & 0.3951(11) & 1.104(11) & 2.42(16) & 3.108(77) & 62.1(1.2) \\ 
\cline{2-8}
& \multirow{4}{*}{$\frac{1}{10}$} & 48 & 0.3952(11) & 1.107(11) & 2.863(86) & 3.136(74) & 72.3(1.1) \\ 
& \multirow{4}{*}{} & 64 & 0.3952(11) & 1.108(11) & 2.78(11) & 3.138(75) & 73.0(1.1) \\ 
& \multirow{4}{*}{} & 96 & 0.3952(11) & 1.107(11) & 2.88(15) & 3.130(75) & 74.9(1.2) \\ 
& \multirow{4}{*}{} & 128 & 0.3952(11) & 1.108(11) & 2.58(23) & 3.138(77) & 77.8(1.2) \\ 
\hline
\hline
\multirow{12}{*}{6} & \multirow{5}{*}{$\frac{1}{4}$} & 32 & 0.3925(12) & 1.013(14) &  2.755(20) & 2.14(11) & 3.057(37) \\ 
& \multirow{5}{*}{} & 48 & 0.3925(12) & 1.011(14) & 2.791(23) & 2.12(11) & 2.565(57) \\ 
& \multirow{5}{*}{} & 64 & 0.3925(12) & 1.011(15) & 2.792(29) & 2.12(11) & 2.241(57) \\ 
& \multirow{5}{*}{} & 96 & 0.3924(12) & 1.011(14) & 2.798(37) & 2.12(11) & 2.170(62) \\ 
& \multirow{5}{*}{} & 128 & 0.3925(12) & 1.012(14) & 2.770(55) & 2.14(11) & 2.039(42) \\ 
\cline{2-8}
& \multirow{4}{*}{$\frac{1}{8}$} & 48 & 0.3923(11) & 1.007(14) & 3.076(68) & 2.086(99) & 1.55(10) \\ 
& \multirow{4}{*}{} & 64 & 0.3923(11) & 1.007(14) & 3.080(79) & 2.086(99) & 1.56(10) \\ 
& \multirow{4}{*}{} & 96 & 0.3922(11) & 1.005(14) & 3.19(10) & 2.072(99) & 1.38(11) \\ 
& \multirow{4}{*}{} & 128 & 0.3921(12) & 1.003(14) & 3.31(15) & 2.05(10) & 1.35(13) \\ 
\cline{2-8}
& \multirow{3}{*}{$\frac{1}{10}$} & 64 & 0.3922(11) & 1.005(13) & 3.25(13) & 2.075(96) & 1.73(14) \\ 
& \multirow{3}{*}{} & 96 & 0.3922(11) & 1.005(13) & 3.30(15) & 2.073(96) & 1.73(14) \\ 
& \multirow{3}{*}{} & 128 & 0.3921(12) & 1.003(14) & 3.49(22) & 2.06(10) & 1.64(15) \\ 
\hline
\hline
\multirow{10}{*}{8} & \multirow{5}{*}{$\frac{1}{4}$} & 32 & 0.3923(13) & 1.002(19) &  2.756(23) & 2.00(18) & 2.676(38) \\ 
& \multirow{5}{*}{} & 48 & 0.3922(13) & 1.001(19) & 2.789(25) & 1.99(18) & 2.240(15) \\ 
& \multirow{5}{*}{} & 64 & 0.3922(13) & 0.999(20) & 2.802(32) & 1.96(18) & 1.982(15) \\ 
& \multirow{5}{*}{} & 96 & 0.3922(13) & 0.998(20) & 2.817(37) & 1.95(18) & 1.890(11) \\ 
& \multirow{5}{*}{} & 128 & 0.3922(13) & 1.000(20) & 2.799(54) & 1.98(18) & 1.790(19) \\ 
\cline{2-8}
& \multirow{3}{*}{$\frac{1}{8}$} & 64 & 0.3918(12) & 0.987(18) & 3.152(89) & 1.85(16) & 0.700(12) \\ 
& \multirow{3}{*}{} & 96 & 0.3918(12) & 0.987(18) & 3.22(10) & 1.85(16) & 0.5839(64) \\ 
& \multirow{3}{*}{} & 128 & 0.3917(13) & 0.984(19) & 3.36(15) & 1.81(17) & 0.4473(36) \\ 
\cline{2-8}
& \multirow{2}{*}{$\frac{1}{10}$} & 96 & 0.3916(12) & 0.982(18) & 3.39(17) & 1.79(16) & 0.526(11) \\ 
& \multirow{2}{*}{} & 128 & 0.3916(13) & 0.981(19) & 3.53(23) & 1.79(16) & 0.438(14) \\ 
\end{tabular}
\end{ruledtabular}
\label{tab:ordinary_sigma-phi}
\end{table*}


\begin{table*}
\caption{
Fits of the two-point function of surface order parameter $S$ at the special UC to Eq.~(\ref{2pt_fit_Ssurf}), as a function of the minimum distance \xmin, the maximum value \xLmax, and the minimum lattice size \Lmin considered in the fits.
In the quoted error bars we sum the statistical error of the fit, the variation coming from $\hat{\Delta}_\sigma$ and the variation coming from $\hat{\Delta}_\varepsilon$.
As a reference, a fit employing only the central values of $\hat{\Delta}_\sigma$ and $\hat{\Delta}_\varepsilon$ for $\xmin = 4$, $\xLmax=1/4$, $\Lmin=48$ results in ${\cal N}_\sigma^2=0.354240(14)$.
}
\begin{ruledtabular}
\begin{tabular}{cclwmmq}
\xmin & \xLmax & \Lmin & \multicolumn{1}{c}{${\cal N}_\sigma^2$} & \multicolumn{1}{c}{$B_{\sigma\sigma}$} & \multicolumn{1}{c}{$C$} & \multicolumn{1}{c}{\chidof} \\
\hline\multirow{10}{*}{4} & \multirow{5}{*}{$\frac{1}{4}$} & 32 & 0.35424(56) & 1.509(12) & -0.031(14) & 15.0(1.4) \\ 
& \multirow{5}{*}{} & 48 & 0.35424(56) & 1.509(13) & -0.031(13) & 14.3(1.3) \\ 
& \multirow{5}{*}{} & 64 & 0.35424(55) & 1.509(13) & -0.031(13) & 13.7(1.2) \\ 
& \multirow{5}{*}{} & 96 & 0.35425(54) & 1.508(14) & -0.031(13) & 12.8(1.1) \\ 
& \multirow{5}{*}{} & 128 & 0.35425(54) & 1.508(15) & -0.031(12) & 11.71(88) \\ 
\cline{2-7}
 & \multirow{5}{*}{$\frac{1}{8}$} & 32 & 0.35440(54) & 1.489(15) & -0.038(12) & 2.37(93) \\ 
& \multirow{5}{*}{} & 48 & 0.35441(54) & 1.488(16) & -0.038(12) & 2.33(96) \\ 
& \multirow{5}{*}{} & 64 & 0.35441(53) & 1.488(17) & -0.038(12) & 2.32(99) \\ 
& \multirow{5}{*}{} & 96 & 0.35441(52) & 1.488(18) & -0.038(12) & 2.27(95) \\ 
& \multirow{5}{*}{} & 128 & 0.35441(50) & 1.488(20) & -0.038(11) & 1.95(72) \\ 
\hline
\hline
\multirow{9}{*}{6} & \multirow{5}{*}{$\frac{1}{4}$} & 32 & 0.35417(66) & 1.5119(99) & -0.019(33) & 13.73(79) \\ 
& \multirow{5}{*}{} & 48 & 0.35418(66) & 1.511(10) & -0.021(32) & 13.22(76) \\ 
& \multirow{5}{*}{} & 64 & 0.35419(65) & 1.511(11) & -0.021(31) & 12.73(73) \\ 
& \multirow{5}{*}{} & 96 & 0.35419(64) & 1.510(12) & -0.021(30) & 11.98(65) \\ 
& \multirow{5}{*}{} & 128 & 0.35419(63) & 1.510(13) & -0.019(28) & 10.97(52) \\ 
\cline{2-7}
& \multirow{4}{*}{$\frac{1}{8}$} & 48 & 0.35444(64) & 1.487(12) & -0.043(29) & 1.63(81) \\ 
& \multirow{4}{*}{} & 64 & 0.35445(63) & 1.486(13) & -0.043(29) & 1.53(79) \\ 
& \multirow{4}{*}{} & 96 & 0.35445(63) & 1.486(14) & -0.044(28) & 1.44(76) \\ 
& \multirow{4}{*}{} & 128 & 0.35445(61) & 1.486(16) & -0.044(26) & 1.35(70) \\ 
\hline
\hline
\multirow{8}{*}{8} & \multirow{5}{*}{$\frac{1}{4}$} & 32 & 0.35406(74) & 1.5152(86) & 0.017(61) & 12.02(51) \\ 
& \multirow{5}{*}{} & 48 & 0.35407(73) & 1.5146(88) & 0.014(60) & 11.69(51) \\ 
& \multirow{5}{*}{} & 64 & 0.35408(73) & 1.5139(92) & 0.012(59) & 11.33(50) \\ 
& \multirow{5}{*}{} & 96 & 0.35409(72) & 1.5135(99) & 0.012(57) & 10.74(45) \\ 
& \multirow{5}{*}{} & 128 & 0.35407(70) & 1.514(11) & 0.018(52) & 9.68(37) \\ 
\cline{2-7}
& \multirow{3}{*}{$\frac{1}{8}$} & 64 & 0.35441(71) & 1.488(11) & -0.034(54) & 1.08(46) \\ 
& \multirow{3}{*}{} & 96 & 0.35442(71) & 1.487(11) & -0.035(54) & 1.06(46) \\ 
& \multirow{3}{*}{} & 128 & 0.35442(70) & 1.487(13) & -0.035(51) & 1.01(44) \\
\end{tabular}
\end{ruledtabular}
\label{tab:special_sigma-sigma}
\end{table*}

\begin{table*}
\caption{
Same as Table \ref{tab:special_sigma-sigma} setting $C=0$.
In the quoted error bars we sum the statistical error of the fit, the variation coming from $\hat{\Delta}_\sigma$ and the variation coming from $\hat{\Delta}_\varepsilon$.
As a reference, a fit employing only the central values of $\hat{\Delta}_\sigma$ and $\hat{\Delta}_\varepsilon$ for $\xmin = 4$, $\xLmax=1/8$, $\Lmin=48$ results in ${\cal N}_\sigma^2=0.3538288(55)$.
}
\begin{ruledtabular}
\begin{tabular}{cclwmq}
\xmin & \xLmax & \Lmin & \multicolumn{1}{c}{${\cal N}_\sigma^2$} & \multicolumn{1}{c}{$B_{\sigma\sigma}$} & \multicolumn{1}{c}{\chidof} \\
\hline\multirow{10}{*}{4} & \multirow{5}{*}{$\frac{1}{4}$} & 32 & 0.35383(38) & 1.524(18) & 50.3(3.8) \\ 
& \multirow{5}{*}{} & 48 & 0.35383(38) & 1.524(19) & 49.5(3.6) \\ 
& \multirow{5}{*}{} & 64 & 0.35383(38) & 1.524(20) & 49.4(3.4) \\ 
& \multirow{5}{*}{} & 96 & 0.35383(38) & 1.524(21) & 49.1(3.2) \\ 
& \multirow{5}{*}{} & 128 & 0.35384(38) & 1.525(22) & 49.2(3.0) \\ 
\cline{2-6}
 & \multirow{5}{*}{$\frac{1}{8}$} & 32 & 0.35383(35) & 1.524(26) & 97.9(5.9) \\ 
& \multirow{5}{*}{} & 48 & 0.35383(35) & 1.525(27) & 96.6(5.7) \\ 
& \multirow{5}{*}{} & 64 & 0.35382(34) & 1.526(28) & 95.9(5.3) \\ 
& \multirow{5}{*}{} & 96 & 0.35382(34) & 1.529(30) & 94.1(4.9) \\ 
& \multirow{5}{*}{} & 128 & 0.35382(34) & 1.532(32) & 92.2(4.1) \\ 
\hline
\hline
\multirow{9}{*}{6} & \multirow{5}{*}{$\frac{1}{4}$} & 32 & 0.35406(47) & 1.515(15) & 15.6(1.4) \\ 
& \multirow{5}{*}{} & 48 & 0.35406(47) & 1.515(16) & 15.4(1.5) \\ 
& \multirow{5}{*}{} & 64 & 0.35406(46) & 1.515(16) & 15.0(1.5) \\ 
& \multirow{5}{*}{} & 96 & 0.35407(46) & 1.514(17) & 14.2(1.4) \\ 
& \multirow{5}{*}{} & 128 & 0.35407(46) & 1.514(18) & 12.8(1.0) \\ 
\cline{2-6}
& \multirow{4}{*}{$\frac{1}{8}$} & 48 & 0.35414(43) & 1.502(22) & 16.2(3.1) \\ 
& \multirow{4}{*}{} & 64 & 0.35414(43) & 1.503(22) & 16.2(3.0) \\ 
& \multirow{4}{*}{} & 96 & 0.35414(42) & 1.503(23) & 15.7(2.8) \\ 
& \multirow{4}{*}{} & 128 & 0.35413(41) & 1.505(26) & 14.3(2.2) \\ 
\hline
\hline
\multirow{8}{*}{8} & \multirow{5}{*}{$\frac{1}{4}$} & 32 & 0.35412(53) & 1.514(14) & 13.1(1.4) \\ 
& \multirow{5}{*}{} & 48 & 0.35412(53) & 1.513(14) & 12.6(1.3) \\ 
& \multirow{5}{*}{} & 64 & 0.35412(52) & 1.513(14) & 12.0(1.2) \\ 
& \multirow{5}{*}{} & 96 & 0.35413(52) & 1.512(15) & 11.4(1.0) \\ 
& \multirow{5}{*}{} & 128 & 0.35414(51) & 1.512(16) & 10.49(98) \\ 
\cline{2-6}
& \multirow{3}{*}{$\frac{1}{8}$} & 64 & 0.35427(49) & 1.494(19) & 4.1(1.8) \\ 
& \multirow{3}{*}{} & 96 & 0.35427(48) & 1.495(20) & 4.1(1.8) \\ 
& \multirow{3}{*}{} & 128 & 0.35427(47) & 1.495(22) & 3.5(1.4) \\
\end{tabular}
\end{ruledtabular}
\label{tab:special_sigma-sigma_noC}
\end{table*}

\begin{table*}
\caption{Fits of the energy profile at the special UC to Eq.~(\ref{fss_1ptE_open}), as a function of the minimum distance \zmin from the surface, the maximum ratio \zLmax, and the minimum lattice size \Lmin considered in the fits.
We employ the fitted value of $E_0$ reported in Table \ref{tab:fitresults}.
In the quoted error bars we sum the statistical error of the fit, the variation coming from $E_0$, the variation coming from $\hat{\Delta}$ and the variation coming from $\hat{\Delta}_\varepsilon$.
}
\begin{ruledtabular}
\begin{tabular}{cclwmmqm}
\zmin & \zLmax & \Lmin & \multicolumn{1}{c}{$A_\epsilon$} & \multicolumn{1}{c}{$z_0$} & \multicolumn{1}{c}{$\mathcal{B}_{\epsilon}$} & \multicolumn{1}{c}{$C$} & \multicolumn{1}{c}{\chidof} \\
\hline\multirow{17}{*}{4} & \multirow{6}{*}{$\frac{1}{4}$} & 32 & 0.8461(33) & 0.535(32) & 5.556(19) & -0.09(13) & 29.71 \\ 
& \multirow{6}{*}{} & 48 & 0.8457(33) & 0.530(32) & 5.564(20) & -0.11(13) & 28.66 \\ 
& \multirow{6}{*}{} & 64 & 0.8466(33) & 0.538(32) & 5.548(21) & -0.09(13) & 26.47 \\ 
& \multirow{6}{*}{} & 96 & 0.8486(32) & 0.557(32) & 5.513(23) & -0.01(13) & 23.13 \\ 
& \multirow{6}{*}{} & 128 & 0.8516(31) & 0.590(31) & 5.460(35) & 0.13(13) & 19.28 \\ 
& \multirow{6}{*}{} & 192 & 0.8577(26) & 0.658(27) & 5.347(50) & 0.43(13) & 12.78 \\ 
\cline{2-8}
 & \multirow{6}{*}{$\frac{1}{8}$} & 32 & 0.8393(29) & 0.510(26) & 5.877(25) & -0.12(10) & 5.26 \\ 
& \multirow{6}{*}{} & 48 & 0.8390(30) & 0.508(27) & 5.885(26) & -0.12(10) & 5.14 \\ 
& \multirow{6}{*}{} & 64 & 0.8389(30) & 0.507(27) & 5.886(29) & -0.13(10) & 5.0 \\ 
& \multirow{6}{*}{} & 96 & 0.8394(31) & 0.511(28) & 5.873(32) & -0.12(11) & 4.62 \\ 
& \multirow{6}{*}{} & 128 & 0.8396(33) & 0.512(29) & 5.870(38) & -0.12(11) & 4.18 \\ 
& \multirow{6}{*}{} & 192 & 0.8423(34) & 0.537(30) & 5.801(52) & -0.03(12) & 2.96 \\ 
\cline{2-8}
& \multirow{5}{*}{$\frac{1}{12}$} & 48 & 0.8379(27) & 0.509(23) & 5.985(30) & -0.105(86) & 1.66 \\ 
& \multirow{5}{*}{} & 64 & 0.8379(28) & 0.509(23) & 5.986(32) & -0.105(86) & 1.68 \\ 
& \multirow{5}{*}{} & 96 & 0.8381(28) & 0.511(24) & 5.977(35) & -0.102(88) & 1.62 \\ 
& \multirow{5}{*}{} & 128 & 0.8378(31) & 0.508(25) & 5.987(45) & -0.115(94) & 1.49 \\ 
& \multirow{5}{*}{} & 192 & 0.8392(36) & 0.519(30) & 5.939(64) & -0.08(11) & 1.23 \\ 
\hline
\hline
\multirow{14}{*}{6} & \multirow{6}{*}{$\frac{1}{4}$} & 32 & 0.8475(47) & 0.518(64) & 5.502(30) & -0.34(37) & 22.93 \\ 
& \multirow{6}{*}{} & 48 & 0.8464(48) & 0.504(64) & 5.520(31) & -0.41(37) & 22.55 \\ 
& \multirow{6}{*}{} & 64 & 0.8464(49) & 0.502(64) & 5.520(34) & -0.44(37) & 21.69 \\ 
& \multirow{6}{*}{} & 96 & 0.8477(50) & 0.518(66) & 5.499(36) & -0.37(38) & 19.84 \\ 
& \multirow{6}{*}{} & 128 & 0.8511(50) & 0.564(68) & 5.450(37) & -0.12(40) & 17.14 \\ 
& \multirow{6}{*}{} & 192 & 0.8602(49) & 0.693(68) & 5.308(48) & 0.65(44) & 11.82 \\ 
\cline{2-8}
& \multirow{5}{*}{$\frac{1}{8}$} & 48 & 0.8404(43) & 0.514(53) & 5.838(38) & -0.14(30) & 3.88 \\ 
& \multirow{5}{*}{} & 64 & 0.8402(43) & 0.512(54) & 5.844(40) & -0.14(30) & 3.87 \\ 
& \multirow{5}{*}{} & 96 & 0.8402(45) & 0.512(55) & 5.841(43) & -0.15(30) & 3.86 \\ 
& \multirow{5}{*}{} & 128 & 0.8399(49) & 0.506(59) & 5.848(54) & -0.19(32) & 3.65 \\ 
& \multirow{5}{*}{} & 192 & 0.8428(55) & 0.538(67) & 5.784(70) & -0.04(37) & 2.73 \\ 
\cline{2-8}
& \multirow{3}{*}{$\frac{1}{12}$} & 96 & 0.8394(41) & 0.526(47) & 5.945(45) & -0.02(25) & 1.38 \\ 
& \multirow{3}{*}{} & 128 & 0.8389(44) & 0.520(49) & 5.960(55) & -0.05(26) & 1.34 \\ 
& \multirow{3}{*}{} & 192 & 0.8404(52) & 0.534(58) & 5.914(82) & 0.007(303) & 1.14 \\ 
\hline
\hline
\multirow{13}{*}{8} & \multirow{6}{*}{$\frac{1}{4}$} & 32 & 0.8479(62) & 0.47(10) & 5.464(40) & -1.04(75) & 17.69 \\ 
& \multirow{6}{*}{} & 48 & 0.8473(62) & 0.46(10) & 5.474(41) & -1.07(75) & 17.49 \\ 
& \multirow{6}{*}{} & 64 & 0.8464(63) & 0.45(10) & 5.486(44) & -1.19(74) & 17.22 \\ 
& \multirow{6}{*}{} & 96 & 0.8469(66) & 0.45(11) & 5.478(48) & -1.23(76) & 16.11 \\ 
& \multirow{6}{*}{} & 128 & 0.8494(69) & 0.49(11) & 5.444(54) & -1.02(82) & 14.22 \\ 
& \multirow{6}{*}{} & 192 & 0.8585(71) & 0.64(12) & 5.316(53) & 0.02(93) & 10.69 \\ 
\cline{2-8}
& \multirow{4}{*}{$\frac{1}{8}$} & 64 & 0.8414(57) & 0.512(88) & 5.800(51) & -0.24(63) & 2.95 \\ 
& \multirow{4}{*}{} & 96 & 0.8414(58) & 0.511(89) & 5.801(53) & -0.24(63) & 2.98 \\ 
& \multirow{4}{*}{} & 128 & 0.8408(63) & 0.500(96) & 5.812(66) & -0.33(67) & 2.92 \\ 
& \multirow{4}{*}{} & 192 & 0.8427(73) & 0.52(11) & 5.768(86) & -0.26(75) & 2.36 \\ 
\cline{2-8}
& \multirow{3}{*}{$\frac{1}{12}$} & 96 & 0.8408(56) & 0.543(83) & 5.909(62) & 0.10(58) & 1.1 \\ 
& \multirow{3}{*}{} & 128 & 0.8406(58) & 0.542(83) & 5.918(68) & 0.10(58) & 1.08 \\ 
& \multirow{3}{*}{} & 192 & 0.8416(65) & 0.549(90) & 5.886(90) & 0.11(61) & 0.97 \\
\end{tabular}
\end{ruledtabular}
\label{tab:special_E}
\end{table*}

\begin{table*}
\caption{Same as Table \ref{tab:special_E} setting $C=0$.}
\begin{ruledtabular}
\begin{tabular}{cclwmmqm}
\zmin & \zLmax & \Lmin & \multicolumn{1}{c}{$A_\epsilon$} & \multicolumn{1}{c}{$z_0$} & \multicolumn{1}{c}{$\mathcal{B}_{\epsilon}$} & \multicolumn{1}{c}{\chidof} \\
\hline\multirow{17}{*}{4} & \multirow{6}{*}{$\frac{1}{4}$} & 32 & 0.8475(16) & 0.5535(61) & 5.541(32) & 29.8 \\ 
& \multirow{6}{*}{} & 48 & 0.8475(16) & 0.5536(58) & 5.544(36) & 28.82 \\ 
& \multirow{6}{*}{} & 64 & 0.8480(15) & 0.5562(56) & 5.531(41) & 26.54 \\ 
& \multirow{6}{*}{} & 96 & 0.8488(14) & 0.5600(51) & 5.510(48) & 23.09 \\ 
& \multirow{6}{*}{} & 128 & 0.8494(12) & 0.5636(45) & 5.491(56) & 19.41 \\ 
& \multirow{6}{*}{} & 192 & 0.8501(10) & 0.5684(36) & 5.460(66) & 14.54 \\ 
\cline{2-7}
 & \multirow{6}{*}{$\frac{1}{8}$} & 32 & 0.8412(15) & 0.5353(53) & 5.855(39) & 5.7 \\ 
& \multirow{6}{*}{} & 48 & 0.8411(15) & 0.5347(53) & 5.861(41) & 5.62 \\ 
& \multirow{6}{*}{} & 64 & 0.8412(15) & 0.5351(53) & 5.858(45) & 5.52 \\ 
& \multirow{6}{*}{} & 96 & 0.8416(14) & 0.5369(51) & 5.842(50) & 5.07 \\ 
& \multirow{6}{*}{} & 128 & 0.8419(13) & 0.5385(48) & 5.830(60) & 4.57 \\ 
& \multirow{6}{*}{} & 192 & 0.8429(11) & 0.5428(40) & 5.790(74) & 2.97 \\ 
\cline{2-7}
& \multirow{5}{*}{$\frac{1}{12}$} & 48 & 0.8397(14) & 0.5319(48) & 5.963(45) & 2.12 \\ 
& \multirow{5}{*}{} & 64 & 0.8397(14) & 0.5318(48) & 5.963(48) & 2.15 \\ 
& \multirow{5}{*}{} & 96 & 0.8399(14) & 0.5326(49) & 5.953(52) & 2.07 \\ 
& \multirow{5}{*}{} & 128 & 0.8400(14) & 0.5331(48) & 5.950(62) & 2.04 \\ 
& \multirow{5}{*}{} & 192 & 0.8410(13) & 0.5367(45) & 5.901(80) & 1.44 \\ 
\hline
\hline
\multirow{14}{*}{6} & \multirow{6}{*}{$\frac{1}{4}$} & 32 & 0.8500(25) & 0.566(13) & 5.481(27) & 23.15 \\ 
& \multirow{6}{*}{} & 48 & 0.8495(25) & 0.563(13) & 5.493(30) & 22.88 \\ 
& \multirow{6}{*}{} & 64 & 0.8499(24) & 0.565(13) & 5.487(35) & 22.07 \\ 
& \multirow{6}{*}{} & 96 & 0.8508(23) & 0.572(12) & 5.467(43) & 20.07 \\ 
& \multirow{6}{*}{} & 128 & 0.8522(21) & 0.582(11) & 5.438(53) & 17.13 \\ 
& \multirow{6}{*}{} & 192 & 0.8544(17) & 0.5988(88) & 5.382(64) & 12.3 \\ 
\cline{2-7}
& \multirow{5}{*}{$\frac{1}{8}$} & 48 & 0.8415(23) & 0.534(11) & 5.828(40) & 3.97 \\ 
& \multirow{5}{*}{} & 64 & 0.8413(23) & 0.533(12) & 5.833(41) & 3.96 \\ 
& \multirow{5}{*}{} & 96 & 0.8415(23) & 0.534(12) & 5.829(45) & 3.95 \\ 
& \multirow{5}{*}{} & 128 & 0.8416(23) & 0.535(12) & 5.826(57) & 3.79 \\ 
& \multirow{5}{*}{} & 192 & 0.8433(21) & 0.545(11) & 5.777(73) & 2.74 \\ 
\cline{2-7}
& \multirow{3}{*}{$\frac{1}{12}$} & 96 & 0.8396(22) & 0.528(10) & 5.944(54) & 1.38 \\ 
& \multirow{3}{*}{} & 128 & 0.8393(23) & 0.527(11) & 5.956(60) & 1.35 \\ 
& \multirow{3}{*}{} & 192 & 0.8403(25) & 0.532(12) & 5.915(81) & 1.14 \\ 
\hline
\hline
\multirow{13}{*}{8} & \multirow{6}{*}{$\frac{1}{4}$} & 32 & 0.8524(34) & 0.580(24) & 5.429(29) & 18.17 \\ 
& \multirow{6}{*}{} & 48 & 0.8520(34) & 0.577(24) & 5.437(29) & 18.0 \\ 
& \multirow{6}{*}{} & 64 & 0.8519(34) & 0.576(23) & 5.441(31) & 17.86 \\ 
& \multirow{6}{*}{} & 96 & 0.8529(33) & 0.585(23) & 5.424(38) & 16.77 \\ 
& \multirow{6}{*}{} & 128 & 0.8548(31) & 0.602(21) & 5.389(50) & 14.61 \\ 
& \multirow{6}{*}{} & 192 & 0.8584(26) & 0.635(18) & 5.317(62) & 10.66 \\ 
\cline{2-7}
& \multirow{4}{*}{$\frac{1}{8}$} & 64 & 0.8426(31) & 0.538(20) & 5.791(50) & 3.0 \\ 
& \multirow{4}{*}{} & 96 & 0.8425(32) & 0.538(20) & 5.792(52) & 3.03 \\ 
& \multirow{4}{*}{} & 128 & 0.8425(33) & 0.538(21) & 5.796(57) & 3.01 \\ 
& \multirow{4}{*}{} & 192 & 0.8442(34) & 0.551(22) & 5.750(72) & 2.41 \\ 
\cline{2-7}
& \multirow{3}{*}{$\frac{1}{12}$} & 96 & 0.8403(30) & 0.532(18) & 5.914(65) & 1.11 \\ 
& \multirow{3}{*}{} & 128 & 0.8400(32) & 0.530(19) & 5.923(72) & 1.09 \\ 
& \multirow{3}{*}{} & 192 & 0.8410(36) & 0.536(22) & 5.892(89) & 0.98 \\ 
\end{tabular}
\end{ruledtabular}
\label{tab:special_E_noC}
\end{table*}

\begin{table*}
\caption{Fits of the $S^2$ profile at the special UC to Eq.~(\ref{fss_1ptS2_open}), as a function of the minimum distance \zmin from the surface, the maximum ratio \zLmax, and the minimum lattice size \Lmin considered in the fits.
We employ the fitted value of $S^2_0$ reported in Table \ref{tab:additional_fitresults}.
In the quoted error bars we sum the statistical error of the fit, the variation coming from $S^2_0$, the variation coming from $\hat{\Delta}$ and the variation coming from $\hat{\Delta}_\varepsilon$.
}
\begin{ruledtabular}
\begin{tabular}{cclwmmqm}
\zmin & \zLmax & \Lmin & \multicolumn{1}{c}{$A_\epsilon$} & \multicolumn{1}{c}{$z_0$} & \multicolumn{1}{c}{$\mathcal{B}_{\epsilon}$} & \multicolumn{1}{c}{$C$} & \multicolumn{1}{c}{\chidof} \\
\hline\multirow{17}{*}{4} & \multirow{6}{*}{$\frac{1}{4}$} & 32 & 0.09339(45) & 0.529(38) & 5.552(35) & 0.04(16) & 31.32 \\ 
& \multirow{6}{*}{} & 48 & 0.09334(45) & 0.524(37) & 5.560(36) & 0.01(15) & 30.23 \\ 
& \multirow{6}{*}{} & 64 & 0.09344(45) & 0.532(37) & 5.545(36) & 0.04(15) & 28.04 \\ 
& \multirow{6}{*}{} & 96 & 0.09366(44) & 0.552(37) & 5.510(38) & 0.12(16) & 24.57 \\ 
& \multirow{6}{*}{} & 128 & 0.09401(42) & 0.586(36) & 5.455(50) & 0.27(16) & 20.58 \\ 
& \multirow{6}{*}{} & 192 & 0.09471(38) & 0.657(33) & 5.335(66) & 0.60(16) & 13.68 \\ 
\cline{2-8}
 & \multirow{6}{*}{$\frac{1}{8}$} & 32 & 0.09263(38) & 0.505(30) & 5.875(50) & 0.01(12) & 5.45 \\ 
& \multirow{6}{*}{} & 48 & 0.09260(39) & 0.503(30) & 5.883(51) & 0.004(122) & 5.3 \\ 
& \multirow{6}{*}{} & 64 & 0.09259(39) & 0.502(31) & 5.885(54) & -0.002(124) & 5.17 \\ 
& \multirow{6}{*}{} & 96 & 0.09264(40) & 0.505(32) & 5.873(57) & 0.006(129) & 4.8 \\ 
& \multirow{6}{*}{} & 128 & 0.09267(42) & 0.507(33) & 5.869(63) & 0.009(132) & 4.37 \\ 
& \multirow{6}{*}{} & 192 & 0.09298(43) & 0.533(34) & 5.798(75) & 0.11(14) & 3.13 \\ 
\cline{2-8}
& \multirow{5}{*}{$\frac{1}{12}$} & 48 & 0.09249(35) & 0.506(26) & 5.983(59) & 0.03(10) & 1.62 \\ 
& \multirow{5}{*}{} & 64 & 0.09249(35) & 0.505(26) & 5.986(61) & 0.03(10) & 1.63 \\ 
& \multirow{5}{*}{} & 96 & 0.09251(36) & 0.507(27) & 5.977(65) & 0.03(10) & 1.58 \\ 
& \multirow{5}{*}{} & 128 & 0.09247(38) & 0.503(29) & 5.987(75) & 0.02(11) & 1.45 \\ 
& \multirow{5}{*}{} & 192 & 0.09263(44) & 0.515(33) & 5.940(94) & 0.06(13) & 1.19 \\ 
\hline
\hline
\multirow{14}{*}{6} & \multirow{6}{*}{$\frac{1}{4}$} & 32 & 0.09351(63) & 0.505(73) & 5.501(46) & -0.27(43) & 24.27 \\ 
& \multirow{6}{*}{} & 48 & 0.09338(64) & 0.492(73) & 5.520(48) & -0.34(42) & 23.83 \\ 
& \multirow{6}{*}{} & 64 & 0.09337(64) & 0.488(73) & 5.520(50) & -0.37(42) & 22.96 \\ 
& \multirow{6}{*}{} & 96 & 0.09353(65) & 0.504(74) & 5.500(52) & -0.30(43) & 21.02 \\ 
& \multirow{6}{*}{} & 128 & 0.09391(66) & 0.552(76) & 5.449(51) & -0.04(45) & 18.24 \\ 
& \multirow{6}{*}{} & 192 & 0.09494(64) & 0.685(78) & 5.303(61) & 0.76(50) & 12.72 \\ 
\cline{2-8}
& \multirow{5}{*}{$\frac{1}{8}$} & 48 & 0.09272(55) & 0.503(60) & 5.840(62) & -0.05(34) & 4.0 \\ 
& \multirow{5}{*}{} & 64 & 0.09270(55) & 0.502(60) & 5.845(64) & -0.05(34) & 3.99 \\ 
& \multirow{5}{*}{} & 96 & 0.09271(57) & 0.501(62) & 5.843(68) & -0.06(35) & 3.98 \\ 
& \multirow{5}{*}{} & 128 & 0.09267(61) & 0.495(67) & 5.850(79) & -0.10(37) & 3.77 \\ 
& \multirow{5}{*}{} & 192 & 0.09299(68) & 0.527(74) & 5.786(94) & 0.05(42) & 2.86 \\ 
\cline{2-8}
& \multirow{3}{*}{$\frac{1}{12}$} & 96 & 0.09263(51) & 0.518(53) & 5.947(74) & 0.10(29) & 1.34 \\ 
& \multirow{3}{*}{} & 128 & 0.09257(54) & 0.512(55) & 5.962(85) & 0.07(30) & 1.3 \\ 
& \multirow{3}{*}{} & 192 & 0.09274(63) & 0.525(64) & 5.92(11) & 0.12(34) & 1.11 \\ 
\hline
\hline
\multirow{13}{*}{8} & \multirow{6}{*}{$\frac{1}{4}$} & 32 & 0.09352(81) & 0.45(12) & 5.465(56) & -1.03(86) & 18.73 \\ 
& \multirow{6}{*}{} & 48 & 0.09345(81) & 0.44(12) & 5.475(57) & -1.07(86) & 18.5 \\ 
& \multirow{6}{*}{} & 64 & 0.09335(82) & 0.43(12) & 5.488(60) & -1.19(85) & 18.23 \\ 
& \multirow{6}{*}{} & 96 & 0.09340(85) & 0.43(12) & 5.481(65) & -1.24(87) & 17.06 \\ 
& \multirow{6}{*}{} & 128 & 0.09369(88) & 0.47(13) & 5.446(69) & -1.02(92) & 15.1 \\ 
& \multirow{6}{*}{} & 192 & 0.09472(90) & 0.62(13) & 5.314(65) & 0.04(1.03) & 11.46 \\ 
\cline{2-8}
& \multirow{4}{*}{$\frac{1}{8}$} & 64 & 0.09281(72) & 0.496(99) & 5.804(75) & -0.20(72) & 3.04 \\ 
& \multirow{4}{*}{} & 96 & 0.09281(73) & 0.50(10) & 5.805(77) & -0.20(72) & 3.07 \\ 
& \multirow{4}{*}{} & 128 & 0.09274(78) & 0.48(11) & 5.816(91) & -0.29(76) & 3.0 \\ 
& \multirow{4}{*}{} & 192 & 0.09295(89) & 0.50(12) & 5.77(11) & -0.22(85) & 2.45 \\ 
\cline{2-8}
& \multirow{3}{*}{$\frac{1}{12}$} & 96 & 0.09278(69) & 0.533(91) & 5.913(90) & 0.19(64) & 1.06 \\ 
& \multirow{3}{*}{} & 128 & 0.09275(71) & 0.531(92) & 5.922(96) & 0.19(64) & 1.04 \\ 
& \multirow{3}{*}{} & 192 & 0.09286(78) & 0.538(99) & 5.89(12) & 0.20(67) & 0.94 \\
\end{tabular}
\end{ruledtabular}
\label{tab:special_S2}
\end{table*}

\begin{table*}
\caption{Same as Table \ref{tab:special_S2} setting $C=0$.}
\begin{ruledtabular}
\begin{tabular}{cclwmmqm}
\zmin & \zLmax & \Lmin & \multicolumn{1}{c}{$A_\epsilon$} & \multicolumn{1}{c}{$z_0$} & \multicolumn{1}{c}{$\mathcal{B}_{\epsilon}$} & \multicolumn{1}{c}{\chidof} \\
\hline\multirow{17}{*}{4} & \multirow{6}{*}{$\frac{1}{4}$} & 32 & 0.09333(20) & 0.5220(66) & 5.557(34) & 31.32 \\ 
& \multirow{6}{*}{} & 48 & 0.09332(19) & 0.5217(64) & 5.562(38) & 30.21 \\ 
& \multirow{6}{*}{} & 64 & 0.09337(18) & 0.5240(60) & 5.552(44) & 28.03 \\ 
& \multirow{6}{*}{} & 96 & 0.09344(16) & 0.5273(55) & 5.535(51) & 24.72 \\ 
& \multirow{6}{*}{} & 128 & 0.09350(14) & 0.5307(47) & 5.518(61) & 21.4 \\ 
& \multirow{6}{*}{} & 192 & 0.09357(11) & 0.5351(37) & 5.491(73) & 17.13 \\ 
\cline{2-7}
 & \multirow{6}{*}{$\frac{1}{8}$} & 32 & 0.09262(18) & 0.5032(59) & 5.877(40) & 5.49 \\ 
& \multirow{6}{*}{} & 48 & 0.09260(18) & 0.5025(58) & 5.884(42) & 5.34 \\ 
& \multirow{6}{*}{} & 64 & 0.09260(18) & 0.5026(57) & 5.885(46) & 5.22 \\ 
& \multirow{6}{*}{} & 96 & 0.09263(17) & 0.5038(55) & 5.874(52) & 4.84 \\ 
& \multirow{6}{*}{} & 128 & 0.09265(16) & 0.5047(51) & 5.872(64) & 4.41 \\ 
& \multirow{6}{*}{} & 192 & 0.09272(13) & 0.5079(42) & 5.845(80) & 3.35 \\ 
\cline{2-7}
& \multirow{5}{*}{$\frac{1}{12}$} & 48 & 0.09243(17) & 0.4991(52) & 5.990(46) & 1.71 \\ 
& \multirow{5}{*}{} & 64 & 0.09243(17) & 0.4989(53) & 5.992(49) & 1.72 \\ 
& \multirow{5}{*}{} & 96 & 0.09245(17) & 0.4995(53) & 5.985(53) & 1.68 \\ 
& \multirow{5}{*}{} & 128 & 0.09243(17) & 0.4992(52) & 5.994(64) & 1.5 \\ 
& \multirow{5}{*}{} & 192 & 0.09249(15) & 0.5012(47) & 5.970(84) & 1.35 \\ 
\hline
\hline
\multirow{14}{*}{6} & \multirow{6}{*}{$\frac{1}{4}$} & 32 & 0.09372(31) & 0.543(15) & 5.484(28) & 24.42 \\ 
& \multirow{6}{*}{} & 48 & 0.09366(30) & 0.540(15) & 5.497(31) & 24.08 \\ 
& \multirow{6}{*}{} & 64 & 0.09370(29) & 0.542(14) & 5.493(37) & 23.25 \\ 
& \multirow{6}{*}{} & 96 & 0.09380(27) & 0.548(13) & 5.474(45) & 21.19 \\ 
& \multirow{6}{*}{} & 128 & 0.09395(24) & 0.558(12) & 5.445(56) & 18.22 \\ 
& \multirow{6}{*}{} & 192 & 0.09419(19) & 0.5748(92) & 5.389(70) & 13.41 \\ 
\cline{2-7}
& \multirow{5}{*}{$\frac{1}{8}$} & 48 & 0.09276(28) & 0.510(13) & 5.837(40) & 4.05 \\ 
& \multirow{5}{*}{} & 64 & 0.09275(28) & 0.509(13) & 5.842(41) & 4.04 \\ 
& \multirow{5}{*}{} & 96 & 0.09276(28) & 0.510(13) & 5.838(45) & 4.03 \\ 
& \multirow{5}{*}{} & 128 & 0.09277(27) & 0.511(13) & 5.839(58) & 3.84 \\ 
& \multirow{5}{*}{} & 192 & 0.09293(25) & 0.520(11) & 5.794(77) & 2.9 \\ 
\cline{2-7}
& \multirow{3}{*}{$\frac{1}{12}$} & 96 & 0.09254(26) & 0.504(11) & 5.954(54) & 1.41 \\ 
& \multirow{3}{*}{} & 128 & 0.09251(27) & 0.503(12) & 5.968(60) & 1.34 \\ 
& \multirow{3}{*}{} & 192 & 0.09261(29) & 0.507(12) & 5.935(83) & 1.2 \\ 
\hline
\hline
\multirow{13}{*}{8} & \multirow{6}{*}{$\frac{1}{4}$} & 32 & 0.09402(42) & 0.560(26) & 5.430(29) & 19.25 \\ 
& \multirow{6}{*}{} & 48 & 0.09397(42) & 0.557(26) & 5.439(30) & 19.06 \\ 
& \multirow{6}{*}{} & 64 & 0.09395(41) & 0.556(26) & 5.443(31) & 18.92 \\ 
& \multirow{6}{*}{} & 96 & 0.09406(40) & 0.565(25) & 5.426(39) & 17.78 \\ 
& \multirow{6}{*}{} & 128 & 0.09428(37) & 0.582(23) & 5.391(52) & 15.53 \\ 
& \multirow{6}{*}{} & 192 & 0.09469(31) & 0.615(19) & 5.317(67) & 11.45 \\ 
\cline{2-7}
& \multirow{4}{*}{$\frac{1}{8}$} & 64 & 0.09292(38) & 0.518(22) & 5.797(50) & 3.1 \\ 
& \multirow{4}{*}{} & 96 & 0.09291(39) & 0.518(23) & 5.797(52) & 3.13 \\ 
& \multirow{4}{*}{} & 128 & 0.09290(40) & 0.517(23) & 5.802(57) & 3.09 \\ 
& \multirow{4}{*}{} & 192 & 0.09309(40) & 0.530(23) & 5.758(74) & 2.51 \\ 
\cline{2-7}
& \multirow{3}{*}{$\frac{1}{12}$} & 96 & 0.09267(36) & 0.511(20) & 5.921(66) & 1.1 \\ 
& \multirow{3}{*}{} & 128 & 0.09264(38) & 0.510(20) & 5.930(73) & 1.08 \\ 
& \multirow{3}{*}{} & 192 & 0.09273(43) & 0.515(23) & 5.901(90) & 0.99 \\
\end{tabular}
\end{ruledtabular}
\label{tab:special_S2_noC}
\end{table*}

\begin{table*}
\caption{
Fits of the surface-bulk two-point function of the observable $S$
at the special UC to Eq.~(\ref{fss_2pt_open}) as a function of the minimum distance \zmin from the surface, the maximum ratio \zLmax, and the minimum lattice size \Lmin considered in the fits.
In the quoted error bars we sum the statistical error of the fit, the variation coming from $\hat{\Delta}_\sigma$ and the variation coming from $\hat{\Delta}_\varepsilon$.
}
\begin{ruledtabular}
\begin{tabular}{cclwmmqm}
\zmin & \zLmax & \Lmin & \multicolumn{1}{c}{$M_{\phi\sigma}$} & \multicolumn{1}{c}{$z_0$} & \multicolumn{1}{c}{$B_{\phi\sigma}$} & \multicolumn{1}{c}{$C$} & \multicolumn{1}{c}{\chidof} \\
\hline\multirow{14}{*}{4} & \multirow{5}{*}{$\frac{1}{4}$} & 32 & 0.32975(53) & 0.572(13) &  1.2924(52) & 0.650(35) & 89.1(6.4) \\ 
& \multirow{5}{*}{} & 48 & 0.32971(53) & 0.570(13) & 1.2934(52) & 0.644(34) & 87.2(6.4) \\ 
& \multirow{5}{*}{} & 64 & 0.32976(53) & 0.572(13) & 1.2920(52) & 0.648(34) & 84.5(6.2) \\ 
& \multirow{5}{*}{} & 96 & 0.32989(53) & 0.578(13) & 1.2877(56) & 0.667(34) & 78.3(5.8) \\ 
& \multirow{5}{*}{} & 128 & 0.33013(52) & 0.591(13) & 1.2802(61) & 0.708(34) & 64.5(4.8) \\ 
\cline{2-8}
 & \multirow{5}{*}{$\frac{1}{8}$} & 32 & 0.32875(50) & 0.543(11) &  1.3635(87) & 0.582(28) & 10.5(1.7) \\ 
& \multirow{5}{*}{} & 48 & 0.32872(50) & 0.543(11) & 1.3653(88) & 0.580(28) & 9.7(1.7) \\ 
& \multirow{5}{*}{} & 64 & 0.32870(50) & 0.542(11) & 1.3663(89) & 0.578(28) & 9.6(1.8) \\ 
& \multirow{5}{*}{} & 96 & 0.32869(50) & 0.541(12) & 1.3666(93) & 0.577(29) & 9.8(1.8) \\ 
& \multirow{5}{*}{} & 128 & 0.32872(51) & 0.543(12) & 1.3651(94) & 0.579(29) & 9.9(1.8) \\ 
\cline{2-8}
& \multirow{4}{*}{$\frac{1}{10}$} & 48 & 0.32866(49) & 0.543(11) & 1.377(10) & 0.583(27) & 6.4(1.1) \\ 
& \multirow{4}{*}{} & 64 & 0.32865(49) & 0.542(11) & 1.377(10) & 0.583(27) & 6.4(1.2) \\ 
& \multirow{4}{*}{} & 96 & 0.32863(49) & 0.542(11) & 1.378(11) & 0.581(27) & 6.5(1.2) \\ 
& \multirow{4}{*}{} & 128 & 0.32862(50) & 0.541(11) & 1.379(11) & 0.579(28) & 6.9(1.3) \\ 
\hline
\hline
\multirow{12}{*}{6} & \multirow{5}{*}{$\frac{1}{4}$} & 32 & 0.32964(61) & 0.554(20) &  1.2906(59) & 0.554(68) & 80.3(4.9) \\ 
& \multirow{5}{*}{} & 48 & 0.32956(60) & 0.550(20) & 1.2928(60) & 0.540(67) & 78.9(5.0) \\ 
& \multirow{5}{*}{} & 64 & 0.32955(60) & 0.549(20) & 1.2928(60) & 0.530(67) & 77.5(4.9) \\ 
& \multirow{5}{*}{} & 96 & 0.32964(61) & 0.553(20) & 1.2901(61) & 0.542(68) & 73.0(4.7) \\ 
& \multirow{5}{*}{} & 128 & 0.32991(61) & 0.570(20) & 1.2833(62) & 0.606(70) & 61.7(4.1) \\ 
\cline{2-8}
& \multirow{4}{*}{$\frac{1}{8}$} & 48 & 0.32841(56) & 0.515(17) & 1.3685(95) & 0.451(54) & 5.32(82) \\ 
& \multirow{4}{*}{} & 64 & 0.32840(56) & 0.515(17) & 1.3692(96) & 0.450(54) & 5.19(84) \\ 
& \multirow{4}{*}{} & 96 & 0.32837(57) & 0.513(17) & 1.370(10) & 0.445(55) & 5.09(85) \\ 
& \multirow{4}{*}{} & 128 & 0.32832(58) & 0.510(18) & 1.372(11) & 0.432(58) & 4.62(83) \\ 
\cline{2-8}
& \multirow{3}{*}{$\frac{1}{10}$} & 64 & 0.32829(55) & 0.514(16) & 1.382(11) & 0.453(50) & 1.83(40) \\ 
& \multirow{3}{*}{} & 96 & 0.32828(55) & 0.513(16) & 1.383(11) & 0.452(50) & 1.80(42) \\ 
& \multirow{3}{*}{} & 128 & 0.32824(57) & 0.511(17) & 1.385(12) & 0.444(53) & 1.67(42) \\ 
\hline
\hline
\multirow{10}{*}{8} & \multirow{5}{*}{$\frac{1}{4}$} & 32 & 0.32967(67) & 0.546(28) &  1.2868(66) & 0.48(12) & 71.5(3.9) \\ 
& \multirow{5}{*}{} & 48 & 0.32963(67) & 0.545(28) & 1.2882(66) & 0.48(12) & 70.2(4.0) \\ 
& \multirow{5}{*}{} & 64 & 0.32957(67) & 0.540(28) & 1.2894(66) & 0.45(12) & 69.8(4.0) \\ 
& \multirow{5}{*}{} & 96 & 0.32959(68) & 0.538(28) & 1.2885(69) & 0.43(12) & 67.2(3.9) \\ 
& \multirow{5}{*}{} & 128 & 0.32978(69) & 0.551(29) & 1.2838(70) & 0.47(13) & 58.1(3.5) \\ 
\cline{2-8}
& \multirow{3}{*}{$\frac{1}{8}$} & 64 & 0.32832(63) & 0.502(24) & 1.368(10) & 0.367(96) & 4.08(53) \\ 
& \multirow{3}{*}{} & 96 & 0.32830(63) & 0.502(24) & 1.368(10) & 0.368(96) & 4.01(54) \\ 
& \multirow{3}{*}{} & 128 & 0.32822(65) & 0.496(25) & 1.371(11) & 0.34(10) & 3.59(53) \\ 
\cline{2-8}
& \multirow{2}{*}{$\frac{1}{10}$} & 96 & 0.32818(62) & 0.501(23) & 1.382(12) & 0.375(91) & 1.21(22) \\ 
& \multirow{2}{*}{} & 128 & 0.32815(63) & 0.499(24) & 1.384(13) & 0.370(92) & 1.11(23) \\
\end{tabular}
\end{ruledtabular}
\end{table*}

\begin{table}
\caption{Fit of $E_{\text{surf}}$ at the special UC to Eq.~(\ref{Esurf_fit}), as a function of the miniumum lattice size \Lmin taken into account.
The quoted error bars are the sum of the statistical uncertainty obtained from the fit procedure and the variation of the fitted parameters on varying the surface exponent $\hat{\Delta}_\varepsilon = 2 - 0.718(6)$ \cite{Hasenbusch-11b} within one error bar.}
\begin{ruledtabular}
\begin{tabular}{lwwm}
\multicolumn{1}{c}{\Lmin} & \multicolumn{1}{c}{$\varEpsilon_0$} & \multicolumn{1}{c}{$U_{\varEpsilon}$} & \multicolumn{1}{c}{\chidof} \\
\hline
32 & 1.322907(11) & 2.121(19) & 0.46(10) \\ 
48 & 1.322908(10) & 2.120(22) & 0.507(78) \\ 
64 & 1.322912(10) & 2.116(22) & 0.178(57) \\ 
\end{tabular}
\end{ruledtabular}
\label{tab:special_surf_energy}
\end{table}
\begin{table}
\caption{Same as Table \ref{tab:special_surf_energy} for the fits of $S^2_{\text{surf}}$ at the special UC to Eq.~(\ref{S2surf_fit}).}
\begin{ruledtabular}
\begin{tabular}{lwwm}
\multicolumn{1}{c}{\Lmin} & \multicolumn{1}{c}{$\Sigma_0$} & \multicolumn{1}{c}{$U_{\sigma^2}$} & \multicolumn{1}{c}{\chidof} \\
\hline
32 & 0.6694465(12) & 0.2310(21) & 0.493(32) \\ 
48 & 0.6694466(11) & 0.2310(23) & 0.584(44) \\ 
64 & 0.6694469(11) & 0.2306(25) & 0.381(79) \\ 
\end{tabular}
\end{ruledtabular}
\label{tab:special_surf_S2}
\end{table}

\begin{table*}
\caption{
Fits of the two-point function of energy observable on the surface at the special UC to Eq.~(\ref{2pt_fit_Esurf}), as a function of the minimum distance \xmin, the maximum value \xLmax, and the minimum lattice size \Lmin considered in the fits.
The variation of the boundary scaling dimension $\hat{\Delta}_\varepsilon = 2 - 0.718(2)$ \cite{Hasenbusch-11b} within one quoted error bar gives the leading contribution to the total uncertainty.
}
\begin{ruledtabular}
\begin{tabular}{cclwmmq}
\xmin & \xLmax & \Lmin & \multicolumn{1}{c}{${\cal N}_\varepsilon^2$} & \multicolumn{1}{c}{$B_{\varepsilon\varepsilon}$} & \multicolumn{1}{c}{$C$} & \multicolumn{1}{c}{\chidof} \\
\hline\multirow{10}{*}{4} & \multirow{5}{*}{$\frac{1}{4}$} & 32 & 1.481(14) & 1.803(17) & 0.828(63) & 4.36(93) \\ 
& \multirow{5}{*}{} & 48 & 1.481(14) & 1.805(21) & 0.829(62) & 4.16(87) \\ 
& \multirow{5}{*}{} & 64 & 1.480(13) & 1.815(29) & 0.829(61) & 3.55(98) \\ 
& \multirow{5}{*}{} & 96 & 1.480(13) & 1.824(42) & 0.829(60) & 3.1(1.0) \\ 
& \multirow{5}{*}{} & 128 & 1.480(13) & 1.826(61) & 0.828(58) & 2.50(90) \\ 
\cline{2-7}
 & \multirow{5}{*}{$\frac{1}{8}$} & 32 & 1.480(14) & 1.824(18) & 0.830(63) & 5.8(2.0) \\ 
& \multirow{5}{*}{} & 48 & 1.480(14) & 1.824(20) & 0.830(62) & 5.8(2.0) \\ 
& \multirow{5}{*}{} & 64 & 1.480(13) & 1.831(27) & 0.831(62) & 4.9(1.8) \\ 
& \multirow{5}{*}{} & 96 & 1.480(13) & 1.840(39) & 0.831(60) & 4.2(2.0) \\ 
& \multirow{5}{*}{} & 128 & 1.480(13) & 1.840(59) & 0.830(58) & 3.6(1.8) \\ 
\hline
\hline
\multirow{9}{*}{6} & \multirow{5}{*}{$\frac{1}{4}$} & 32 & 1.482(16) & 1.774(12) & 0.80(15) & 1.76(24) \\ 
& \multirow{5}{*}{} & 48 & 1.482(16) & 1.771(15) & 0.80(14) & 1.69(26) \\ 
& \multirow{5}{*}{} & 64 & 1.482(16) & 1.780(20) & 0.81(14) & 1.44(20) \\ 
& \multirow{5}{*}{} & 96 & 1.482(16) & 1.788(30) & 0.81(14) & 1.41(23) \\ 
& \multirow{5}{*}{} & 128 & 1.482(16) & 1.792(43) & 0.81(13) & 1.28(22) \\ 
\cline{2-7}
& \multirow{4}{*}{$\frac{1}{8}$} & 48 & 1.482(16) & 1.789(16) & 0.80(14) & 1.94(55) \\ 
& \multirow{4}{*}{} & 64 & 1.481(16) & 1.795(18) & 0.81(14) & 1.74(52) \\ 
& \multirow{4}{*}{} & 96 & 1.481(16) & 1.806(26) & 0.81(14) & 1.43(42) \\ 
& \multirow{4}{*}{} & 128 & 1.481(16) & 1.809(39) & 0.81(14) & 1.38(39) \\ 
\hline
\hline
\multirow{8}{*}{8} & \multirow{5}{*}{$\frac{1}{4}$} & 32 & 1.483(18) & 1.767(12) & 0.77(27) & 1.296(99) \\ 
& \multirow{5}{*}{} & 48 & 1.483(18) & 1.761(13) & 0.76(27) & 1.141(85) \\ 
& \multirow{5}{*}{} & 64 & 1.483(18) & 1.762(18) & 0.76(27) & 1.044(84) \\ 
& \multirow{5}{*}{} & 96 & 1.483(18) & 1.766(25) & 0.77(26) & 1.053(92) \\ 
& \multirow{5}{*}{} & 128 & 1.483(18) & 1.770(36) & 0.77(26) & 1.010(95) \\ 
\cline{2-7}
& \multirow{3}{*}{$\frac{1}{8}$} & 64 & 1.482(18) & 1.779(19) & 0.77(26) & 0.99(13) \\ 
& \multirow{3}{*}{} & 96 & 1.482(18) & 1.787(22) & 0.78(26) & 0.91(13) \\ 
& \multirow{3}{*}{} & 128 & 1.482(18) & 1.788(31) & 0.78(26) & 0.94(14) \\ 
\end{tabular}
\end{ruledtabular}
\label{tab:special_energysurf-energysurf}
\end{table*}

\begin{table*}
\caption{
Fits of the two-point function of $S^2$ on the surface at the special UC to Eq.~(\ref{2pt_fit_S2surf}, as a function of the minimum distance \xmin, the maximum value \xLmax, and the minimum lattice size \Lmin considered in the fits.
The variation of the boundary scaling dimension $\hat{\Delta}_\varepsilon = 2 - 0.718(2)$ \cite{Hasenbusch-11b} within one quoted error bar gives the leading contribution to the total uncertainty.
}
\begin{ruledtabular}
\begin{tabular}{cclwmmq}
\xmin & \xLmax & \Lmin & \multicolumn{1}{c}{${\cal N}_{\sigma^2}^2$} & \multicolumn{1}{c}{$B_{\varepsilon\varepsilon}$} & \multicolumn{1}{c}{$C$} & \multicolumn{1}{c}{\chidof} \\
\hline\multirow{10}{*}{4} & \multirow{5}{*}{$\frac{1}{4}$} & 32 & 0.01760(16) & 1.756(16) & 0.270(60) & 3.9(1.4) \\ 
& \multirow{5}{*}{} & 48 & 0.01760(16) & 1.764(22) & 0.271(60) & 3.3(1.3) \\ 
& \multirow{5}{*}{} & 64 & 0.01760(16) & 1.776(29) & 0.273(59) & 2.5(1.0) \\ 
& \multirow{5}{*}{} & 96 & 0.01760(16) & 1.785(41) & 0.273(57) & 2.4(1.2) \\ 
& \multirow{5}{*}{} & 128 & 0.01760(15) & 1.790(59) & 0.273(55) & 2.2(1.1) \\ 
\cline{2-7}
 & \multirow{5}{*}{$\frac{1}{8}$} & 32 & 0.01760(16) & 1.767(18) & 0.271(60) & 5.7(2.9) \\ 
& \multirow{5}{*}{} & 48 & 0.01760(16) & 1.773(20) & 0.272(60) & 5.1(2.7) \\ 
& \multirow{5}{*}{} & 64 & 0.01760(16) & 1.784(27) & 0.273(59) & 3.9(2.3) \\ 
& \multirow{5}{*}{} & 96 & 0.01759(16) & 1.795(38) & 0.275(58) & 3.3(2.2) \\ 
& \multirow{5}{*}{} & 128 & 0.01759(15) & 1.799(57) & 0.274(56) & 3.1(2.1) \\ 
\hline
\hline
\multirow{9}{*}{6} & \multirow{5}{*}{$\frac{1}{4}$} & 32 & 0.01761(19) & 1.750(11) & 0.25(14) & 1.92(32) \\ 
& \multirow{5}{*}{} & 48 & 0.01761(19) & 1.750(15) & 0.25(14) & 1.77(32) \\ 
& \multirow{5}{*}{} & 64 & 0.01761(19) & 1.760(20) & 0.26(14) & 1.36(25) \\ 
& \multirow{5}{*}{} & 96 & 0.01760(19) & 1.768(29) & 0.26(14) & 1.32(28) \\ 
& \multirow{5}{*}{} & 128 & 0.01760(19) & 1.775(40) & 0.26(13) & 1.24(28) \\ 
\cline{2-7}
& \multirow{4}{*}{$\frac{1}{8}$} & 48 & 0.01761(19) & 1.762(16) & 0.25(14) & 2.07(67) \\ 
& \multirow{4}{*}{} & 64 & 0.01760(19) & 1.770(18) & 0.26(14) & 1.71(62) \\ 
& \multirow{4}{*}{} & 96 & 0.01760(19) & 1.782(25) & 0.27(14) & 1.30(50) \\ 
& \multirow{4}{*}{} & 128 & 0.01760(19) & 1.787(36) & 0.27(13) & 1.26(48) \\ 
\hline
\hline
\multirow{8}{*}{8} & \multirow{5}{*}{$\frac{1}{4}$} & 32 & 0.01761(22) & 1.753(11) & 0.23(27) & 1.35(13) \\ 
& \multirow{5}{*}{} & 48 & 0.01762(22) & 1.748(13) & 0.22(27) & 1.23(12) \\ 
& \multirow{5}{*}{} & 64 & 0.01762(22) & 1.749(17) & 0.23(26) & 1.06(12) \\ 
& \multirow{5}{*}{} & 96 & 0.01761(22) & 1.753(23) & 0.24(26) & 1.05(13) \\ 
& \multirow{5}{*}{} & 128 & 0.01761(21) & 1.760(32) & 0.24(25) & 1.02(14) \\ 
\cline{2-7}
& \multirow{3}{*}{$\frac{1}{8}$} & 64 & 0.01761(22) & 1.762(17) & 0.23(26) & 0.96(16) \\ 
& \multirow{3}{*}{} & 96 & 0.01761(22) & 1.770(20) & 0.24(26) & 0.86(18) \\ 
& \multirow{3}{*}{} & 128 & 0.01761(22) & 1.774(28) & 0.25(26) & 0.89(19) \\ 
\end{tabular}
\end{ruledtabular}
\label{tab:special_S2surf-S2surf}
\end{table*}


\begin{table*}
\caption{Fits of the energy profile at the normal UC realized with $(+,o)$ to Eq.~(\ref{fss_1ptE_open}), as a function of the minimum distance \zmin from the surface, the maximum ratio \zLmax, and the minimum lattice size \Lmin considered in the fits.
We employ the fitted value of $E_0$ reported in Table \ref{tab:fitresults}.
Varying $E_0$ and $\Delta_\epsilon$ within one error bar quoted in Table \ref{tab:fitresults} results in a significant additional uncertainty in the results.
As a reference, a fit employing only the central values of $E_0$ and of $\Delta_\epsilon$ for $\zmin = 8$, $\zLmax=1/4$, $\Lmin=96$ results in $A_\epsilon = 4.8273(19)$.
}
\begin{ruledtabular}
\begin{tabular}{cclwmmqm}
\zmin & \zLmax & \Lmin & \multicolumn{1}{c}{$A_\epsilon$} & \multicolumn{1}{c}{$z_0$} & \multicolumn{1}{c}{$\mathcal{B}_{\epsilon}$} & \multicolumn{1}{c}{$C$} & \multicolumn{1}{c}{\chidof} \\
\hline\multirow{12}{*}{4} & \multirow{6}{*}{$\frac{1}{4}$} & 32 & 4.8317(39) & 1.4260(73) & -2.660(35) & -0.311(34) & 4.04 \\ 
& \multirow{6}{*}{} & 48 & 4.8326(39) & 1.4280(72) & -2.702(42) & -0.302(33) & 3.2 \\ 
& \multirow{6}{*}{} & 64 & 4.8331(38) & 1.4292(71) & -2.729(49) & -0.296(32) & 2.93 \\ 
& \multirow{6}{*}{} & 96 & 4.8339(37) & 1.4311(68) & -2.773(61) & -0.287(31) & 2.48 \\ 
& \multirow{6}{*}{} & 128 & 4.8346(35) & 1.4327(64) & -2.822(75) & -0.278(29) & 2.06 \\ 
& \multirow{6}{*}{} & 192 & 4.8349(34) & 1.4336(60) & -2.86(10) & -0.274(27) & 1.96 \\ 
\cline{2-8}
 & \multirow{6}{*}{$\frac{1}{8}$} & 32 & 4.8346(34) & 1.4332(59) & -2.659(65) & -0.274(26) & 1.15 \\ 
& \multirow{6}{*}{} & 48 & 4.8347(34) & 1.4334(59) & -2.682(78) & -0.274(26) & 1.12 \\ 
& \multirow{6}{*}{} & 64 & 4.8348(34) & 1.4336(59) & -2.698(99) & -0.273(26) & 1.12 \\ 
& \multirow{6}{*}{} & 96 & 4.8352(33) & 1.4343(59) & -2.76(13) & -0.270(26) & 1.06 \\ 
& \multirow{6}{*}{} & 128 & 4.8357(33) & 1.4354(57) & -2.85(16) & -0.265(25) & 1.0 \\ 
& \multirow{6}{*}{} & 192 & 4.8361(31) & 1.4362(54) & -2.94(22) & -0.261(24) & 1.01 \\ 
\hline
\hline
\multirow{11}{*}{6} & \multirow{6}{*}{$\frac{1}{4}$} & 32 & 4.8276(54) & 1.409(14) & -2.651(32) & -0.451(86) & 3.09 \\ 
& \multirow{6}{*}{} & 48 & 4.8287(54) & 1.412(14) & -2.688(38) & -0.431(86) & 2.37 \\ 
& \multirow{6}{*}{} & 64 & 4.8293(54) & 1.414(14) & -2.709(43) & -0.418(85) & 2.2 \\ 
& \multirow{6}{*}{} & 96 & 4.8305(53) & 1.418(13) & -2.746(53) & -0.395(84) & 1.93 \\ 
& \multirow{6}{*}{} & 128 & 4.8316(51) & 1.421(13) & -2.791(66) & -0.371(80) & 1.66 \\ 
& \multirow{6}{*}{} & 192 & 4.8323(48) & 1.423(12) & -2.827(91) & -0.357(73) & 1.64 \\ 
\cline{2-8}
& \multirow{5}{*}{$\frac{1}{8}$} & 48 & 4.8325(46) & 1.425(11) & -2.676(78) & -0.341(66) & 0.72 \\ 
& \multirow{5}{*}{} & 64 & 4.8326(47) & 1.425(11) & -2.679(90) & -0.341(66) & 0.72 \\ 
& \multirow{5}{*}{} & 96 & 4.8329(47) & 1.426(11) & -2.73(11) & -0.336(67) & 0.69 \\ 
& \multirow{5}{*}{} & 128 & 4.8334(48) & 1.427(11) & -2.78(14) & -0.329(68) & 0.67 \\ 
& \multirow{5}{*}{} & 192 & 4.8336(48) & 1.428(11) & -2.82(20) & -0.325(68) & 0.72 \\ 
\hline
\hline
\multirow{10}{*}{8} & \multirow{6}{*}{$\frac{1}{4}$} & 32 & 4.8245(69) & 1.392(22) & -2.661(33) & -0.64(18) & 2.42 \\ 
& \multirow{6}{*}{} & 48 & 4.8253(70) & 1.394(22) & -2.683(36) & -0.62(18) & 1.93 \\ 
& \multirow{6}{*}{} & 64 & 4.8260(70) & 1.397(22) & -2.700(40) & -0.60(18) & 1.79 \\ 
& \multirow{6}{*}{} & 96 & 4.8273(70) & 1.402(22) & -2.732(48) & -0.56(18) & 1.61 \\ 
& \multirow{6}{*}{} & 128 & 4.8288(69) & 1.407(22) & -2.771(59) & -0.52(18) & 1.43 \\ 
& \multirow{6}{*}{} & 192 & 4.8296(65) & 1.410(21) & -2.801(84) & -0.49(17) & 1.47 \\ 
\cline{2-8}
& \multirow{4}{*}{$\frac{1}{8}$} & 64 & 4.8311(60) & 1.418(18) & -2.689(91) & -0.41(14) & 0.62 \\ 
& \multirow{4}{*}{} & 96 & 4.8313(61) & 1.418(18) & -2.72(11) & -0.41(14) & 0.59 \\ 
& \multirow{4}{*}{} & 128 & 4.8318(62) & 1.420(18) & -2.76(13) & -0.40(14) & 0.59 \\ 
& \multirow{4}{*}{} & 192 & 4.8317(66) & 1.419(20) & -2.77(19) & -0.40(15) & 0.62 \\ 
\end{tabular}
\end{ruledtabular}
\label{tab:normal_po_E}
\end{table*}

\begin{table*}
\caption{
Same as Table \ref{tab:normal_po_E} for the normal UC realized with $(+,+)$ BCs.
Varying $E_0$ and $\Delta_\epsilon$ within one error bar quoted in Table \ref{tab:fitresults} results in a significant additional uncertainty in the results.
As a reference, a fit employing only the central values of $E_0$ and of $\Delta_\epsilon$ for $\zmin = 6$, $\zLmax=1/4$, $\Lmin=128$ results in $A_\epsilon = 4.8364(14)$.
}
\begin{ruledtabular}
\begin{tabular}{cclwmmqm}
\zmin & \zLmax & \Lmin & \multicolumn{1}{c}{$A_\epsilon$} & \multicolumn{1}{c}{$z_0$} & \multicolumn{1}{c}{$\mathcal{B}_{\epsilon}$} & \multicolumn{1}{c}{$C$} & \multicolumn{1}{c}{\chidof} \\
\hline\multirow{12}{*}{4} & \multirow{6}{*}{$\frac{1}{4}$} & 32 & 4.8397(38) & 1.4448(73) & 3.591(29) & -0.217(35) & 4.92 \\ 
& \multirow{6}{*}{} & 48 & 4.8386(37) & 1.4426(71) & 3.651(36) & -0.227(34) & 3.04 \\ 
& \multirow{6}{*}{} & 64 & 4.8379(36) & 1.4409(68) & 3.692(44) & -0.236(32) & 2.37 \\ 
& \multirow{6}{*}{} & 96 & 4.8370(34) & 1.4388(64) & 3.747(57) & -0.246(30) & 1.51 \\ 
& \multirow{6}{*}{} & 128 & 4.8368(33) & 1.4383(61) & 3.763(74) & -0.249(28) & 1.45 \\ 
& \multirow{6}{*}{} & 192 & 4.8368(31) & 1.4380(57) & 3.78(10) & -0.251(26) & 1.45 \\ 
\cline{2-8}
 & \multirow{6}{*}{$\frac{1}{8}$} & 32 & 4.8387(32) & 1.4420(57) & 3.425(61) & -0.233(26) & 1.63 \\ 
& \multirow{6}{*}{} & 48 & 4.8385(32) & 1.4416(57) & 3.477(76) & -0.235(26) & 1.45 \\ 
& \multirow{6}{*}{} & 64 & 4.8383(32) & 1.4412(56) & 3.511(93) & -0.236(26) & 1.42 \\ 
& \multirow{6}{*}{} & 96 & 4.8380(32) & 1.4406(56) & 3.57(13) & -0.238(26) & 1.36 \\ 
& \multirow{6}{*}{} & 128 & 4.8381(31) & 1.4408(56) & 3.55(16) & -0.238(25) & 1.44 \\ 
& \multirow{6}{*}{} & 192 & 4.8382(30) & 1.4411(52) & 3.53(22) & -0.236(24) & 1.62 \\ 
\hline
\hline
\multirow{11}{*}{6} & \multirow{6}{*}{$\frac{1}{4}$} & 32 & 4.8407(55) & 1.449(15) & 3.597(23) & -0.176(95) & 4.36 \\ 
& \multirow{6}{*}{} & 48 & 4.8396(55) & 1.447(14) & 3.647(28) & -0.193(94) & 2.91 \\ 
& \multirow{6}{*}{} & 64 & 4.8385(53) & 1.443(14) & 3.689(35) & -0.214(91) & 2.28 \\ 
& \multirow{6}{*}{} & 96 & 4.8369(51) & 1.438(13) & 3.749(45) & -0.248(86) & 1.47 \\ 
& \multirow{6}{*}{} & 128 & 4.8364(49) & 1.437(13) & 3.769(61) & -0.260(81) & 1.39 \\ 
& \multirow{6}{*}{} & 192 & 4.8361(46) & 1.436(12) & 3.788(87) & -0.270(75) & 1.34 \\ 
\cline{2-8}
& \multirow{5}{*}{$\frac{1}{8}$} & 48 & 4.8384(45) & 1.442(11) & 3.507(71) & -0.232(68) & 1.32 \\ 
& \multirow{5}{*}{} & 64 & 4.8383(45) & 1.441(11) & 3.525(81) & -0.234(67) & 1.31 \\ 
& \multirow{5}{*}{} & 96 & 4.8380(45) & 1.441(11) & 3.57(11) & -0.238(68) & 1.29 \\ 
& \multirow{5}{*}{} & 128 & 4.8381(46) & 1.441(11) & 3.56(14) & -0.237(69) & 1.37 \\ 
& \multirow{5}{*}{} & 192 & 4.8383(46) & 1.442(11) & 3.53(19) & -0.231(70) & 1.51 \\ 
\hline
\hline
\multirow{10}{*}{8} & \multirow{6}{*}{$\frac{1}{4}$} & 32 & 4.8413(73) & 1.455(24) & 3.622(19) & -0.11(20) & 3.37 \\ 
& \multirow{6}{*}{} & 48 & 4.8407(72) & 1.453(24) & 3.646(22) & -0.12(20) & 2.73 \\ 
& \multirow{6}{*}{} & 64 & 4.8396(71) & 1.449(24) & 3.684(28) & -0.15(20) & 2.19 \\ 
& \multirow{6}{*}{} & 96 & 4.8374(69) & 1.441(23) & 3.745(37) & -0.22(19) & 1.41 \\ 
& \multirow{6}{*}{} & 128 & 4.8366(67) & 1.438(22) & 3.768(52) & -0.25(18) & 1.33 \\ 
& \multirow{6}{*}{} & 192 & 4.8357(63) & 1.434(20) & 3.794(78) & -0.29(17) & 1.27 \\ 
\cline{2-8}
& \multirow{4}{*}{$\frac{1}{8}$} & 64 & 4.8386(59) & 1.443(18) & 3.539(81) & -0.21(14) & 1.24 \\ 
& \multirow{4}{*}{} & 96 & 4.8384(59) & 1.443(18) & 3.57(10) & -0.22(14) & 1.22 \\ 
& \multirow{4}{*}{} & 128 & 4.8385(61) & 1.443(18) & 3.56(13) & -0.21(14) & 1.27 \\ 
& \multirow{4}{*}{} & 192 & 4.8387(64) & 1.444(19) & 3.53(18) & -0.21(15) & 1.39 \\ 
\end{tabular}
\end{ruledtabular}
\label{tab:normal_pp_E}
\end{table*}

\begin{table*}
\caption{
Fits of the $S^2$ profile at the normal UC realized with $(+,o)$ BCs to Eq.~(\ref{fss_1ptS2_open}), as a function of the minimum distance \zmin from the surface, the maximum ratio \zLmax, and the minimum lattice size \Lmin considered in the fits.
We employ the fitted value of $S^2_0$ reported in Table \ref{tab:additional_fitresults}.
Varying $S^2_0$ and $\Delta_\epsilon$ within one error bar quoted in Table \ref{tab:additional_fitresults} results in a significant additional uncertainty in the results.
As a reference, a fit employing only the central values of $S^2_0$  and of $\Delta_\epsilon$ for $\zmin = 4$, $\zLmax=1/8$, $\Lmin=192$ results in $A_{S^2} = 0.53396(12)$.
}
\begin{ruledtabular}
\begin{tabular}{cclwmmqm}
\zmin & \zLmax & \Lmin & \multicolumn{1}{c}{$A_{S^2}$} & \multicolumn{1}{c}{$z_0$} & \multicolumn{1}{c}{$\mathcal{B}_{\epsilon}$} & \multicolumn{1}{c}{$C$} & \multicolumn{1}{c}{\chidof} \\
\hline\multirow{12}{*}{4} & \multirow{6}{*}{$\frac{1}{4}$} & 32 & 0.53347(49) & 1.4217(83) & -2.677(38) & -0.226(39) & 4.32 \\ 
& \multirow{6}{*}{} & 48 & 0.53356(49) & 1.4235(82) & -2.716(46) & -0.217(38) & 3.54 \\ 
& \multirow{6}{*}{} & 64 & 0.53362(47) & 1.4246(80) & -2.741(54) & -0.212(37) & 3.29 \\ 
& \multirow{6}{*}{} & 96 & 0.53370(46) & 1.4264(77) & -2.784(67) & -0.202(35) & 2.86 \\ 
& \multirow{6}{*}{} & 128 & 0.53378(44) & 1.4281(72) & -2.835(83) & -0.194(33) & 2.39 \\ 
& \multirow{6}{*}{} & 192 & 0.53383(42) & 1.4291(67) & -2.88(11) & -0.189(31) & 2.25 \\ 
\cline{2-8}
 & \multirow{6}{*}{$\frac{1}{8}$} & 32 & 0.53381(42) & 1.4289(66) & -2.689(67) & -0.188(29) & 1.16 \\ 
& \multirow{6}{*}{} & 48 & 0.53381(42) & 1.4291(66) & -2.707(82) & -0.188(29) & 1.14 \\ 
& \multirow{6}{*}{} & 64 & 0.53382(42) & 1.4292(66) & -2.72(10) & -0.187(30) & 1.15 \\ 
& \multirow{6}{*}{} & 96 & 0.53386(41) & 1.4299(66) & -2.78(13) & -0.184(29) & 1.11 \\ 
& \multirow{6}{*}{} & 128 & 0.53391(40) & 1.4309(63) & -2.87(17) & -0.179(28) & 1.05 \\ 
& \multirow{6}{*}{} & 192 & 0.53396(38) & 1.4317(60) & -2.95(24) & -0.176(26) & 1.07 \\ 
\hline
\hline
\multirow{11}{*}{6} & \multirow{6}{*}{$\frac{1}{4}$} & 32 & 0.53302(68) & 1.405(15) & -2.667(35) & -0.366(98) & 3.33 \\ 
& \multirow{6}{*}{} & 48 & 0.53314(68) & 1.408(15) & -2.701(41) & -0.347(98) & 2.67 \\ 
& \multirow{6}{*}{} & 64 & 0.53320(67) & 1.410(15) & -2.721(47) & -0.335(97) & 2.51 \\ 
& \multirow{6}{*}{} & 96 & 0.53332(66) & 1.413(15) & -2.757(58) & -0.312(95) & 2.26 \\ 
& \multirow{6}{*}{} & 128 & 0.53345(63) & 1.416(14) & -2.803(72) & -0.288(90) & 1.96 \\ 
& \multirow{6}{*}{} & 192 & 0.53353(59) & 1.419(13) & -2.84(10) & -0.272(82) & 1.91 \\ 
\cline{2-8}
& \multirow{5}{*}{} & 48 & 0.53358(57) & 1.421(12) & -2.697(80) & -0.253(74) & 0.76 \\ 
& \multirow{5}{*}{} & 64 & 0.53358(58) & 1.421(12) & -2.699(93) & -0.253(74) & 0.77 \\ 
& \multirow{5}{*}{} & 96 & 0.53361(58) & 1.422(12) & -2.74(12) & -0.249(75) & 0.74 \\ 
& \multirow{5}{*}{} & 128 & 0.53367(59) & 1.423(12) & -2.80(14) & -0.241(76) & 0.73 \\ 
& \multirow{5}{*}{} & 192 & 0.53369(58) & 1.424(12) & -2.84(21) & -0.237(75) & 0.78 \\ 
\hline
\hline
\multirow{10}{*}{8} & \multirow{6}{*}{$\frac{1}{4}$} & 32 & 0.53265(87) & 1.386(25) & -2.674(36) & -0.57(20) & 2.61 \\ 
& \multirow{6}{*}{} & 48 & 0.53274(87) & 1.389(25) & -2.695(39) & -0.55(20) & 2.15 \\ 
& \multirow{6}{*}{} & 64 & 0.53281(87) & 1.391(25) & -2.712(43) & -0.53(20) & 2.02 \\ 
& \multirow{6}{*}{} & 96 & 0.53295(87) & 1.396(25) & -2.742(52) & -0.49(20) & 1.86 \\ 
& \multirow{6}{*}{} & 128 & 0.53311(85) & 1.401(24) & -2.782(64) & -0.44(20) & 1.67 \\ 
& \multirow{6}{*}{} & 192 & 0.53322(81) & 1.405(23) & -2.816(91) & -0.41(18) & 1.69 \\ 
\cline{2-8}
& \multirow{4}{*}{} & 64 & 0.53340(74) & 1.413(20) & -2.707(94) & -0.33(15) & 0.64 \\ 
& \multirow{4}{*}{} & 96 & 0.53342(75) & 1.414(20) & -2.73(11) & -0.33(15) & 0.61 \\ 
& \multirow{4}{*}{} & 128 & 0.53347(76) & 1.415(20) & -2.78(13) & -0.32(16) & 0.61 \\ 
& \multirow{4}{*}{} & 192 & 0.53347(80) & 1.415(21) & -2.78(20) & -0.32(16) & 0.64 \\ 
\end{tabular}
\end{ruledtabular}
\label{tab:normal_po_S2}
\end{table*}

\begin{table*}
\caption{
Same as Table \ref{tab:normal_po_S2} for the normal UC realized with $(+,+)$ BCs.
Varying $S^2_0$ within one error bar quoted in Table \ref{tab:additional_fitresults} results in a significant additional uncertainty in the results.
As a reference, a fit employing only the central values of $S^2_0$  and of $\Delta_\epsilon$ for $\zmin = 6$, $\zLmax=1/4$, $\Lmin=128$ results in $A_{S^2} = 0.53398(15)$.
}
\begin{ruledtabular}
\begin{tabular}{cclwmmqm}
\zmin & \zLmax & \Lmin & \multicolumn{1}{c}{$A_{S^2}$} & \multicolumn{1}{c}{$z_0$} & \multicolumn{1}{c}{$\mathcal{B}_{\epsilon}$} & \multicolumn{1}{c}{$C$} & \multicolumn{1}{c}{\chidof} \\
\hline\multirow{12}{*}{4} & \multirow{6}{*}{$\frac{1}{4}$} & 32 & 0.53431(48) & 1.4395(83) & 3.603(32) & -0.135(39) & 4.47 \\ 
& \multirow{6}{*}{} & 48 & 0.53421(47) & 1.4375(81) & 3.658(40) & -0.145(38) & 2.84 \\ 
& \multirow{6}{*}{} & 64 & 0.53414(45) & 1.4360(77) & 3.696(49) & -0.153(37) & 2.26 \\ 
& \multirow{6}{*}{} & 96 & 0.53405(43) & 1.4340(73) & 3.746(63) & -0.162(34) & 1.5 \\ 
& \multirow{6}{*}{} & 128 & 0.53403(41) & 1.4337(69) & 3.757(83) & -0.165(32) & 1.52 \\ 
& \multirow{6}{*}{} & 192 & 0.53403(39) & 1.4336(64) & 3.77(11) & -0.165(30) & 1.54 \\ 
\cline{2-8}
 & \multirow{6}{*}{$\frac{1}{8}$} & 32 & 0.53422(40) & 1.4371(64) & 3.460(62) & -0.149(29) & 1.56 \\ 
& \multirow{6}{*}{} & 48 & 0.53420(40) & 1.4367(64) & 3.507(79) & -0.151(29) & 1.4 \\ 
& \multirow{6}{*}{} & 64 & 0.53419(39) & 1.4364(63) & 3.536(97) & -0.152(29) & 1.4 \\ 
& \multirow{6}{*}{} & 96 & 0.53416(39) & 1.4359(63) & 3.59(13) & -0.154(29) & 1.36 \\ 
& \multirow{6}{*}{} & 128 & 0.53417(39) & 1.4362(62) & 3.56(17) & -0.153(28) & 1.44 \\ 
& \multirow{6}{*}{} & 192 & 0.53419(37) & 1.4365(58) & 3.53(23) & -0.151(26) & 1.61 \\ 
\hline
\hline
\multirow{11}{*}{6} & \multirow{6}{*}{$\frac{1}{4}$} & 32 & 0.53440(69) & 1.443(16) & 3.610(24) & -0.10(11) & 4.01 \\ 
& \multirow{6}{*}{} & 48 & 0.53429(68) & 1.441(16) & 3.655(30) & -0.12(11) & 2.72 \\ 
& \multirow{6}{*}{} & 64 & 0.53418(66) & 1.438(16) & 3.693(37) & -0.14(10) & 2.19 \\ 
& \multirow{6}{*}{} & 96 & 0.53402(64) & 1.433(15) & 3.748(49) & -0.169(97) & 1.45 \\ 
& \multirow{6}{*}{} & 128 & 0.53398(61) & 1.432(14) & 3.763(67) & -0.178(91) & 1.44 \\ 
& \multirow{6}{*}{} & 192 & 0.53396(57) & 1.431(13) & 3.776(96) & -0.185(83) & 1.44 \\ 
\cline{2-8}
& \multirow{5}{*}{} & 48 & 0.53419(56) & 1.437(12) & 3.531(72) & -0.149(75) & 1.31 \\ 
& \multirow{5}{*}{} & 64 & 0.53418(56) & 1.437(12) & 3.546(83) & -0.150(75) & 1.3 \\ 
& \multirow{5}{*}{} & 96 & 0.53416(56) & 1.436(12) & 3.58(11) & -0.154(76) & 1.29 \\ 
& \multirow{5}{*}{} & 128 & 0.53417(57) & 1.436(12) & 3.57(14) & -0.152(76) & 1.37 \\ 
& \multirow{5}{*}{} & 192 & 0.53420(57) & 1.437(12) & 3.53(20) & -0.146(77) & 1.52 \\ 
\hline
\hline
\multirow{10}{*}{8} & \multirow{6}{*}{$\frac{1}{4}$} & 32 & 0.53444(91) & 1.447(27) & 3.634(19) & -0.06(22) & 3.16 \\ 
& \multirow{6}{*}{} & 48 & 0.53438(90) & 1.445(27) & 3.656(23) & -0.07(22) & 2.58 \\ 
& \multirow{6}{*}{} & 64 & 0.53426(89) & 1.442(26) & 3.690(29) & -0.09(22) & 2.11 \\ 
& \multirow{6}{*}{} & 96 & 0.53405(86) & 1.434(26) & 3.747(39) & -0.16(21) & 1.42 \\ 
& \multirow{6}{*}{} & 128 & 0.53397(83) & 1.432(24) & 3.764(56) & -0.18(20) & 1.39 \\ 
& \multirow{6}{*}{} & 192 & 0.53390(78) & 1.428(23) & 3.783(85) & -0.21(18) & 1.35 \\ 
\cline{2-8}
& \multirow{4}{*}{} & 64 & 0.53421(73) & 1.438(20) & 3.557(81) & -0.14(16) & 1.23 \\ 
& \multirow{4}{*}{} & 96 & 0.53419(73) & 1.438(20) & 3.58(10) & -0.14(16) & 1.21 \\ 
& \multirow{4}{*}{} & 128 & 0.53420(74) & 1.438(20) & 3.57(13) & -0.14(16) & 1.27 \\ 
& \multirow{4}{*}{} & 192 & 0.53424(77) & 1.439(21) & 3.54(18) & -0.13(17) & 1.39 \\  
\end{tabular}
\end{ruledtabular}
\label{tab:normal_pp_S2}
\end{table*}

\begin{table*}
\caption{
Fits of the profile of the order-parameter $S$ at the normal UC realized with $(+,o)$ to Eq.~(\ref{fss_1ptS_normal}), as a function of the minimum distance \zmin from the surface, the maximum ratio \zLmax, and the minimum lattice size \Lmin considered in the fits.
A variation of the critical exponent $\Delta_\phi=\numprint{0.5181489}(10)$ \cite{KPSDV-16} within one error bar gives a negligible increase in the uncertainty of the fitted parameters.
}
\begin{ruledtabular}
\begin{tabular}{cclwmmqm}
\zmin & \zLmax & \Lmin & \multicolumn{1}{c}{$A_\phi$} & \multicolumn{1}{c}{$z_0$} & \multicolumn{1}{c}{$B_\phi$} & \multicolumn{1}{c}{$C$} & \multicolumn{1}{c}{\chidof} \\
\hline\multirow{12}{*}{4} & \multirow{6}{*}{$\frac{1}{4}$} & 32 & 1.126327(63) & 1.4363(15) & -1.0123(39) & -0.0870(27) & 2.77 \\ 
& \multirow{6}{*}{} & 48 & 1.126369(64) & 1.4375(15) & -1.0190(45) & -0.0848(28) & 1.91 \\ 
& \multirow{6}{*}{} & 64 & 1.126399(64) & 1.4383(15) & -1.0241(50) & -0.0832(28) & 1.52 \\ 
& \multirow{6}{*}{} & 96 & 1.126440(65) & 1.4395(15) & -1.0313(56) & -0.0809(28) & 0.98 \\ 
& \multirow{6}{*}{} & 128 & 1.126469(63) & 1.4404(15) & -1.0378(59) & -0.0791(27) & 0.55 \\ 
& \multirow{6}{*}{} & 192 & 1.126478(63) & 1.4407(15) & -1.0412(71) & -0.0786(26) & 0.48 \\ 
\cline{2-8}
 & \multirow{6}{*}{$\frac{1}{8}$} & 32 & 1.126439(62) & 1.4399(14) & -0.995(12) & -0.0798(24) & 0.61 \\ 
& \multirow{6}{*}{} & 48 & 1.126445(64) & 1.4400(14) & -0.999(14) & -0.0796(25) & 0.58 \\ 
& \multirow{6}{*}{} & 64 & 1.126450(66) & 1.4401(15) & -1.003(17) & -0.0795(26) & 0.55 \\ 
& \multirow{6}{*}{} & 96 & 1.126465(69) & 1.4405(16) & -1.013(21) & -0.0788(27) & 0.51 \\ 
& \multirow{6}{*}{} & 128 & 1.126486(70) & 1.4410(16) & -1.027(23) & -0.0779(28) & 0.45 \\ 
& \multirow{6}{*}{} & 192 & 1.126499(73) & 1.4413(17) & -1.038(28) & -0.0773(29) & 0.44 \\ 
\hline
\hline
\multirow{11}{*}{6} & \multirow{6}{*}{$\frac{1}{4}$} & 32 & 1.126195(84) & 1.4293(27) & -1.0108(40) & -0.1098(69) & 2.13 \\ 
& \multirow{6}{*}{} & 48 & 1.126245(86) & 1.4310(28) & -1.0166(46) & -0.1056(71) & 1.43 \\ 
& \multirow{6}{*}{} & 64 & 1.126283(88) & 1.4324(29) & -1.0210(51) & -0.1020(73) & 1.14 \\ 
& \multirow{6}{*}{} & 96 & 1.126341(91) & 1.4346(30) & -1.0278(58) & -0.0963(75) & 0.74 \\ 
& \multirow{6}{*}{} & 128 & 1.126391(90) & 1.4366(29) & -1.0344(62) & -0.0910(74) & 0.4 \\ 
& \multirow{6}{*}{} & 192 & 1.126409(89) & 1.4373(29) & -1.0375(75) & -0.0890(73) & 0.36 \\ 
\cline{2-8}
& \multirow{5}{*}{$\frac{1}{8}$} & 48 & 1.126377(89) & 1.4367(27) & -1.000(16) & -0.0897(63) & 0.35 \\ 
& \multirow{5}{*}{} & 64 & 1.126377(91) & 1.4367(27) & -1.000(18) & -0.0896(64) & 0.36 \\ 
& \multirow{5}{*}{} & 96 & 1.126391(97) & 1.4371(29) & -1.007(22) & -0.0888(68) & 0.33 \\ 
& \multirow{5}{*}{} & 128 & 1.12641(10) & 1.4378(31) & -1.018(26) & -0.0872(73) & 0.3 \\ 
& \multirow{5}{*}{} & 192 & 1.12642(11) & 1.4380(35) & -1.022(34) & -0.0869(83) & 0.31 \\ 
\hline
\hline
\multirow{10}{*}{8} & \multirow{6}{*}{$\frac{1}{4}$} & 32 & 1.12611(11) & 1.4229(46) & -1.0125(42) & -0.138(15) & 1.62 \\ 
& \multirow{6}{*}{} & 48 & 1.12614(11) & 1.4242(46) & -1.0157(46) & -0.134(15) & 1.19 \\ 
& \multirow{6}{*}{} & 64 & 1.12619(11) & 1.4262(48) & -1.0198(52) & -0.128(15) & 0.94 \\ 
& \multirow{6}{*}{} & 96 & 1.12626(12) & 1.4294(51) & -1.0261(60) & -0.117(16) & 0.62 \\ 
& \multirow{6}{*}{} & 128 & 1.12633(12) & 1.4328(52) & -1.0326(65) & -0.106(17) & 0.33 \\ 
& \multirow{6}{*}{} & 192 & 1.12636(13) & 1.4341(54) & -1.0353(82) & -0.102(18) & 0.32 \\ 
\cline{2-8}
& \multirow{4}{*}{$\frac{1}{8}$} & 64 & 1.12634(12) & 1.4345(46) & -1.003(20) & -0.099(14) & 0.32 \\ 
& \multirow{4}{*}{} & 96 & 1.12635(12) & 1.4347(47) & -1.007(22) & -0.098(14) & 0.3 \\ 
& \multirow{4}{*}{} & 128 & 1.12637(13) & 1.4355(50) & -1.016(27) & -0.096(15) & 0.27 \\ 
& \multirow{4}{*}{} & 192 & 1.12637(16) & 1.4356(61) & -1.017(38) & -0.096(18) & 0.28 \\ 
\end{tabular}
\end{ruledtabular}
\label{tab:normal_po_S}
\end{table*}

\begin{table*}
\caption{
Same as Table \ref{tab:normal_po_S} for the normal UC realized with $(+,+)$ BCs.
}
\begin{ruledtabular}
\begin{tabular}{cclwmmqm}
\zmin & \zLmax & \Lmin & \multicolumn{1}{c}{$A_\phi$} & \multicolumn{1}{c}{$z_0$} & \multicolumn{1}{c}{$B_\phi$} & \multicolumn{1}{c}{$C$} & \multicolumn{1}{c}{\chidof} \\
\hline\multirow{12}{*}{4} & \multirow{6}{*}{$\frac{1}{4}$} & 32 & 1.126914(62) & 1.4536(16) & 1.3235(33) & -0.0535(30) & 16.57 \\ 
& \multirow{6}{*}{} & 48 & 1.126823(61) & 1.4512(16) & 1.3415(37) & -0.0578(29) & 9.56 \\ 
& \multirow{6}{*}{} & 64 & 1.126755(60) & 1.4492(15) & 1.3546(40) & -0.0616(29) & 6.62 \\ 
& \multirow{6}{*}{} & 96 & 1.126672(60) & 1.4468(15) & 1.3714(43) & -0.0663(28) & 3.28 \\ 
& \multirow{6}{*}{} & 128 & 1.126632(59) & 1.4456(15) & 1.3817(53) & -0.0688(28) & 2.34 \\ 
& \multirow{6}{*}{} & 192 & 1.126607(58) & 1.4447(15) & 1.3908(62) & -0.0704(27) & 1.61 \\ 
\cline{2-8}
 & \multirow{6}{*}{$\frac{1}{8}$} & 32 & 1.126733(56) & 1.4476(13) & 1.285(12) & -0.0657(25) & 2.77 \\ 
& \multirow{6}{*}{} & 48 & 1.126711(57) & 1.4472(14) & 1.304(14) & -0.0664(25) & 2.22 \\ 
& \multirow{6}{*}{} & 64 & 1.126690(57) & 1.4467(14) & 1.319(16) & -0.0671(25) & 2.02 \\ 
& \multirow{6}{*}{} & 96 & 1.126664(60) & 1.4461(14) & 1.338(20) & -0.0682(26) & 1.83 \\ 
& \multirow{6}{*}{} & 128 & 1.126663(63) & 1.4460(15) & 1.339(23) & -0.0683(27) & 1.95 \\ 
& \multirow{6}{*}{} & 192 & 1.126647(65) & 1.4456(16) & 1.353(27) & -0.0690(29) & 2.11 \\ 
\hline
\hline
\multirow{11}{*}{6} & \multirow{6}{*}{$\frac{1}{4}$} & 32 & 1.127119(90) & 1.4645(32) & 1.3191(35) & -0.0181(83) & 14.67 \\ 
& \multirow{6}{*}{} & 48 & 1.127024(90) & 1.4616(32) & 1.3348(39) & -0.0245(82) & 8.58 \\ 
& \multirow{6}{*}{} & 64 & 1.126931(89) & 1.4582(31) & 1.3479(42) & -0.0329(82) & 5.9 \\ 
& \multirow{6}{*}{} & 96 & 1.126798(89) & 1.4531(31) & 1.3658(46) & -0.0462(82) & 2.88 \\ 
& \multirow{6}{*}{} & 128 & 1.126726(89) & 1.4502(31) & 1.3769(56) & -0.0542(81) & 2.09 \\ 
& \multirow{6}{*}{} & 192 & 1.126668(89) & 1.4478(31) & 1.3872(65) & -0.0610(81) & 1.44 \\ 
\cline{2-8}
& \multirow{5}{*}{$\frac{1}{8}$} & 48 & 1.126789(84) & 1.4511(28) & 1.306(15) & -0.0542(69) & 1.77 \\ 
& \multirow{5}{*}{} & 64 & 1.126776(85) & 1.4508(28) & 1.314(17) & -0.0547(70) & 1.65 \\ 
& \multirow{5}{*}{} & 96 & 1.126751(89) & 1.4500(29) & 1.329(21) & -0.0562(72) & 1.51 \\ 
& \multirow{5}{*}{} & 128 & 1.126755(96) & 1.4502(31) & 1.327(25) & -0.0559(76) & 1.61 \\ 
& \multirow{5}{*}{} & 192 & 1.12674(11) & 1.4497(35) & 1.335(32) & -0.0571(88) & 1.75 \\ 
\hline
\hline
\multirow{10}{*}{8} & \multirow{6}{*}{$\frac{1}{4}$} & 32 & 1.12727(12) & 1.4758(53) & 1.3218(41) & 0.031(17) & 11.1 \\ 
& \multirow{6}{*}{} & 48 & 1.12722(12) & 1.4744(53) & 1.3302(42) & 0.029(17) & 7.73 \\ 
& \multirow{6}{*}{} & 64 & 1.12712(12) & 1.4705(53) & 1.3423(46) & 0.018(17) & 5.23 \\ 
& \multirow{6}{*}{} & 96 & 1.12695(12) & 1.4629(54) & 1.3605(51) & -0.007(18) & 2.44 \\ 
& \multirow{6}{*}{} & 128 & 1.12685(12) & 1.4576(55) & 1.3719(63) & -0.024(18) & 1.79 \\ 
& \multirow{6}{*}{} & 192 & 1.12674(13) & 1.4524(56) & 1.3835(74) & -0.043(18) & 1.23 \\ 
\cline{2-8}
& \multirow{4}{*}{$\frac{1}{8}$} & 64 & 1.12686(12) & 1.4557(49) & 1.315(19) & -0.035(15) & 1.28 \\ 
& \multirow{4}{*}{} & 96 & 1.12684(12) & 1.4554(49) & 1.323(21) & -0.036(15) & 1.19 \\ 
& \multirow{4}{*}{} & 128 & 1.12685(13) & 1.4555(52) & 1.321(26) & -0.036(16) & 1.25 \\ 
& \multirow{4}{*}{} & 192 & 1.12683(15) & 1.4549(63) & 1.326(36) & -0.038(19) & 1.37 \\ 
\end{tabular}
\end{ruledtabular}
\label{tab:normal_pp_S}
\end{table*}

\begin{table*}
\caption{
Same as Table \ref{tab:normal_pp_S}, fixing $C=0$.
}
\begin{ruledtabular}
\begin{tabular}{cclwmqm}
\zmin & \zLmax & \Lmin & \multicolumn{1}{c}{$A_\phi$} & \multicolumn{1}{c}{$z_0$} & \multicolumn{1}{c}{$B_\phi$} & \multicolumn{1}{c}{\chidof} \\
\hline\multirow{12}{*}{4} & \multirow{6}{*}{$\frac{1}{4}$} & 32 & 1.127577(29) & 1.47748(28) & 1.2932(36) & 32.31 \\ 
& \multirow{6}{*}{} & 48 & 1.127548(29) & 1.47711(28) & 1.3056(39) & 28.09 \\ 
& \multirow{6}{*}{} & 64 & 1.127534(28) & 1.47694(27) & 1.3121(43) & 27.79 \\ 
& \multirow{6}{*}{} & 96 & 1.127522(28) & 1.47677(26) & 1.3189(47) & 28.12 \\ 
& \multirow{6}{*}{} & 128 & 1.127517(28) & 1.47672(26) & 1.3201(57) & 30.08 \\ 
& \multirow{6}{*}{} & 192 & 1.127510(28) & 1.47665(26) & 1.3215(67) & 33.11 \\ 
\cline{2-7}
 & \multirow{6}{*}{$\frac{1}{8}$} & 32 & 1.127647(27) & 1.47810(25) & 1.163(12) & 44.98 \\ 
& \multirow{6}{*}{} & 48 & 1.127645(27) & 1.47807(25) & 1.166(13) & 45.33 \\ 
& \multirow{6}{*}{} & 64 & 1.127655(26) & 1.47818(24) & 1.150(15) & 45.91 \\ 
& \multirow{6}{*}{} & 96 & 1.127674(27) & 1.47838(24) & 1.120(18) & 46.49 \\ 
& \multirow{6}{*}{} & 128 & 1.127704(27) & 1.47874(25) & 1.060(20) & 45.79 \\ 
& \multirow{6}{*}{} & 192 & 1.127723(27) & 1.47898(24) & 0.998(23) & 46.37 \\ 
\hline
\hline
\multirow{11}{*}{6} & \multirow{6}{*}{$\frac{1}{4}$} & 32 & 1.127242(40) & 1.47041(54) & 1.3145(34) & 14.89 \\ 
& \multirow{6}{*}{} & 48 & 1.127193(40) & 1.46959(54) & 1.3282(37) & 9.07 \\ 
& \multirow{6}{*}{} & 64 & 1.127159(39) & 1.46903(53) & 1.3382(40) & 6.82 \\ 
& \multirow{6}{*}{} & 96 & 1.127126(38) & 1.46845(52) & 1.3498(45) & 4.72 \\ 
& \multirow{6}{*}{} & 128 & 1.127115(38) & 1.46827(51) & 1.3550(54) & 4.66 \\ 
& \multirow{6}{*}{} & 192 & 1.127108(38) & 1.46814(50) & 1.3593(64) & 4.85 \\ 
\cline{2-7}
& \multirow{5}{*}{$\frac{1}{8}$} & 48 & 1.127218(37) & 1.46971(48) & 1.260(14) & 6.1 \\ 
& \multirow{5}{*}{} & 64 & 1.127212(37) & 1.46962(48) & 1.266(15) & 6.1 \\ 
& \multirow{5}{*}{} & 96 & 1.127208(38) & 1.46956(50) & 1.269(19) & 6.23 \\ 
& \multirow{5}{*}{} & 128 & 1.127227(39) & 1.46985(51) & 1.247(21) & 6.19 \\ 
& \multirow{5}{*}{} & 192 & 1.127244(39) & 1.47012(50) & 1.222(24) & 6.29 \\ 
\hline
\hline
\multirow{10}{*}{8} & \multirow{6}{*}{$\frac{1}{4}$} & 32 & 1.127132(52) & 1.46772(91) & 1.3264(34) & 11.26 \\ 
& \multirow{6}{*}{} & 48 & 1.127093(52) & 1.46694(91) & 1.3344(36) & 7.87 \\ 
& \multirow{6}{*}{} & 64 & 1.127043(51) & 1.46589(90) & 1.3451(39) & 5.28 \\ 
& \multirow{6}{*}{} & 96 & 1.126983(50) & 1.46461(88) & 1.3593(43) & 2.44 \\ 
& \multirow{6}{*}{} & 128 & 1.126960(50) & 1.46411(87) & 1.3666(53) & 1.91 \\ 
& \multirow{6}{*}{} & 192 & 1.126947(49) & 1.46379(86) & 1.3727(62) & 1.61 \\ 
\cline{2-7}
& \multirow{4}{*}{$\frac{1}{8}$} & 64 & 1.127042(49) & 1.46546(82) & 1.300(17) & 1.69 \\ 
& \multirow{4}{*}{} & 96 & 1.127033(51) & 1.46529(85) & 1.307(19) & 1.63 \\ 
& \multirow{4}{*}{} & 128 & 1.127042(54) & 1.46545(89) & 1.300(22) & 1.68 \\ 
& \multirow{4}{*}{} & 192 & 1.127050(55) & 1.46561(93) & 1.293(26) & 1.83 \\ 
\end{tabular}
\end{ruledtabular}
\label{tab:normal_pp_S_noC}
\end{table*}

\end{document}